\documentclass[12pt]{article}

\usepackage{amsmath}
\usepackage{multirow}
\usepackage{amsfonts}
\usepackage{amssymb}
\usepackage{amscd}

\def\hybrid{\topmargin -20pt    \oddsidemargin 0pt
        \headheight 0pt \headsep 0pt
        \textwidth 6.25in       
        \textheight 9.5in       
        \marginparwidth .875in
        \parskip 5pt plus 1pt   \jot = 1.5ex}

\hybrid

\numberwithin{equation}{section}
\numberwithin{table}{section}

\newcommand{\beq}{\begin{equation}}
\newcommand{\eeq}{\end{equation}}
\newcommand{\bi}{\begin{itemize}}
\newcommand{\ei}{\end{itemize}}
\newcommand{\bea}{\begin{eqnarray}}
\newcommand{\eea}{\end{eqnarray}}
\newcommand{\ba}{\begin{array}}
\newcommand{\ea}{\end{array}}
\newcommand{\bt}{\begin{tabular}}
\newcommand{\et}{\end{tabular}}
\newcommand{\bc}{\begin{center}}
\newcommand{\ec}{\end{center}}

\newcommand{\dstyle}{\displaystyle}

\newcommand{\Gx}{\Gamma}

\newcommand{\cN}{\mathcal{N}}

\newcommand{\cF}{\mathcal{F}}

\newcommand{\cM}{\mathcal M}

\newcommand{\IM}{\textrm{Im} \,}

\newcommand{\cref}{{\bf [check ref]}}

\newcommand{\M}{M}

\newcommand{\Y}{Y}

\newcommand{\La}{\Theta}

\newcommand{\flux}{{\rm fl}}

\newcommand{\ef}{{\tilde e}}
\newcommand{\mf}{{\tilde m}}
\newcommand{\phione}{{\alpha}}
\newcommand{\phitwo}{{\beta}}

\newcommand{\pure}{\Phi}

\newcommand{\id}{\mathbf{1}}

\newcommand{\cP}{\mathcal{P}}

\newcommand{\dd}{\mathrm{d}}
\newcommand{\ee}{\mathrm{e}}
\newcommand{\ii}{\mathrm{i}}

\newcommand{\der}{\partial}

\newcommand{\bbR}{\mathbb{R}}
\newcommand{\bbC}{\mathbb{C}}

\DeclareMathOperator{\SL}{\mathit{SL}}
\DeclareMathOperator{\GL}{\mathit{GL}}
\DeclareMathOperator{\SU}{\mathit{SU}}
\DeclareMathOperator{\SO}{\mathit{SO}}
\DeclareMathOperator{\Symp}{\mathit{Sp}}
\DeclareMathOperator{\Spin}{\mathit{Spin}}

\DeclareMathOperator{\Int}{Int}

\newcommand{\rep}[1]{\mathbf{#1}}

\DeclareMathOperator{\re}{Re}
\DeclareMathOperator{\im}{Im}

\newcommand{\sym}[2]{\omega\!\left(#1,#2\right)}
\newcommand{\mukai}[2]{\left<{#1},{#2}\right>}
\newcommand{\revmukai}[2]{\left<{#2},{#1}\right>}

\newcommand{\Lodd}{\Lambda^\text{odd}}
\newcommand{\Leven}{\Lambda^\text{even}}
\newcommand{\Leo}{\Lambda^\text{even/odd}}

\newcommand{\chis}{\chi_\epsilon}
\newcommand{\chisb}{\bar{\chi}_\epsilon}
\newcommand{\psis}{\psi_\epsilon}
\newcommand{\chihs}{\hat{\chi}_\epsilon}

\newcommand{\Omegan}{\Omega_{\eta}}
\newcommand{\Omeganb}{\bar{\Omega}_{\eta}}
\newcommand{\cscale}{n}
\newcommand{\Lt}{\Lambda_\text{finite}}  
\newcommand{\Lteven}{\Lambda_\text{finite}^\text{even}}   
  
\newcommand{\Lteo}{\Lambda_\text{finite}^\text{even/odd}}  
\newcommand{\bJ}{b_J}
\newcommand{\br}{b_\rho}

\newcommand{\hfunc}{H}
\newcommand{\hfuncs}{H_\epsilon}
\newcommand{\hfuncY}{H_\Y}
\newcommand{\qs}{q_\epsilon}
\newcommand{\Us}{U_\epsilon}
\newcommand{\pures}{\pure_\epsilon}
\newcommand{\puresb}{\bar{\pure}_\epsilon}

\newcommand{\UtJ}{U^\text{finite}_J} 
\newcommand{\Utr}{U^\text{finite}_\rho} 
\newcommand{\MtJ}{\mathcal{M}^\text{finite}_J} 
\newcommand{\Mtr}{\mathcal{M}^\text{finite}_\rho} 
\newcommand{\Mt}{\mathcal{M}^\text{finite}}  

\begin{document}


\begin{titlepage}
\begin{center}

\hfill hep-th/0505264

\rightline{\small LPTENS-05/14}
\rightline{\small ZMP-HH/05-07}
\rightline{\small Imperial/TP/050401}
\vskip 1cm

{\Large \bf Hitchin Functionals in $N=2$ Supergravity}

\vskip 0.8cm

{\bf Mariana Gra{\~n}a$^{a,b}$, Jan Louis$^{c,d}$ and
Daniel Waldram$^{e,f}$ }
\
\vskip 0.6cm
{}$^{a}${\em Laboratoire de Physique Th\'eorique de l'Ecole Normale
Sup\'erieure\\
24 rue Lhomond, 75231 Paris Cedex, France}\\
{\tt mariana@lpt.ens.fr}
\vskip 0.3cm

{}$^{b}${\em Centre de Physique Th{\'e}orique, Ecole
Polytechnique \\
91128 Palaiseau Cedex, France}
\vskip 0.3cm

{}$^{c}${\em II. Institut f{\"u}r Theoretische Physik der Universit{\"a}t Hamburg\\
Luruper Chaussee 149,  D-22761 Hamburg, Germany}\\
 {\tt jan.louis@desy.de} 
\vskip 0.3cm

{}$^{d}${\em Zentrum f\"ur Mathematische Physik, 
Universit\"at Hamburg,\\
Bundesstrasse 55, D-20146 Hamburg}

\vskip 0.3cm

{}$^{e}${\em Blackett Laboratory, Imperial College\\
London, SW7 2AZ, U.K.}\\
 {\tt d.waldram@imperial.ac.uk}

\vskip 0.3cm

{}$^{f}${\em The Institute for Mathematical Sciences, Imperial College\\
London, SW7 2AZ, U.K.}\\

\end{center}

\vskip 1cm

\begin{center} {\bf ABSTRACT } \end{center}

\noindent

We consider type II string theory in space-time backgrounds which
admit eight supercharges. Such backgrounds are characterized
by the existence of a (generically non-integrable) generalized
$\SU(3)\times\SU(3)$ structure. We demonstrate how the corresponding
ten-dimensional supergravity theories can in part be rewritten using 
generalised $O(6,6)$-covariant fields, in a form that strongly
resembles that of four-dimensional $N=2$ supergravity, and precisely
coincides with such after an appropriate Kaluza--Klein
reduction. Specifically we demonstrate that the NS sector admits a
special K\"ahler geometry with K\"ahler potentials given by the
Hitchin functionals. Furthermore we explicitly compute the $N=2$
version  of the superpotential from the transformation law of the
gravitinos, and find its $N=1$ counterpart.

\vfill

\noindent May 2005

\end{titlepage}

\tableofcontents

\newpage

\section{Introduction}


The interplay between supersymmetry and 
geometry has been very fruitful in the past. 
For example, compactifications of ten-dimensional type II
supergravities on Calabi--Yau threefolds $Y$  preserve eight
supercharges and yield four-dimensional $N=2$ (ungauged)
supergravities as effective low energy field theories~\cite{CFG,FS,BCF,BGHL}. 
The spectrum and couplings of these $N=2$ supergravities are in turn
determined  by geometrical  (and topological)  
properties of the Calabi--Yau manifolds.
Supersymmetry strongly constrains the couplings 
and thus also constrains the Calabi--Yau geometry.  
For example,  it implies that the moduli space of metric
deformations of a Calabi--Yau manifold is the
product of two special K\"ahler manifold characterized by two
holomorphic prepotentials \cite{CFG,Strominger,DKL,CdO}. The
Calabi--Yau moduli space indeed satisfies this property and
furthermore one can use geometrical methods together with mirror
symmetry to compute both prepotentials exactly~\cite{CDGP,HKT}.

Expanding on earlier work in refs.~\cite{Rocek,StromingerT,hullT},
there has recently been much interest in a generalized class of
backgrounds where the Calabi--Yau manifold is
replaced by a manifold $\Y$ which is no longer
Ricci-flat~\cite{Vafa}--\cite{GMPT2}. One way such generalized
compactifications arise is when localized sources (D-branes,
orientifold  planes)  and/or background fluxes are present and the
solution of the equations of motion forces the geometry to back-react
to the additional background energy density. A certain class of
manifolds, called `half-flat manifolds' \cite{CS}, also appeared as
mirror symmetric backgrounds of type II Calabi--Yau compactifications
with background fluxes~\cite{GLMW,GM,FMT}.  

Within this generalized set-up one is particularly interested in 
backgrounds which continue to preserve some of the supercharges 
or more  generally where a number of supercurrents exist
but the associated supercharges are spontaneously broken. 
The latter case includes examples which do not
satisfy the equations of motion, such as the classical example of a
Calabi--Yau manifold with generic background fluxes. 
In either case, the existence of the supercurrents requires that a set
of spinors are globally well defined on the manifold $\Y$ which in
turn implies that the structure group has to be reduced. In the
mathematical literature manifolds with a reduced structure group $G$
are called manifolds with
$G$-structure~\cite{salamonb,joyce}. Generically $G$ does not coincide 
with the holonomy group since the spinors are not necessarily
covariantly constant with respect to the Levi-Civita connection. The
degree to which they fail to be covariantly constant is measured by a
quantity known as the intrinsic torsion and can be used to classify
the $G$-structure.  

{}From a particle physics point of view preserving the minimal
amount of supersymmetry is the most interesting case. 
On a six-manifold the existence of a single  globally
defined spinor $\eta$ requires the reduction of the structure group
from  $\Spin(6)$ to $\SU(3)$  and therefore manifolds with 
$\SU(3)$ structure play a special role. They can be characterized by
the invariant spinor on the manifold or, more conveniently, by a
real two-form $J$ and a complex three-form $\Omega$. Since $\eta$ is
not covariantly constant neither $J$ nor $\Omega$ are closed. Instead
$\dd J$ and $\dd \Omega$ decompose into $SU(3)$ representations. These
define the intrinsic torsion and can be used to classify the different
$SU(3)$ structures~\cite{CS}. For Calabi--Yau threefolds $J$ and
$\Omega$ are closed, $\eta$ is covariantly constant and the holonomy
group is $\SU(3)$.

Here, we will focus on type II supergravities which have $N=2$
supersymmetry in ten space-time dimensions. 
Decomposing the spinor representation in ten dimensions under
$\Spin(1,3)\times\Spin(6)$ and requiring
$N=2$ supersymmetry in four dimensions implies that there are two
non-vanishing spinors on $\Y$, one for each of the original
ten-dimensional spinors. Each defines an $\SU(3)$ structure. 
Locally, the two $SU(3)$ structures define an $SU(2)$ structure,
which survives globally as an $SU(2)$ structure if the spinors
never become parallel. If the spinors are always parallel we
just have a single $\SU(3)$ structure. 

One way to characterize this structure mathematically is 
in terms of ``generalised geometry'', first introduced by
Hitchin~\cite{GCY}. One considers the sum of the tangent and 
cotangent bundle of $\Y$, $T\Y \oplus T^*\Y$ on which there is a
natural $O(6,6)$ structure. The two six-dimensional spinors transform
under a $\Spin(6)\times\Spin(6)$ subgroup defined by the metric and NS
$B$-field, and being globally defined, imply that the structure group
of $T\Y\oplus T^*\Y$ actually reduces to
$\SU(3)\times\SU(3)$~\cite{JW} (see~\cite{Gualtieri} for the original,
related discussion of $U(n)\times U(n)$ structures). In this formulation, the
$\SU(3)\times\SU(3)$ structure can be defined by a sum of odd forms
$\pure^-$ and a sum of even forms $\pure^+$, each built out of spinor
bilinears~\cite{GMPT,JW} (see also~\cite{Witt} for the 
construction in the case of $G_2 \times G_2$ structures). 
From the point of view of the $T\Y\oplus T^*\Y$ bundle these forms
correspond to a pair of $\Spin(6,6)$
spinors~\cite{3form,HitchinHF,GCY}.

Since we are interested in backgrounds with supercurrents but, in
general, spontaneously broken supersymmetry, we do not require the
$\SU(3)$ structures to be integrable. Enforcing preserved
supersymmetry (and the equations of motion) would impose integrability
constrains~\cite{GMPT,JW,GMPT2}. The geometric structures used  
throughout this paper are therefore ``almost'' (or not necessarily
integrable) structures. However, in order to avoid cumbersome wording,
we will typically drop the ``almost'' when referring to them. 

The description of backgrounds in terms of $\SU(3)$ structures and the
generalization to $\SU(3)\times\SU(3)$ structures has also recently
played an important role in topological string theories. In
particular, it has been argued~\cite{Dijkgraaf,GS,Nekrasov} that the
target space theory of the A and B model topological strings can be
defined in terms of a functional of the structures $J$ or $\Omega$
first considered by Hitchin~\cite{3form,HitchinHF,GCY}. More
generally~\cite{PW} one must consider the corresponding functional for
the $\Spin(6,6)$ spinors $\pure^\pm$. Similarly it has been
possible~\cite{KL,Zucchini} to generalize the notion of topological
strings away from backgrounds with $\SU(3)$-structure (such as
Calabi--Yau manifolds) to more general spaces with
$\SU(3)\times\SU(3)$ structure, again using the spinors $\pure^\pm$.

Returning to the physical string, for Calabi--Yau compactifications the
$N=2$ low energy effective action in four space-time dimensions can be
derived by a standard Kaluza--Klein reduction where only the massless
modes corresponding to harmonic forms on $Y$ are
kept~\cite{CFG,FS,BCF,BGHL}. This procedure is valid whenever $Y$ is
large and the supergravity approximation can be used reliably. In the
presence of background fluxes the same method has been applied for
example in refs.~\cite{PS}-\cite{KK}.
One chooses the fluxes to be small, the compactification manifold to
be large and hence consistently neglects the back-reaction of the
geometry.  One finds that the kinetic terms are unaltered and the flux
parameters appear as gauge couplings and/or mass parameters which turn
the supergravity into a gauged or massive supergravity. However, when
dealing with manifolds with non-integrable $SU(3)$ structure, this
procedure is a bit more tricky since generically it is harder to
specify in what sense one is making a small deformation. For instance,
turning on $H$-flux on a Calabi--Yau manifold can map to a change in
topology of the mirror manifold. Thus one cannot treat the intrinsic
torsion easily as a simple deformation of the supergravity as was done
for the fluxes. 

The goal of this paper is to study type II supergravity in generic
backgrounds with $\SU(3)$ (the case where the two spinors are always
parallel) or, more generally, $\SU(3)\times\SU(3)$ structure. 
Our motivation is to define a `rule' for deriving the
low-energy four-dimensional effective theory and to uncover the role
of the torsion in supergravity. As in Calabi--Yau compactifications
this might lead to interesting insights into the interplay of geometry and
supersymmetry of the effective theory. However, we begin with a
more general set-up. We do not immediately confine our interest to the
low energy effective action or performing a Kaluza--Klein reduction.
This leads us to a reformulation of the full ten-dimensional theory,
abandoning manifest ten-dimensional Lorentz invariance, but with
bosonic fields transforming in $\Spin(1,3)\times O(6,6)$
multiplets. This is similar to and inspired by the approach pioneered
in ref.~\cite{deWN}, which considered a related reformulation of
eleven-dimensional supergravity. Although we provide no direct
evidence, we expect the reformulation has a local
$\Spin(1,3)\times\SU(3)\times\SU(3)$ symmetry. 
More specifically, we first demand that the tangent space of the
ten-dimensional background is a direct sum  $T^{1,3} \oplus F$ 
where $T^{1,3}$ is a $\Spin(1,3)$ bundle while $F$ is a $\Spin(6)$
bundle. Then we further require that structure group of $F$ reduces
admitting an $\SU(3)$ structure. In fact we also consider the more
general situation where the sum of the tangent plus the cotangent
bundle admits $\SU(3)\times\SU(3)$ structure. In both cases eight of
the original 32 supercharges are singled out and we can rewrite
the ten-dimensional supergravity with 32 supercharges in a form as if
it had only eight supercharges. 
 
In this framework,  the supermultiplet structure and
action follow the form of four-dimensional $N=2$ supergravity although
the theory remains fully ten-dimensional. In particular, concentrating
on the bosonic fields which are scalars under $\Spin(1,3)$, we define
a space of (not necessarily integrable) $\SU(3)\times\SU(3)$
structures and show, following refs.~\cite{3form,HitchinHF,GCY}, that
it admits a special K\"ahler geometry with a K\"ahler potential given by a
Hitchin functional. Restricting to the particular case 
of a single $\SU(3)$ structure, we furthermore rewrite
the supersymmetry transformation law of the eight gravitinos in a form
analogous to the transformation law of the four-dimensional $N=2$
gravitinos. This allows us to read off the three `Killing
prepotentials' or momentum maps $\cP^x, x=1,2,3$ which are the $N=2$
equivalent of the superpotential and the $D$-term. In this
ten-dimensional theory they turn out to be determined by the
background fluxes and the intrinsic torsion. 

In the same spirit we can continue the decomposition keeping only
four supercharges. In this way  we find the most general $N=1$
superpotential induced by the fluxes and torsion. We find that this
generalized superpotential contains all previously known cases
in appropriate limits when either torsion, NS or RR  fluxes are set to
zero. For example, in the torsionless case we recover the
Gukov-Taylor-Vafa-Witten superpotential \cite{GVW,gukov,TV}. 

After having rewritten the ten-dimensional theory in an `$N=2$ form'
it is straightforward to perform a KK-reduction. We choose the
background to be a product $\M^{1,3} \times \Y$
where $\M^{1,3}$ is a four-dimensional manifold with Minkowskian
signature while $\Y$ is a compact manifold with
$SU(3)$ structure. (The more general case of compactifications 
with $\SU(3)\times\SU(3)$ structure will be discussed elsewhere.)
In the KK-reduction one conventionally keeps the light modes and
integrates out the heavy ones. However, backgrounds with a generic
$\Y$ do not necessarily have a flat Minkowskian ground state and the
distinction between heavy and light is not
straightforward.
Therefore we do not specify the precise form of the truncation, which
would depend on the particular choice of background, but
instead leave it generic, extracting the set of conditions that such a
reduction must satisfy to be self-consistent.
The truncation is defined by extracting from the infinite tower of
KK-modes only a finite subset. However, we impose one further
condition in that we only keep the two gravitini in the gravitational
multiplet but project out all gravitini which reside in their own
(massive) spin-$\frac{3}{2}$ multiplets. This ensures that the resulting
low effective action contains apart form the gravitational
multiplet only $N=2$ vector, tensor and hypermultiplets.

Once the ten-dimensional spectrum is truncated the gauge invariance
of the original ten-dimensional theory is no longer automatically maintained.
Instead, as we will see, gauge invariance
imposes additional constraints on the truncation 
which also have been observed in~\cite{DFTV}. 
Imposing these constraints, the $N=2$ action takes a  standard form 
\cite{dWvP,N=2review} --  possibly with massive
tensor multiplets \cite{LM,DSV,Kuzenko}.
This enables us to discuss in detail the supergravity/geometry
correspondence. We find, as expected, that 
the torsion (as well as the fluxes) deform the $N=2$ supergravity
and turn it into a gauged or massive supergravity. 
The gauge charges and mass parameters are directly related to the
fluxes and torsion and we derive the precise relationship by computing
the supersymmetry transformations of the gravitino. 

This paper is organized into two main sections.
In section~2 we discuss the reformulation of
the ten-dimensional 
type II supergravity theory in terms of $N=2$--like
structures while in section~\ref{d=4N=2} we perform the Kaluza--Klein
reduction and compute some of the couplings in the low energy
effective theory. More specifically in \ref{compact} we first 
show that eight linearly realized supercharges require 
that the theory has a  
$\Spin(1,3)\times\SU(3)\times\SU(3)$ structure. 
In section~\ref{fields}  we then show how the ten-dimensional fields
decompose into $N=2$ multiplets for the case of a single $SU(3)$ structure.
After reviewing a few facts about $\SU(3)$ structures, 
we give the part of the action for the deformations of the
NS fields in section~\ref{kinetic}. In section~\ref{SKM} we
show that their kinetic
terms form a product of two 
special K\"ahler geometries in exact analogy with the moduli space of
Calabi--Yau compactifications.  Furthermore the K\"ahler potential is
determined by the sum of two Hitchin functionals both of which can be
derived from a universal expression given in terms of a pure 
$Spin(6,6)$ spinor \cite{3form,GCY}. 
In section~\ref{N=2W} we compute the scalar part of the 
supersymmetry transformations of the gravitinos and determine 
the ten-dimensional analog of the Killing prepotential $\cP^x$. 
By an appropriate further reduction we
compute the $N=1$ superpotential in section~\ref{N=1W}.
In section~\ref{d=4N=2} we perform the KK-truncation.  We first 
define the `rules' for the reduction in \ref{defreduction}. We
project out all
${\bf 3} \oplus {\bf \bar 3}$ representations and then impose local
$p$-form gauge invariance on the remaining spectrum. 
In \ref{NSred} we discuss the reduction of the common NS-sector and
show that the resulting K\"ahler potentials precisely coincide with
the analogous K\"ahler potentials of Calabi--Yau manifolds.
In \ref{IIAreduc} and \ref{IIBreduc} we perform the reduction of the
RR-sector for type IIA and type IIB. In \ref{sugra} we 
check that the `proper' $N=2$
Killing prepotential $\cP^x$ obtained by truncation from its
higher-dimensional `father' agrees with the generic 
form dictated by $N=2$ gauged supergravity.
Finally in \ref{mirror} we briefly discuss mirror symmetry for these
generalized compactifications and \ref{conclusions} contains our conclusions.


\section{Type II supergravity with $\SU(3)\times\SU(3)$ structure} 


The goal in this section is to understand some of the details of how
we can reformulate the ten-dimensional type II supergravity theory in
terms structures analogous to $N=2$ four-dimensional supergravity. In
doing so we lose manifest $\Spin(1,9)$ Lorentz symmetry, and instead
arrange the fields in $\Spin(1,3)\times O(6,6)$ multiplets. We will
concentrate on the scalar field part of the theory, that is multiplets
which contain fields which are singlets under $\Spin(1,3)$. In an
$N=2$ language these are the vector, tensor and hypermultiplets. In
particular, we will show that there are special K\"ahler geometries on
the spaces of scalar fields describing their kinetic
terms. Furthermore, we will show how the ten-dimensional theory gauges
a set of isometries on these spaces, described by a set of Killing
prepotentials again just as in four-dimensional $N=2$
supergravity. We find that all these objects can be written in a
simple way in terms of generalised geometrical structures, invariant
under $O(6,6)$ transformations. In particular, the K\"ahler potential
of the special K\"ahler geometry is given by the Hitchin functional. 

Let us start by discussing the relation between rewriting the theory
in terms of eight linearly realized supercharges ($N=2$) and the
existence of generalised $\SU(3)\times\SU(3)$ structures.

\subsection{$N=2$ and $\SU(3)\times\SU(3)$ structures}
\label{compact}

\subsubsection{Effective theories and $G$-structures}

One motivation for this paper is to consider the general low-energy
gauged supergravity theory that arises when type II string theory (or
rather type II supergravity) is compactified on the space-time
background 
\beq\label{spacetime}
   \M^{1,9} =  \M^{1,3} \times \Y\ .
\eeq
Here $\M^{1,3}$ is the four-dimensional, physical space-time while
$\Y$ is a six-dimensional compact manifold.\footnote{In this paper
we do not consider the possibility of a warped background but
leave the study of this class of compactification to a separate publication.}
The product structure of the space-time background~\eqref{spacetime} 
implies a decomposition of the Lorentz group $\Spin(1,9)\supset
\Spin(1,3)\times\Spin(6)$ and an associated decomposition of the
spinor representation  ${\bf 16}\in\Spin(1,9)$ according to 
${\bf 16}\to ({\bf 2},{\bf 4}) \oplus ({\bf \bar 2},{\bf \bar 4})$.

We are interested in the situation where the effective theory on
$\M^{1,3}$ has the minimal $N=2$ supersymmetry. In other words, we
need to single out eight particular type II supersymmetries which
descend to the effective theory. 
For type IIA we start with two supersymmetry parameters of opposite
ten-dimensional chirality. Using a standard decomposition of the
ten-dimensional gamma matrices $\Gamma^M = (\Gamma^\mu,\Gamma^m)$ as
\begin{equation}\label{gammadef}
   \Gamma^\mu=\gamma^\mu\otimes\id\ ,\quad \mu=0,1,2,3\ ,\qquad
   \Gamma^m=\gamma_5\otimes\gamma^m\ ,\quad m=1,\ldots,6\ , 
\end{equation}
where
$\gamma_5=\ii\gamma^0\gamma^1\gamma^2\gamma^3$, we can write 
\begin{equation}\begin{aligned}
\label{decompepsilon}
\varepsilon^1_{\text{IIA}} &= \varepsilon_+^1 \otimes \eta^1_+
   + \varepsilon_-^1 \otimes \eta^1_- \ , \\
\varepsilon^2_{\text{IIA}} &= \varepsilon_+^2 \otimes \eta^2_- 
   + \varepsilon_-^2 \otimes \eta^2_+ \ ,
\end{aligned}\end{equation}
where $\varepsilon_-^{1,2}=(\varepsilon_+^{1,2})^c$ and
$\eta^{1,2}_-=(\eta_+^{1,2})^c$. (Here as usual
$\eta^c=D\eta^*$, where $D$ is the intertwiner giving
$-\gamma^{m*}=D^{-1}\gamma^m D$. We also have $\bar{\eta}=\eta^\dag A$, 
where $\gamma^{m\dag}=A\gamma^m A^{-1}$. By a slight abuse of notation
we use plus and minus to indicate both four-dimensional and
six-dimensional chiralities.) For type IIB both spinors have negative
chirality resulting in the decomposition
\beq
\label{decompepIIB}
 \varepsilon^A_{\text{IIB}} = \varepsilon_+^A \otimes \eta^A_- 
   + \varepsilon_-^A \otimes \eta^A_+ \ , \qquad A=1,2 \ .
\eeq
In each case we have a pair of spinors
$\varepsilon_+^A$ in $\M^{1,3}$ parameterizing the $N=2$ supersymmetries. In
addition, we have two spinors $\eta^A_+$ of $\Y$  fixing
precisely which of the ten-dimensional supersymmetries descend to four
dimensions. Note that generically these can be different for the two
ten-dimensional supersymmetry parameters $\varepsilon^A$. 

For such a reduction to work, the spinors $\eta_+^A$ must be
globally defined and nowhere vanishing on $\Y$ and hence the
structure group of the tangent space of $\Y$ has to reduce. Consider one
such global spinor. It has to transform as a singlet under the
structure group. Decomposing under $\SU(3)\subset\Spin(6)$, the
complex spinor representation splits as
$\rep{4}=\rep{3}\oplus\rep{1}$. Thus if the structure group is
contained in $\SU(3)$ we indeed get a spinor singlet. Manifolds with
this property are known as `manifolds with $\SU(3)$ structure' in the
mathematical literature~\cite{CS}.  Since we do not
require the background to be supersymmetric, only that the
four-dimensional effective action has a set of $N=2$ supercurrents,
there are generically no differential conditions on the spinors
$\eta_+^A$. In the mathematical literature this means we have an
``almost'' or not necessarily integrable $\SU(3)$ structure. 

{}From eqs.~\eqref{decompepsilon} and \eqref{decompepIIB} we see that
in general we have a pair $\eta_+^A$ of such
spinors, each of which defines an $\SU(3)$ structure. 
In summary 
\begin{equation}
   \text{$d=4$, $N=2$ effective theory} \quad \Leftrightarrow \quad
   \text{$\Y$ admits a pair of $\SU(3)$ structures}\  .
\end{equation}
Locally the two spinors $\eta_+^1$ and $\eta_+^2$ span a
two-dimensional subspace of the four-dimensional space of positive
chirality $\Spin(6)\cong\SU(4)$ spinors. This space is invariant under
$\SU(2)\subset\SU(4)$ rotations, under which both spinors are
singlets. Thus locally the presence of two $\SU(3)$ structures
actually implies that we have an $\SU(2)$ structure. However, globally
there can be points where the spinors become parallel, and hence at
these points no $\SU(2)$ structure is defined. The extreme case where
the two spinors are parallel everywhere is allowed, and in this case
the  two SU(3) structures coincide, leading to a single  SU(3)
structure.

As we discuss in more detail below, a special case of a supersymmetric
compactification is where $\Y$ is a Calabi--Yau manifold and
$\eta^1_+=\eta^2_+$. In this case, in deriving the low-energy
effective theory, one keeps only the massless modes and disregards all
heavy Kaluza--Klein modes (together with all heavy string
states). However, for compactifications on generic manifolds with
a pair of $\SU(3)$ structures the distinction between heavy and light
modes is not straightforward. It appears that we have to define a
`rule' for the reduction to decide which modes we keep in the
four-dimensional effective action and which modes we truncate away. In
fact, as we now discuss, we can actually start by doing something more
general, where we do not truncate the theory at all. 

\subsubsection{A $d=10$ reformulation and generalized structures}


The previous discussion was based on the assumption that we had a
product manifold~\eqref{spacetime}. However it is not really
necessary to make such an assumption. In general, if we break the local
$\Spin(1,9)$-invariance one can always rewrite the full $d=10$ type II
supergravity theory  as though it were a theory with only eight
supercharges. The structure of the theory is then analogous to
${N}=2$ in four dimensions, but no Kaluza--Klein expansion is
made and instead we work in ten space-time dimensions keeping all the
degrees of freedom. A similar reorganization of eleven-dimensional
supergravity was done in ref.~\cite{deWN} in order to understand the
origin of hidden symmetries in lower dimensional supergravities. 

More precisely, we require only that the ten-dimensional manifold
$\M^{1,9}$ admits a pair of $\SU(3)$ structures. This
means that the tangent space $T\M^{1,9}$ splits as  
\begin{equation}
\label{Tdecomp}
   T\M^{1,9} = T^{1,3} \oplus F \ ,
\end{equation}
where $T^{1,3}$ is a real $\SO(1,3)$ vector bundle and $F$ is a
$\SO(6)$ vector bundle which admits  a pair of
$\SU(3)$ structures. In other words we have two different 
decompositions of the complex vector bundle $F_\bbC=F\otimes\bbC$,
that is 
\begin{equation}
   F_\bbC = E^1 \oplus \bar{E}^1 
      = E^2 \oplus \bar{E}^2  \ ,
\end{equation}
where each $E^A$ is a complex $\SU(3)$ vector bundle corresponding to the
$\SU(3)$ structure defined by $\eta_+^A$. Equivalently, recall that
the original type II theory is 
formulated on a supermanifold $M^{1,9|16+16}$ of bosonic dimension
$(1,9)$, with a manifest local $\SO(1,9)$ invariance and with the
Grassmann variables transforming as a pair of 16-dimensional spinor
representations. The requirement that we have a pair of
$\SU(3)$ structures means there is a
sub-supermanifold 
\begin{equation}
   N^{1,9|4+4} \subset \M^{1,9|16+16}
\end{equation}
still with bosonic dimension $(1,9)$, but now with only eight
Grassmann variables transforming as spinors of $\Spin(1,3)$ and
singlets of one or the other of the $\SU(3)$ groups. It is natural to
reformulate the $d=10$ supergravity in this sub-superspace. As such,
although the theory is still defined in ten-dimensions (though without
manifest $\SO(1,9)$ invariance), it will have structures analogous to
those of $d=4$, $N=2$ supergravity, such as special K\"ahler
moduli spaces and Killing prepotentials.

Let us now turn to a second key point. The pair of $\SU(3)$
structures are actually most naturally described as a single
``generalized structure'', a notion first introduced by
Hitchin~\cite{GCY}. One starts by considering the sum of the tangent
and cotangent bundles $T\Y\oplus T^*\Y$, or more generally $F\oplus
F^*$. If $v\in F$ and $\xi\in F^*$, one can see that there is a
natural $O(6,6)$ metric on this space, defined by  
\begin{equation}
\label{Odd-metric}
   \left(v+\xi,v+\xi\right) = \xi(v) \equiv \xi_m v^m\ , 
\end{equation}
which makes no reference to any additional structure (such as a
metric) on $F$. Note that the metric is invariant under the
diffeomorphism group $\GL(6,\bbR)\subset O(6,6)$ acting on
$F$ and $F^*$. The choice of metric $g$ and NS two-form $B$ can be
shown to correspond to fixing an $O(6)\times O(6)$ substructure. Given
this substructure, the two spinors $\eta_+^A$ transform separately
under the two different $\Spin(6)$ groups and the pair of $\SU(3)$
structures is actually equivalent to a (not necessarily integrable)
$\SU(3)\times\SU(3)$ structure on $F\oplus F^*$, as discussed in 
ref.~\cite{JW,Witt}. In summary, we have argued that 
\begin{equation}
   \text{\parbox{5cm}{\center{$N=2$-like reformulation of type II }}}
      \quad \Leftrightarrow \quad
   \text{\parbox{7cm}{\center{$F\oplus F^*$ admits a (not necessarily
         integrable) $\SU(3)\times\SU(3)$ structure}}}   
\end{equation}

We expect that this  $\SU(3)\times\SU(3)$ structure is actually
promoted to a local symmetry of the reformulated theory, in analogy
with~\cite{deWN}. Suppose, for instance we had compactified
on a torus $\Y=T^6$. It is then a familiar result that the low-energy
theory has a local $O(6)\times O(6)$ symmetry and a global
$O(6,6)$ symmetry, concomitant with the fact that the string theory
has a T-duality symmetry~\cite{Narain}. For instance the scalar
degrees of freedom coming from the internal metric and $B$-field
arrange themselves into a $O(6,6)/O(6)\times O(6)$ coset. The two
$\Spin(6)$ groups act separately on each spinor $\eta_+^A$. 
On $T^6$ any pair of constant spinors $(\eta^1_+,\eta^2_+)$ parameterizes a
pair of preserved supersymmetries in four dimensions and hence
compactification gives an $N=8$ effective theory. If we isolate a single
pair, this can be reformulated as an $N=2$ theory. The local $O(6)\times
O(6)$ symmetry should then reduce to those symmetries that leave the pair
invariant, namely a local $\SU(3)\times\SU(3)$ symmetry.
 Thus, generically we
expect that the effective theory on $N^{1,9|4+4}$ will have a local
$\Spin(1,3)\times\SU(3)\times\SU(3)$ symmetry. In what
follows we will however concentrate on the analog of the scalar sector
of the theory and do not provide any direct evidence for this local
symmetry. 

In order to simplify the discussion we will frequently 
specialize to the case where the
$\SU(3)\times\SU(3)$ structure defines a \emph{global} $\SU(3)$
structure (rather than some local $\SU(2)$ structure). In other words
we assume $\eta_+^1=\eta_+^2$. This is mostly for convenience and 
it also allows us to connect with the existing literature on
compactifications on spaces with $\SU(3)$-structure. We stress
nonetheless that from a supergravity perspective the natural framework
for $N=2$ theories and truncations is actually a generic
$\SU(3)\times\SU(3)$ structure.


\subsection{Field decompositions}
\label{fields}

Let us first look at the decomposition of the fields of type II
supergravities in the sub-supermanifold 
$N^{1,9|4+4}$. 
Let us use the same notion as the previous section even though we no
longer necessarily have a product manifold. A $\mu,\nu,\dots$
index denotes the representation of a field as a tensor of $T^{1,3}$
while a $m,n,\dots$ index denotes the representation as a tensor of
$F$. In addition, we specialize to the case of a global $\SU(3)$
structure. This means we can futher decompose the $F$-tensors into
into $\SU(3)$ representations.
 
The common NS sector contains the metric $g_{MN}$, an antisymmetric
tensor $B_{MN}$ and the dilaton $\phi$. Their decomposition into
$\SU(3)$ representation is displayed in table~\ref{NS}.
The notation $\rep{a}_\rep{b}$ denotes a field in the $\SU(3)$
representation $\rep{a}$ and with four-dimensional spin $\rep{b}$,
with $\rep{T}$ denoting an antisymmetric tensor or pseudo-scalar. The
representations are real except for $\rep{6}$ and $\rep{3}$ and their
conjugates. 
\begin{table}[h]
\begin{center}
\begin{tabular}{|c|c|l|} \hline
\rule[-0.3cm]{0cm}{0.8cm}
\multirow{3}{10mm}[-3.5mm]{$ g_{MN}$}&$
 g_{\mu\nu}$& $ \rep{1}_\rep{2}$ \\ 
 \cline{2-3}
\rule[-0.3cm]{0cm}{0.8cm}
   & $ g_{\mu m}$ & $(\rep{3}+\rep{\bar{3}})_\rep{1}$ \\\cline{2-3}
\rule[-0.3cm]{0cm}{0.8cm} 
 &$ g_{mn}$ & $\rep{1}_\rep{0}+(\rep{6}+\rep{\bar{6}})_\rep{0}+\rep{8}_\rep{0}$\\
  \hline
\rule[-0.3cm]{0cm}{0.8cm}
\multirow{3}{12mm}[-3.5mm]{$ B_{MN}$}& $
 B_{\mu\nu}$ & $ \rep{1}_\rep{T}$ \\ \cline{2-3}
\rule[-0.3cm]{0cm}{0.8cm}
   & $ B_{\mu m}$ & $(\rep{3}+\rep{\bar{3}})_\rep{1}$ \\ \cline{2-3}
\rule[-0.3cm]{0cm}{0.8cm} 
 &$ B_{mn}$ & $\rep{1}_\rep{0}+(\rep{3}+\rep{\bar{3}})_\rep{0}+\rep{8}_\rep{0}$\\
  \hline\rule[-0.3cm]{0cm}{0.8cm}
$  \phi $&& $\rep{1}_\rep{0}$\\
\hline
\end{tabular}
\caption{\small 
\textit{ Decomposition of the NS sector in $SU(3)$ representations}}
\label{NS}
\end{center}
\end{table}

The RR sector of type IIA supergravity features a vector $C_M$ and a
three-form $C_{MNP}$; their decompositions are given in table~\ref{RRIIA}.
\begin{table}[htb]
\begin{center}
\begin{tabular}{|c|c|l|} \hline
\rule[-0.3cm]{0cm}{0.8cm}
\multirow{2}{10mm}[-3.5mm]{$ C_{M}$}&$
 C_{\mu}$& $ \rep{1}_\rep{1}$ \\ 
 \cline{2-3}
\rule[-0.3cm]{0cm}{0.8cm}
   & $ C_{m}$ & $(\rep{3}+\rep{\bar{3}})_\rep{0}$ \\
  \hline
\rule[-0.3cm]{0cm}{0.8cm}
\multirow{3}{15mm}[-3.5mm]{$ C_{MNP}$}& $
 C_{\mu\nu p}$ & $(\rep{3}+\rep{\bar{3}})_\rep{T}$ \\ \cline{2-3}
\rule[-0.3cm]{0cm}{0.8cm}
   & $ C_{\mu np}$ & $\rep{1}_\rep{1}
+(\rep{3}+\rep{\bar{3}})_\rep{1} +\rep{8}_\rep{1}$ \\ \cline{2-3}
\rule[-0.3cm]{0cm}{0.8cm} 
 &$ C_{mnp}$ & $(\rep{1} + \rep{1})_\rep{0} +(\rep{3}+\rep{\bar{3}})_\rep{0}
        +(\rep{6}+\rep{\bar{6}})_\rep{0} $\\
\hline
\end{tabular}
\caption{\small 
\textit{ Type IIA decomposition of the RR sector in $SU(3)$
  representations}}
\label{RRIIA}
\end{center}
\end{table}

In type IIB one has a scalar $C_0$, a two-form $C_2$ and a
four-form $C_4$ with a self-dual field strength $F_5$. Their
decompositions are recorded in table~\ref{RRIIB}.  
The self-duality of $F_5$ relates
$ C_{\mu \nu mn}$ to $ C_{mnpq}$, and leaves only half of the
representations in $ C_{\mu npq}$ as independent fields. 
\begin{table}[htb]
\begin{center}
\begin{tabular}{|c|c|l|} \hline
\rule[-0.3cm]{0cm}{0.8cm}
$C^{(0)}$&&$\rep{1}_\rep{0}$\\ \hline\rule[-0.3cm]{0cm}{0.8cm}
\multirow{3}{12mm}[-3.5mm]{$ C_{MN}$}& $
 C_{\mu\nu}$ & $ \rep{1}_\rep{T}$ \\ \cline{2-3}
\rule[-0.3cm]{0cm}{0.8cm}
   & $ C_{\mu m}$ & $(\rep{3}+\rep{\bar{3}})_\rep{1}$ \\ \cline{2-3}
\rule[-0.3cm]{0cm}{0.8cm} 
 &$ C_{mn}$ & $\rep{1}_\rep{0}+(\rep{3}+\rep{\bar{3}})_\rep{0}+\rep{8}_\rep{0}$\\
  \hline\rule[-0.3cm]{0cm}{0.8cm}
\multirow{2}{12mm}[-3.5mm]{$ C_{MNPQ}$}&$
 C_{\mu npq}$& $  \tfrac{1}{2}\left[ 
     (\rep{1}+ \rep{1})_\rep{1} +  
     (\rep{3}+ \rep{\bar{3}})_\rep{1} + 
     (\rep{6}+ \rep{\bar{6}})_\rep{1}\right] $ \\ 
 \cline{2-3}
\rule[-0.3cm]{0cm}{0.8cm}
   & $  C_{mnpq}/ C_{\mu\nu mn}$ & $\rep{1}_\rep{0}+(\rep{3}+\rep{\bar{3}})_\rep{0}+\rep{8}_\rep{0}$ \\
  \hline
\end{tabular}
\caption{\small 
\textit{ Type IIB decomposition of the RR sector in $SU(3)$
  representations}}
\label{RRIIB}
\end{center}
\end{table}

Finally let us turn to the fermionic sector which contains two
gravitinos $\Psi_M$ and two dilatinos $\lambda$. In type IIA they have
opposite ten-dimensional chirality while in IIB they have the same
chirality. The 16-dimensional spinor representation decomposes according to
\begin{equation} 
\label{repdecomp}
   \rep{16} \to (\rep{2},\rep{1}) \oplus (\rep{2},\rep{3})
      \oplus (\bar{\rep{2}},\rep{1}) \oplus
      (\bar{\rep{2}},\bar{\rep{3}})\ .
\end{equation}
This in turn leads to the  decompositions displayed in 
table~\ref{FII}. (Here all the representations are assumed to be
complex.) 
\begin{table}[h]
\begin{center}
\begin{tabular}{|c|c|l|} \hline
\rule[-0.3cm]{0cm}{0.8cm}
\multirow{2}{10mm}[-3.0mm]{$ \Psi_{M}$}& $
 \Psi_{\mu}$ & $ \rep{1}_\rep{3/2} + \rep{3}_\rep{3/2}$ \\ \cline{2-3}
\rule[-0.3cm]{0cm}{0.8cm}
   & $ \Psi_{m}$ & 
$\rep{1}_\rep{1/2}+ \rep{3}_\rep{1/2}+2\,\rep{\bar{3}}_\rep{1/2} + 
\rep{6}_\rep{1/2}+\rep{{8}}_\rep{1/2}$ \\ 
  \hline\rule[-0.3cm]{0cm}{0.8cm}
$ \lambda$&& $ \rep{1}_ \rep{1/2}+ \rep{{3}}_\rep{1/2} $ \\
  \hline
\end{tabular}
\caption{\small 
\textit{ Decomposition of the fermions in $SU(3)$ representations}}
\label{FII}
\end{center}
\end{table}

Altogether these fields can be assembled into $N=2$ multiplets.
In both theories one finds  a gravitational multiplet, six
spin-$\frac32$ multiplets, 15 vector multiplets, nine hypermultiplets
and one tensor multiplet. (Of course, altogether these $N=2$
multiplets  precisely fit into
a single $N=8$ gravitational multiplet.)
The distribution of the fields into $N=2$ 
multiplets is not uniquely determined  by their $SU(3)$
representation. However, we are mostly interested in the case where
only the two gravitinos in the gravitational multiplet are kept while
the six extra spin-$\frac32$ multiplets are projected out or become 
massive.\footnote{Recall that a massless $N=2$ spin-$\frac32$ multiplet
  contains a spin-$\frac32$ gravitino, two vectors and a Weyl fermion,
  while a massive spin-$\frac32$ multiplet contains a spin-$\frac32$
  gravitino, four vectors, six Weyl fermions and four scalars.} 
In this case one has a `standard' $N=2$ theory in that only vector,
tensor and hypermultiplets coupled to the gravitational multiplet
with known couplings occur. This situation can be arranged if one removes all
triplets from the spectrum. In this case the distribution of the
fields into $N=2$ multiplets is uniquely determined by their $\SU(3)$
quantum numbers.

For type IIA we display  the multiplets in table~\ref{N=2multipletsA}
while the type IIB multiplets are given 
in table~\ref{N=2multipletsB}.
We see that the $SU(3)$ representations are permuted between type IIA
and type IIB which expresses the fact   vector- and
hypermultiplets  are exchanged under mirror
symmetry.\footnote{Strictly speaking in type IIB one has the choice to
  assemble the spectrum either in 
  tensor or hypermultiplets. If they are massless one can always
  dualize the tensor to a hypermultiplet. However for massive
  multiplets such a procedure is not straightforward and it often more
  convenient to keep the tensors in the spectrum \cite{LM,DSV}.}
\begin{table}[h]
\begin{center}
\begin{tabular}{|c|c|c|} \hline
 \rule[-0.3cm]{0cm}{0.8cm}
multiplet & $SU(3)$ rep. & field content\\ \hline
\rule[-0.3cm]{0cm}{0.8cm}
 gravity multiplet&$\rep 1$ & $(g_{\mu \nu}, C_\mu, \Psi_\mu)$ \\ \hline
\rule[-0.3cm]{0cm}{0.8cm} 
{tensor multiplet} & $\rep 1$& $(B_{\mu\nu}, \phi, C_{mnp}, \lambda) $ \\ \hline
\rule[-0.3cm]{0cm}{0.8cm}
 vector multiplets& $\rep{8} + \rep {1}$ &  $(C_{\mu np},
 g_{mn},B_{mn}, \Psi_m) $\\ \hline
\rule[-0.3cm]{0cm}{0.8cm}
{hypermultiplets } & $\rep{6}$&   $ (g_{mn}, C_{mnp}, \Psi_m) $ \\
\hline
\end{tabular}
\caption{\small 
\textit{N=2 multiplets in type IIA}}\label{N=2multipletsA}
\end{center}
\end{table}
\begin{table}[h]
\begin{center}
\begin{tabular}{|c|c|c|} \hline
 \rule[-0.3cm]{0cm}{0.8cm}
multiplet & $SU(3)$ rep. & field content\\ \hline
\rule[-0.3cm]{0cm}{0.8cm}
 gravity multiplet&$\rep 1$ & $(g_{\mu \nu},C_{\mu npq}, \Psi_\mu)$ \\ \hline
\rule[-0.3cm]{0cm}{0.8cm} 
{double tensor multiplet} & $\rep 1$& 
$(B_{\mu\nu}, C_{\mu\nu}, \phi, C^{(0)}, \lambda) $ \\ \hline
\rule[-0.3cm]{0cm}{0.8cm}
 vector multiplets& $\rep{6}  $ &  $(C_{\mu npq}, g_{mn}, \Psi_m) $\\ \hline
\rule[-0.3cm]{0cm}{0.8cm}
{hypermultiplets } & $\rep{8} + \rep {1}$ &   $ (g_{mn}, B_{mn},
C_{mn}, C_{mnpq}, \Psi_m) $ \\
\hline
\end{tabular}
\caption{\small 
\textit{N=2 multiplets in type IIB}}\label{N=2multipletsB}
\end{center}
\end{table}


\subsection{Scalar action: kinetic terms}
\label{kinetic}

From now on, we will concentrate on the scalar part of the generalized
action. As we have seen in the previous section, the relevant NS
components are $\phi$, $B_{\mu\nu}$, $g_{mn}$ and $B_{mn}$. Note that
$g_{mn}$ enters both the vector multiplets and hypermultiplets. To
distinguish the two contributions we first need to understand the
geometry on the space of metrics on manifolds with $\SU(3)\times\SU(3)$  
structure. As before we will actually specialize to the $\SU(3)$
case. Thus as a prerequisite it is useful to recall some facts about
$\SU(3)$ structures. In what follows we will consider the analog of
the four-dimensional kinetic terms for these scalar degrees of
freedom. The analog of the potential term will be discussed in
section~\ref{N=2W}.

\subsubsection{Geometry of $\SU(3)$ structures}
\label{SU3Geo}

One way to define manifolds with $G$-structure is to demand
the existence of a $G$-invariant tensor or spinor on the
manifold~\cite{salamonb,joyce}. 
We have argued that an invariant spinor $\eta_+$
corresponds to picking out a particular supersymmetry in the type II
theory. Given $\eta_+$ defining an $\SU(3)$-structure we can also
build a set of $\SU(3)$-invariant forms. These are constructed as
follows. 
Using the six-dimensional
gamma-matrices $\gamma^m$ defined in \eqref{gammadef} 
we can construct a globally defined two-form
$J$ and a complex three-form $\Omegan$ given by 
\beq
\label{Jeta}
   \bar{\eta}_{\pm} \gamma^{mn} \eta_{\pm} = 
      \pm \tfrac{1}{2}\ii \, J^{mn} \ , \qquad
   \bar{\eta}_{-} \gamma^{mnp} \eta_{+} =  
      \tfrac{1}{2}\ii \, \Omegan^{mnp} \ , \qquad
   \bar{\eta}_{+} \gamma^{mnp} \eta_{-} =  
      \tfrac{1}{2}\ii \, {\Omeganb}^{mnp}\ .
\eeq
%

Here $\eta_{\pm}$ are normalized so that $\bar{\eta}_{\pm}
\eta_{\pm}=\frac{1}{2}$ and $\gamma^{m_1\dots m_p} =
\gamma^{[m_1}\gamma^{m_2}\dots\gamma^{m_p]}$ are antisymmetrized 
products of six-dimensional $\gamma$-matrices.\footnote{By $\Omegan$
  we denote the three-form defined in \eqref{Jeta} which is built from
  normalized spinors $\eta$. Later on in this paper we will also
  introduce the three-form $\Omega$ which obeys a different normalization.}
Using appropriate Fierz identities one shows that 
with this normalization for the spinors, $J$  and $\Omegan$
are not independent but satisfy 
\beq\label{JOconstraint}
J\wedge J\wedge J  = \tfrac{3}{4}\ii\, \Omegan\wedge\Omeganb\ , 
   \qquad
   J\wedge \Omegan = 0 \ .
\eeq
Furthermore, raising an index on $J$ defines an almost complex
structure $I$ in that it satisfies $I^2 = - \id$. With respect to this
almost complex structure $J$ is a $(1,1)$-form while $\Omegan$ is a
$(3,0)$-form. 

It is helpful in what follows to note that one can actually 
\emph{define} the $\SU(3)$ structure in terms of a pair of \emph{real}
forms $(J,\rho)$ where $\rho=\re\Omegan$.
The forms cannot be
arbitrary but must be \emph{stable}~\cite{HitchinHF}. This means that they live
in an open orbit under the action of general transformations $\GL(6,\bbR)$
in the tangent space at each point. A stable two-form $J$ then defines a 
$\Symp(6,\bbR)$ structure while a stable real form $\rho$ defines a
$\SL(3,\bbC)$ structure. Together they define an $\SU(3)$ structure
provided the embeddings of $\Symp(6,\bbR)$ and $\SL(3,\bbC)$ in
$\GL(6,\bbR)$ are compatible. This requires
\begin{equation}
\label{Jrconds}
 J\wedge J\wedge J  =  \tfrac{3}{2} \rho \wedge \hat{\rho}\ , \qquad 
J \wedge \rho = 0 \ , 
\end{equation}
where $\hat{\rho}=\im\Omegan=*\rho$. As we will discuss in
section~\ref{rhoSK} $\hat{\rho}$ is actually determined by $\rho$,
independent of $J$. Note that since $\SU(3)\subset\SO(6)$ the pair
$(J,\rho)$ satisfying~\eqref{Jrconds} also defines an $\SO(6)$
structure and hence a metric.

Returning to the spinor $\eta_+$, if the corresponding supercharge is
conserved in a space-time background where all fields vanish other
than the metric, then the Killing spinor equations imply that $\eta_+$
is covariantly constant with respect to the Levi-Civita
connection. Geometrically this says that the holonomy of $\M^{1,9}$ is
in $\SU(3)$. This means we have a metric product of the
form~\eqref{spacetime} where $\M^{1,3}$ is flat $\bbR^{1,3}$ and $\Y$
is a Calabi--Yau manifold.
 
However, if one also allows for the possibility of spontaneously
broken supercharges, while we still have an $\SU(3)$ structure,
$\eta_+$ is no longer required to be covariantly
constant. Nonetheless, for any $\eta_+$, one can always find a
torsionful connection $\nabla^{(T)}$ on the six-dimensional vector
bundle $F$, which is compatible with the metric to $g_{mn}$ and which obeys
$\nabla^{(T)}\eta =0$. Calabi--Yau manifolds are thus special cases of
manifolds with $SU(3)$ structure for which the torsion vanishes. The
part of the torsion which is independent of the choice of
$\nabla^{(T)}$ is known as the ``intrinsic torsion'' and can be used to
classify the types of $\SU(3)$ structure. Since $\eta$ is no longer
covariantly constant both $J$ and $\Omegan$ are no longer closed but
instead they obey~\cite{CS}
\begin{equation}
\label{dJdOmega}
\begin{aligned}
   \dd J &= \tfrac{3}{4}\ii \left(
      W_1\Omeganb - \bar W_1\Omegan\right) + W_4\wedge J + W_3\ , \\
   \dd\Omegan &= W_1 J^2 +W_2\wedge J +\bar W_5\wedge \Omegan\ ,
\end{aligned}
\end{equation}
with
\begin{equation}\label{constraints}
   W_3\wedge J\ =\ W_3\wedge\Omegan\ =\ W_2\wedge J^2\ =\ 0\ .
\end{equation}
One finds that the five different $W$'s completely
determine the intrinsic torsion. Note that $W_1$ is a zero-form,
$W_4, W_5$ are one-forms, $W_2$ is a two-form and $W_3$ is a
three-form and each can be characterized by its $\SU(3)$ transformation
properties. Calabi--Yau manifolds are manifolds of $\SU(3)$ structure 
where all five torsion classes vanish. Any subset of
vanishing torsion classes on the other hand define specific classes of
$\SU(3)$ structure manifolds. 

Now we turn to the effective action for the metric degrees of freedom
or, as we will see, the effective action for $J$ and $\rho$.


\subsubsection{Kinetic terms for NS deformations}
\label{modulispace}

In order to separate the vector multiplet and hypermultiplet degrees
of freedom it is better to work in terms of the $\SU(3)$-structure
$(J,\rho)$ rather than the metric $g_{mn}$. Decomposing deformations of the
structure into $\SU(3)$-representations, given the
constraints~\eqref{Jrconds}, one can write 
\begin{equation}
\label{delJr}
\begin{aligned}
   \delta J &= \lambda J + i_v\rho + K\ , \\
   \delta \rho &= \tfrac{3}{2}\lambda \rho + \gamma \hat{\rho}
      - v \wedge J + M\ .
\end{aligned}
\end{equation}
where $\lambda$ and $\gamma$ are scalars, $v$ is a real vector,
transforming as $\rep{3}+\rep{\bar{3}}$, $K$ a is real two-form
transforming as $\rep{8}$ and $M$ is a (primitive) three-form
transforming as $\rep{6}+\rep{\bar{6}}$. This implies that
\begin{equation}
\begin{aligned}
  \rho \wedge K\ &=\ J \wedge J \wedge K\ =\ 0\ , \\
  J \wedge M\ &=\ \rho \wedge M\ =\ \hat{\rho} \wedge M\ =\ 0\ .
\end{aligned}
\end{equation}
Recalling that the $\SU(3)$ structure defines the metric
$g_{mn}$, it is easy to show that the two sets of deformations
are related by  
\begin{equation}\label{metricJrho}
   \delta g_{mn} = \lambda g_{mn} - {J_m}^pK_{pn}
      - \tfrac{1}{2}{\rho_m}^{pq}M_{pqn}\ .
\end{equation}
Here we see explicitly that $\delta g_{mn}$
contains scalar, $\rep{8}$ and $\rep{6}+\rep{\bar{6}}$ 
deformations as we already noted in table~\ref{NS}. 
(By definition ${J_m}^pK_{pn}$ and
${\rho_m}^{pq}M_{npq}$ are symmetric on $m$ and $n$.) 

Clearly, there are more degrees of freedom in the $\SU(3)$-structure
than in the metric. This is not surprising since the metric
parameterizes the coset $\GL(6,\bbR)/\SO(6)$ which is 21-dimensional,
while the pair $(J,\rho)$ parameterizes $\GL(6,\bbR)/\SU(3)$ which is
28-dimensional. The vector $v$ and scalar $\gamma$ represent the extra
seven parameters: deformations which change the $\SU(3)$-structure but
leave the metric invariant. It is natural, since we have local
$\SU(3)$ symmetry, to formulate the theory in terms of $J$ and
$\rho$. However, we expect to find a local symmetry removing the
non-metric degrees of freedom represented by $v$ and $\gamma$. This is
the remnant of the local $\Spin(6)\subset\Spin(9,1)$ rotational
symmetry of the vierbein formulation of gravity. 

Note, in addition, that the vector deformation $v$ transforms as
$\rep{3}+\rep{\bar{3}}$ as do the additional spin-$\tfrac32$ degrees
of freedom coming form $\Psi_\mu$ discussed in
section~\ref{fields}. If we want to consistently restrict to a
`standard' $N=2$ theory and ignore these additional spin-$\tfrac32$
fields we must drop all such triplet representations and hence set
$v$ to zero. We will often do this in what follows 
and only in section~\ref{metricSK}
discuss some properties of the more general case. 

Now let us  finally turn to the kinetic terms for $g_{mn}$,
$B_{mn}$ and $\phi$.  The point is that, given the
split~\eqref{Tdecomp}, we can always decompose the derivative operator
$\der_M$ into a part on $T^{1,3}$ and a part on $F$ labeled $\der_\mu$
and $\der_m$ respectively,  
\begin{equation}
   \der_M = (\der_\mu, \der_m) . 
\end{equation}
Any term in the ten-dimensional theory with two $\der_\mu$ operators
we denote as kinetic, while any scalar field term with no such
operators we denote as contributing to the potential. 

 As when conpactifying, we need to rescale the
four-dimensional part of the metric $g_{\mu\nu}$ and also define a new
``four-dimensional'' dilaton in order to diagonalize the kinetic terms
and get a conventional Einstein term. In analogy with the case of
compactification on a Calabi--Yau manifolds we write
\begin{equation}
\label{4dNS}
   g^{(4)}_{\mu\nu} = \ee^{-2\phi^{(4)}} g_{\mu\nu}\ , \qquad
   \phi^{(4)} = \phi - \tfrac{1}{4}\ln\det g_{mn}\ ,
\end{equation}
where $g_{mn}$ is not rescaled but taken in the ten-dimensional string
frame. Note that these definitions imply that 
\begin{equation}
\begin{aligned}
   \ee^{-2\phi^{(4)}} &\in \det F^* = \Lambda^6F^* \ , \\
   g^{(4)}_{\mu\nu} &\in T^{1,3\,*}\otimes T^{1,3\,*}
      \otimes \Lambda^6F^* \ .
\end{aligned}
\end{equation}
This means for instance that the exponential of the four-dimensional
dilaton transforms as a six-form under $\GL(6,\bbR)$ transformations
on $F$. 

The bosonic NS part of the ten-dimensional type II action
reads  
\begin{equation}
\label{NSaction}
   S_{\text{NS}} = \int\dd^{10}x \sqrt{g}\, \ee^{-2\phi}\left[
        R + 4(\der\phi)^2 - \tfrac{1}{12}H^2 \right]\ .
\end{equation}
Keeping only terms with $\der_\mu$ and also only $\Spin(1,3)$-scalar
fields we find, given the redefinitions~\eqref{4dNS},
\begin{multline}
\label{KEterms}
   S_{\text{NS}} = \int\dd^{10}x \sqrt{g^{(4)}}\big[
        R^{(4)} - 2(\der\phi^{(4)})^2 
        - \tfrac{1}{12}\ee^{-4\phi^{(4)}}H_{(4)}^2 \\
        - \tfrac{1}{4}g^{mp}g^{nq}(
           \der_\mu g_{mn}\der^\mu g_{pq}
           + \der_\mu B_{mn}\der^\mu B_{pq} ) + \ldots
     \big] \ ,\qquad
\end{multline}
where $H^{(4)}_{\mu\nu\rho}=3\der_{[\mu}B_{\nu\rho]}$ and all
contractions are with $g^{(4)}_{\mu\nu}$. Note that, for instance,  
$\sqrt{g^{(4)}}R^{(4)}\in\det T^{1,3\,*}\otimes\det F^*$ and
hence~\eqref{KEterms} does transform properly as a 
ten-dimensional generally covariant expression. The first three
terms give the usual kinetic terms of the four-dimensional metric
$g^{(4)}$ and the $B_{\mu\nu}$--$\phi$ part of the tensor or double
tensor multiplet. The last two terms in~\eqref{KEterms} define a
metric of the space of metric and $B$-field deformations
\cite{CdO}. We write, given the expansions~\eqref{delJr}
\begin{equation}
\label{KEmetric}
\begin{aligned}
   \dd s^2 &= \tfrac{1}{8}g^{mp}g^{nq}(
           \delta g_{mn}\delta g_{pq}
           + \delta B_{mn}\delta B_{pq} ) \\
      &= \left[\tfrac{3}{4}\delta\lambda\delta\lambda
         + \tfrac{1}{8}g^{mp}g^{nq}\delta K_{mn}\delta K_{pq}
         + \tfrac{1}{8}g^{mp}g^{nq}\delta B_{mn}\delta B_{pq}\right]
           \\ & \qquad \qquad \qquad
         + \tfrac{1}{24}g^{mr}g^{ns}g^{pt}\delta M_{mnp}\delta M_{rst} \\
      &\equiv \dd s^2(J,B) + \dd s^2(\rho) .
\end{aligned}
\end{equation}
Note that without any truncation this metric 
is precisely the metric on the Narain coset $O(6,6)/O(6)\times O(6)$
\cite{Narain}.
The vector $\delta v_m$ and scalar $\delta\gamma$
deformations of $J$ and $\rho$ do not enter~\eqref{KEmetric} 
because they represent non-metric degrees of freedom, that is
deformations which change the $\SU(3)$ structure but leave the metric
unchanged. 

The derivation of~\eqref{KEterms} and~\eqref{KEmetric} is completely
analogous to the derivation given for example in refs.~\cite{BCF,CdO}
for Calabi--Yau compactifications. The difference here is that we do
not assume any compactification and keep the dependence on all ten
space-time coordinates. Furthermore, the background is not a
Calabi--Yau but only constrained to have $\SU(3)$ or
$\SU(3)\times\SU(3)$ structure. 

The exact same decomposition of the ten-dimensional action can be
done for the RR-part of the respective actions for type IIA and type
IIB. Again this is in complete analogy with the derivation of the
four-dimensional action for Calabi--Yau compactifications  performed in
\cite{BCF,BGHL}. This results in the kinetic terms for the gauge
bosons and the kinetic terms for the RR scalars and tensors. We will
not do this explicitly here since the couplings to the $\SU(3)$
structure $J$ and $\rho$ is exactly the same as for Calabi--Yau
compactifications and thus we can borrow the results from the
literature \cite{BCF,BGHL}.

Instead in the following sections we will show that
$\dd s^2(J,B)$ and $\dd s^2(\rho)$ correspond to special K\"ahler
metrics on the moduli space of $B+\ii J$ and $\rho$ respectively, and,
in addition, how these structures are intimately related to
$\Spin(6,6)$ spinors and Hitchin functionals. 


\subsection{Special K\"ahler manifolds and stable forms}
\label{SKM}

In this section we review, essentially following Hitchin~\cite{HitchinHF}, how
the spaces of forms $J$ and $\rho$ defining the $\SU(3)$
structure each separately admit a natural special K\"ahler metric. We
then discuss the corresponding structure on the space of $\SU(3)$
metrics $g_{mn}$ (that is metrics which are compatible with some
$\SU(3)$ structure). This is done by first noting that 
both $\ee^{\ii J}$ and $\rho+\ii\hat{\rho}$ are pure
$\Spin(6,6)$ spinors and using
Hitchin's result that there is a natural special K\"ahler metric on
the space of (stable) real $\Spin(6,6)$ spinors. The special K\"ahler 
metrics on the spaces of $J$ and $\rho$ then arise as special cases.
In each case they agree with the metrics~\eqref{KEmetric} we found
directly from rewriting the type II supergravity theory.

Concretely this section is organized as follows. We first briefly review the
notion of special
K\"ahler geometry in \ref{SKG}. Then  in section \ref{SU3SU3}
we discuss in more detail the
properties of $\SU(3)\times\SU(3)$
structures and in particular show that they 
can be conveniently expressed in terms of pure
$\Spin(6,6)$ spinors.
In \ref{genSK} we 
review (in a slightly different language)
Hitchin's result that there is a natural special K\"ahler metric on
the space of (stable) real $\Spin(6,6)$ spinors. The special K\"ahler
metrics for $\rho$ and $J$ will then arise as special cases in
sections~\ref{JSK} and~\ref{rhoSK}. Finally in~\ref{metricSK} we
discuss the geometry of the corresponding constrained space of
$\SU(3)$ metrics.


\subsubsection{Review of special K\"ahler geometry}
\label{SKG}

First let us briefly recall the structure of special K\"ahler
geometry~\cite{dWvP,Strominger,CRTV,freed,N=2review}. 
There are two types of special K\"ahler structures which we will now
summarize. 

In globally supersymmetric $N=2$ theories the scalar fields in the
vector multiplets can be viewed as coordinates of a \emph{rigid}
special K\"ahler manifold. This implies one has $(2n)$-dimensional
K\"ahler manifold $U$ with a \emph{flat} holomorphic $\Symp(2n,\bbR)$
bundle $\mathcal{G}$ with a holomorphic section $\pure$ such that the
K\"ahler potential is given by 
\begin{equation}
\label{rigidSK}
   K_\text{rigid} = \ii\sym{\pure}{\bar{\pure}} , \qquad 
   \sym{\der \pure}{\der \pure}=0\ .
\end{equation}
where $\sym{\cdot}{\cdot}$ is the symplectic product on $\mathcal{G}$
and $\partial$ is the holomorphic derivative on $U$. One can typically
introduce holomorphic coordinates $Z^I$ on $U$ and holomorphic
functions $F_I(Z)$ such that  
\begin{equation}
\label{prepot}
   \sym{\pure}{\bar{\pure}} = \bar{Z}^I F_I - Z^I\bar{F}_I \ ,
\end{equation}
with $I=1,\dots,n$. The second condition in~\eqref{rigidSK} implies
that locally we can write $F_I=\der F/\der Z^I$ where the holomorphic
function $F(Z)$ is known as the \emph{prepotential}. 

In $N=2$ supergravity the same scalar fields in the vector multiplets
are coordinates of a \emph{local} special K\"ahler manifold. The
latter is a Hodge--K\"ahler manifold $\mathcal{M}$ of real dimension
$2n$ together with a line bundle $\mathcal{L}$ and a holomorphic
$\Symp(2n+2,\bbR)$ vector bundle $\mathcal{H}$ over $\mathcal{M}$. One
requires that $\mathcal{L}$ embeds holomorphically in
$\mathcal{H}$. In addition there exists a holomorphic section $\pure$
of $\mathcal{L}$ such that the K\"ahler potential is given by 
\begin{equation}\label{Kdef}
   K\ =\ - \ln\, \ii\sym{\pure}{\bar{\pure}}\ , \qquad 
   \sym{\pure}{\partial{\pure}}\ =\ 0 \ ,
\end{equation}
where now $\sym{\cdot}{\cdot}$ is the symplectic product on 
$\mathcal{H}$ and $\partial$ is the holomorphic derivative on
$\mathcal{M}$. The section $\pure$ is not unique but can be shifted by
a holomorphic gauge transformation on $\mathcal{L}$ corresponding to a 
K\"ahler transformation. Just as in the rigid case one can
introduce holomorphic coordinates $Z^I$ on $\mathcal{H}$
together with a set of holomorphic functions ${F}_I(Z)$
such that~\eqref{prepot} holds, though now with $I=0,1,\dots,n$. 
Again, the second condition in~\eqref{Kdef} implies $F_I=\der F/\der
Z^I$. However, the prepotential $F(Z)$ is now homogeneous of degree
two. Locally one can introduce special holomorphic coordinates
$z^i=Z^i/Z^0, i=1,\ldots,n$ on $\mathcal{M}$. In these coordinates the
prepotential has the form 
\begin{equation}
\label{prepotf}
   F(Z)=(Z^0)^2\,f(z^i)\ ,
\end{equation}
where $f(z^i)$ is a function of the $z^i$.


\subsubsection{$O(6,6)$ spinors and $\SU(3)\times\SU(3)$ structures} 
\label{SU3SU3}

We argued in section \ref{compact}
that in general $N=2$ supersymmetry (or eight linearly
realized supercharges) leads to a theory with
$\SU(3)\times\SU(3)$ structure, by which we mean a pair of $\SU(3)$
structures. From one perspective this simply says that we have a pair of
spinors $\eta_+^A$, $A=1,2$, or equivalently a pair of structures
$(J^A,\rho^A)$ which are given in terms of the $\eta_+^A$ exactly as
in section~\ref{SU3Geo}. However, it turns out that it is convenient
to reformulate these structures in terms of $\Spin(6,6)$
spinors~\cite{Gualtieri,GCY} since it makes the local
$\SU(3)\times\SU(3)$ symmetry of the theory
manifest~\cite{JW,Witt}. In this section we briefly review this  
reformulation as it will be essential for showing the special K\"ahler
structure of the metric~\eqref{KEmetric}. 

Let $X=v+\xi$ be an element of $F\oplus F^*$ where $v\in F$ and
$\xi\in F^*$. Recall that there is a natural $O(6,6)$ metric on
$F\oplus F^*$ given by~\eqref{Odd-metric}. The basic
$\Spin(6,6)$ spinor representations are Majorana--Weyl. In fact, the
two chiral spinor bundles $S^+$ and $S^-$ are isomorphic to the space
of even and odd forms respectively
\begin{equation}
   S^+\simeq\Leven F^* , \qquad 
   S^-\simeq\Lodd F^* . 
\end{equation}
Under this isomorphism, the Clifford action is realized by 
\begin{equation}
   (v+\xi)\cdot\chi^\pm = i_v \chi^\pm + \xi\wedge\chi^\pm
\end{equation}
for $\chi^\pm\in\Leo F^*$ and where $i_v$ denotes contraction with the
vector $v$. More explicitly, let $f_m$ with $m=1,\dots,6$ be a basis
on $F$ and let $e^m$ be the dual basis on $F^*$ (that is
$e^m(f_n)=\delta^m{}_n$). We can write the $O(6,6)$ gamma matrices
$\Gamma^\Sigma$ where $\Sigma=1,\dots,12$ as  
\begin{equation}
\label{Gamma-def}
   X_\Sigma\Gamma^\Sigma = v^m\Gamma_m+\xi_m\Gamma^m 
   \quad \text{with} \quad 
   \Gamma_m = i_m\ , \quad \Gamma^m = e^m\wedge\ , 
\end{equation}
where we use the shorthand $i_m$ for $i_{f_m}$. 

The isomorphism between spinors and forms is not canonical. If one 
considers carefully how the spinors transform under the $\GL(6,\bbR)$
subgroup one finds\footnote{Note that there is actually always a second
spin structure, twisted not by $|\det F|^{1/2}$ but by $\det F|\det
F|^{-1/2}$. These differ by a sign in the action of elements of
$\GL(6,\bbR)$ with negative determinant.} 
\begin{equation}
   S^\pm \simeq \Leo F^*\otimes|\det F|^{1/2}\ . 
\end{equation}
To specify the isomorphism one needs to choose an element of $|\det
F|^{1/2}$. Since one takes the absolute value this bundle is
trivial. If the manifold is orientable (as will be the case for us)
the isomorphism can equivalently be fixed by choosing a particular
volume form  $\epsilon\in\Lambda^6F^*=\det F^*$. Specifically we
define the isomorphism 
\begin{equation}
\label{isomorph}
\begin{aligned}
   f_\epsilon : S^\pm &\to \Leo F^* , \\
   u^\pm &\mapsto \chi^\pm = u^\pm \sqrt{\epsilon} .
\end{aligned}
\end{equation}
In what follows we will actually be interested in forms
$\chi^\pm\in\Leo F^*$. We will use 
\begin{equation}
   \chis = f_\epsilon^{-1}(\chi^\pm)
\end{equation}
to denote the corresponding element in $S^\pm$. We will also often
refer to the forms $\chi^\pm$ as ``spinors'' assuming that the
isomorphism~\eqref{isomorph} is understood. 

As usual, the intertwiner $C$ between $\Gamma^\Sigma$ and its transpose
representation $-(\Gamma^\Sigma)^\mathrm{T}=C^{-1}\Gamma^\Sigma C$
defines  a bilinear form on $S^\pm$. Explicitly if
$\alpha$, $\beta$ are $\Spin(6,6)$ spinor indices and $u,v\in S^\pm$
then the inner product is given 
by 
\begin{equation}\label{normC}
   \bar{u}^\pm v^\pm \equiv 
      (C^{-1})_{\alpha\beta}\,u^{\pm\,\alpha}v^{\pm\,\beta}\ .
\end{equation}
The inner prduct between spinors in $S^+$ and $S^-$ vanishes because the spinors
have different chirality.  Since here $C^\mathrm{T}=-C$ 
the bilinear form actually defines a symplectic structure on
$S^\pm$. Thus we will also often write it as 
\begin{equation}
\label{sym-def}
   \sym{u^\pm}{v^\pm} \equiv \bar{u}^\pm v^\pm ,
\end{equation}
so that in components $\omega_{\alpha\beta}=C^{-1}_{\alpha\beta}$. 

Given the isomorphism between $S^\pm$ and $\Leo$ there is a
corresponding inner product $\mukai{\psi^\pm}{\chi^\pm}$ on the space of
forms known as the Mukai pairing. If the isomorphism is defined by the
volume form $\epsilon$ one defines
\begin{equation}
\label{mukai}
\begin{aligned}
  \mukai{\psi^+}{\chi^+} 
     \equiv \sym{\psis^+}{\chis^+}\epsilon
     &= \psi^+_0\wedge\chi^+_6 - \psi^+_2\wedge\chi^+_4
        + \psi^+_4\wedge\chi^+_2 - \psi^+_6\wedge\chi^+_0 \ , \\
  \mukai{\psi^-}{\chi^-}
     \equiv \sym{\psis^-}{\chis^-} \epsilon
     &= - \psi^-_1\wedge\chi^-_5 + \psi^-_3\wedge\chi^-_3 
        - \psi^-_5\wedge\chi^-_1\ ,
\end{aligned}
\end{equation}
where the subscripts denote the degree of the component forms in
$\Leo F^*$. Note that the Mukai pairing is independent of the
particular choice of volume form $\epsilon$. However it is also not
quite a symplectic structure on $\Leo F^*$ since it is a map to
$\Lambda^6F^*$ and not to $\bbR$.

{}From the metric $g_{mn}$ one identifies the subgroup of
diffeomorphisms which preserves $g_{mn}$
as $O(6)\subset\GL(6,\bbR)\subset\Spin(6,6)$. Given a spin
structure, we can then decompose
$u^+\in S^+$ into representations of the corresponding $\Spin(6)$
group according to 
\begin{equation}
   u^+ = \zeta_+\otimes \bar{\zeta}'_+ 
      + \zeta_-\otimes \bar{\zeta}'_- \ ,
\end{equation}
where $\zeta_+$ and $\zeta'_+$ are positive chirality (complex)
spinors of $\Spin(6)$. For $u^-\in S^-$ one finds
\begin{equation}
   u^- = \zeta_+\otimes\bar{\zeta}'_-
      + \zeta_-\otimes \bar{\zeta}'_+\ .
\end{equation}
In both cases the $\Spin(6)$ acts on both the left and the right. 

The volume form $\epsilon_g$ defined by the metric and spin structure,
provides a natural isomorphism with $\Leo F^*$. This can be 
seen directly by taking Fierz identities. In particular, if $\gamma^m$ are
the gamma matrices for the $\Spin(6)$, one has 
\begin{equation}
\label{fierz}
   \zeta_{+}\otimes\bar{\zeta}'_\pm =
      \frac{1}{4} \sum_{k=0}^6 \frac{1}{k!}
      \left(\bar{\zeta}'_\pm\gamma_{m_1\dots m_k}\zeta_+\right)
         \gamma^{m_k\dots m_1} ,
\end{equation}
showing that any given $\Spin(6,6)$ spinor is equivalent to a set of
$k$-forms 
$\bar{\zeta}'_\pm\gamma_{m_1\dots m_k}\zeta_+$. Explicitly we have the
isomorphism 
\begin{equation}
   \zeta_{+}\otimes\bar{\zeta}'_\pm \sqrt{\epsilon_g} =
      \frac{1}{4} \bigoplus_{k=0}^6 \frac{1}{k!}
      \left(\bar{\zeta}'_\pm\gamma_{m_1\dots m_k}\zeta_+\right)
         e^{m_k}\wedge\dots\wedge e^{m_1} .
\end{equation}

Let us now rewrite a given $\SU(3)$ structure in this
formalism. Recall the $\SU(3)$ structure was defined by a spinor
$\eta_+$ with $\eta_-=\eta_+^c$. This allows us to define two complex
$\Spin(6,6)$ spinors as~\cite{GMPT} 
\begin{equation}
\label{SU3purespinors}
\begin{aligned}
   \eta_+\otimes\bar{\eta}_+ \sqrt{\epsilon_g}  
      &= \tfrac{1}{8}\ee^{-\ii J} 
         \in \Leven F^*_\bbC \ , \\
   \eta_+ \otimes \bar{\eta}_- \sqrt{\epsilon_g} 
      &= -\tfrac{1}{8}\ii \Omegan 
         \in \Lodd F^*_\bbC \ .
\end{aligned}
\end{equation}
$\ee^{-\ii J}$ and $\Omegan$ are known as ``pure
spinors''~\cite{Gualtieri} 
since their annihilator is a maximal isotropic subspace 
(for the case at hand, the annihilators of $e^{-\ii J}$ and $\Omegan$
are six-dimensional, and are given in~\cite{FMT,GMPT}). 
In terms of the $O(6,6)$ structure group on $F\oplus F^*$ each complex
pure spinor defines a $\SU(3,3)$ sub-bundle. Together the common
sub-bundle of the two $\SU(3,3)$ structures defined by $\ee^{\ii J}$
and $\Omegan$ is a $\SU(3)\times\SU(3)$ bundle. Within this there is an
$\SU(3)$ subgroup of diffeomorphisms $\GL(6,\bbR)\subset\SO(6,6)$
which leave $J$ and $\rho$ invariant and defines the original $\SU(3)$
structure. From this perspective, the $\SU(3)$ structure is defined by
the pair of pure spinors $\ee^{\ii J}$ and $\Omegan$.   

Now consider the case of a pair of $\SU(3)$ structures, given by
$\eta^A_+$. As discussed above, this is the generic situation for the
reformulated supergravity theory. Again one can construct two pairs
of pure complex spinors
$\eta_+^A\otimes\bar{\eta}^A_+\sqrt{\epsilon_g}=\ee^{\ii J^A}$ and 
$\eta^A_+\otimes\bar{\eta}_-^A\sqrt{\epsilon_g}=\Omegan^A$ with $A=1,2$. It
is actually more natural (and equivalent) to define,
following~\cite{JW,Witt}, the pure spinors 
\begin{equation}
\label{su2pure}
\begin{aligned}
   \pure^+ &= \eta_+^1\otimes \bar{\eta}_+^2\sqrt{\epsilon_g} 
      \in \Leven F^*_\bbC , \\
   \pure^- &= \eta_+^1\otimes \bar{\eta}_-^2\sqrt{\epsilon_g} 
       \in \Lodd F^*_\bbC .
\end{aligned}
\end{equation}
Their expression in terms of the local $SU(2)$ structure is given
in~\cite{JW,GMPT2}. Again here each pure spinor given
in~\eqref{su2pure} defines a $\SU(3,3)$ structure. Since as $\Spin(6)$
spinors $\bar{\eta}^A_\pm\gamma_m\eta_+^A=0$, it is easy to show that
$\bar{\pure}_{\epsilon_g}^-\Gamma^\Sigma\pure_{\epsilon_g}^+=0$. This
implies that together they define an $\SU(3)\times\SU(3)$
structure on $F\oplus F^*$. (This is discussed in more detail in
section~\ref{metricSK}.) Writing the structure in terms of $\pure^\pm$
the $\SU(3)\times\SU(3)$ action is immediately apparent: it
corresponds to two independent  $\SU(3)\subset\Spin(6)$ groups acting
separately on $\eta_\pm^1$ and $\eta_\pm^2$. In the generic case,
globally there is no common subgroup of $\SU(3)\times\SU(3)$ and we 
simply have a pair of $\SU(3)$ structures. Locally there is a common
$\SU(2)$ group, but this does not, generically, survive globally (this
is the case when the spinors $\eta^1$ and $\eta^2$ become parallel at
one or more points on the  manifold.) Nonetheless, globally, the pair
of $\SU(3)$ structures is equivalent to the pair of complex pure
spinors given in eq.~\eqref{su2pure}. In the special case where
$\eta^1_+=\eta^2_+=\eta_+$, we have a single $\SU(3)$ structure with 
\begin{equation}
   \pure^+ = \tfrac{1}{8}\ee^{-\ii J}\ , \qquad
   \pure^- = -\tfrac{1}{8}\ii \Omegan\ .
\end{equation}
This is the case with which we will be most concerned. However let us
continue a little further in the more general setting of
$\SU(3)\times\SU(3)$ structures and show that stable $\Spin(6,6)$
spinors define a special K\"ahler geometry following \cite{3form}.


\subsubsection{Special K\"ahler structure for stable $\Spin(6,6)$
  spinors}  
\label{genSK}

Consider a general odd or even form $\chi$, that is a section either
of $\Leven F^*$ or $\Lodd F^*$. Let $\chis\in S^\pm$ be the
corresponding spinor defined using the volume form
$\epsilon$. Following Hitchin~\cite{GCY,3form} we will show, first, that
there is a natural special K\"ahler structure on the space of so called
``stable'' spinors $\chis$, and, second, how this also gives a special
K\"ahler structure on the space of stable forms $\chi$. 

Just as a nowhere vanishing $\Spin(6)$ spinor defines an $\SU(3)$ structure, a 
nowhere vanishing $\Spin(6,6)$ spinor $\chis$ defines an $\SU(3,3)$
structure. That is to say a generic $\chis$ is invariant under
$\SU(3,3)\subset\Spin(6,6)$ rotations. As shown in~\cite{GCY}, not
quite all spinors define an $\SU(3,3)$ structure, but rather only an
open subset of $S^\pm$ corresponding to so-called ``stable spinors''. 
To see how the stable spinors are defined it is useful to start by
noting that, in analogy to the $\SU(3)$ case discussed in
section~\ref{SU3Geo}, one can construct $\SU(3,3)$-invariant forms 
out of the spinor bilinears. In particular, one can define a
fundamental two-form $\mathcal{J}\in\Lambda^2(F\oplus F^*)$ given by  
\begin{equation}\label{cJdef}
   \mathcal{J}_{\Pi\Sigma} =
      \chisb\,\Gamma_{\Pi\Sigma}\,\chis\ ,
   \qquad \Pi, \Sigma = 1, \ldots, 12\ ,
\end{equation}
where $ \Gamma_{\Pi\Sigma}$ is the antisymmetrized product of two 
$\SO(6,6)$ $\Gamma$-matrices. Using the $O(6,6)$ metric one can raise
one index forming $\mathcal{J}^\Pi{}_\Sigma$ which generically defines
an almost complex structure. The only caveat is that
$\mathcal{J}_{\Pi\Sigma}$ will not be properly normalized to be
compatible with the $\SO(6,6)$ metric, that is
$\mathcal{J}^\Pi{}_\Omega\mathcal{J}^\Omega{}_\Sigma=
-k^2\delta^\Pi{}_\Sigma$ but $k\neq1$. This is because the
normalization of $\chis$ is not fixed, and there is no simple way to
fix it, since $\chisb\chis=0$ identically as can be seen, for example,
from~\eqref{normC} or~\eqref{mukai}. Following Hitchin~\cite{3form},
one can instead introduce a quartic function of $\chis$, given by the
square of $\mathcal{J}_{\Pi\Sigma}$. One defines  
\begin{equation}
\label{qsdef}
   \qs(\chis) = -\tfrac{1}{4} 
      (\chisb\Gamma_{\Pi\Sigma}\chis)
      (\chisb\Gamma^{\Pi\Sigma}\chis)\ ,
\end{equation}
together with the homogeneous Hitchin function of degree two
\begin{equation}\label{phisdef}
   \hfuncs(\chis)\ =\ \sqrt{-\tfrac{1}{3}\,\qs(\chis)} \
       =\  \sqrt{\tfrac{1}{12}\,{\cal J}^2}
\end{equation}
Note that when $\mathcal{J}$ is correctly normalized,
$\hfuncs(\chis)=1$. 

With the help of these functions one can define the notion of stable
spinors. If $\mathcal{J}$ does define an almost complex structure then
clearly $\qs(\chis)<0$. In fact this is also sufficient for there to
be an $\SU(3,3)$ structure. 
Hitchin defines the set of \emph{stable} real $\Spin(6,6)$ spinors 
\begin{equation}
   \Us = \left\{ \chis\in S^\pm : \qs(\chis) < 0\right\}\  .
\end{equation}
By definition $\qs(\chis)\leq0$, and hence one can see that $\Us$ is 
an open subset of $S^\pm$, consisting of all spinors such that
$\mathcal{J}\neq 0$, and so is 32-dimensional. This is the statement
that a generic spinor $\chis$ defines a $\SU(3,3)$ structure. Clearly
$k\chi^\pm$, for any  non-zero $k\in\bbR-\{0\}$, defines
the same structure. Hence the set of stable forms $\Us$ is a
homogeneous space given by the set of $\SU(3,3)$ structures compatible
with the $O(6,6)$ metric together with an overall scale~\cite{GCY}   
\begin{equation}
   \Us \simeq O(6,6)\otimes\bbR^+/\SU(3,3) .
\end{equation}
In what follows it is useful to note that since $S^\pm$ is
a vector space and $\Us$ is an open subset of $S^\pm$, there is a
natural isomorphism between $T_{\chis}\Us$ at any point $\chis\in\Us$
and $S^\pm$. (For instance $\chis$ can be viewed either as a
coordinate on $\Us$ or the ``position vector field'' in $T\Us$.) 

On manifolds with $\SU(3)$ structure there is a pair of real
invariant spinors, or equivalently a complex invariant spinor 
(and its complex conjugate). The same is true for $\SU(3,3)$
structures. The second real spinor $\chihs$ can be written in
terms of $\chis$ by acting with the correctly normalized operator 
$\mathcal{J}^{\Pi\Sigma}\Gamma_{\Pi\Sigma}$. Explicitly one has
\begin{equation}\label{hatphi}
   \chihs = - \frac{1}{6\,\sqrt{\mathcal{J}^2/12}}\
        \mathcal{J}^{\Pi\Sigma}\Gamma_{\Pi\Sigma}\,\chis\ .
\end{equation}
This second spinor can also be defined in terms of the function
$\hfuncs(\chis)$. Recall that the spinor inner product defined a symplectic
structure $\sym{\xi}{\eta}=\bar{\xi}\eta$. The function
$\hfuncs(\chis)$ defines a Hamiltonian vector field $\chihs\in
T\Us\simeq S^\pm$
\begin{equation}
\label{hatdef}
   i_{\chihs}\omega = - \dd\hfuncs\ ,
\end{equation}
where $\dd$ is the exterior derivative on $\Us$, so $\dd\hfuncs\in
T^*\Us$.  In components we have
\begin{equation}\label{phicomp}
   \chihs^\alpha = 
      - (\omega^{-1})^{\alpha\beta}\der_\beta\hfuncs(\chis) , 
\end{equation}
where $\der_\alpha=\der/\der\chis^\alpha$.  
Using~\eqref{phisdef} and~\eqref{cJdef} it is straightforward to
see that~\eqref{phicomp} coincides with~\eqref{hatphi}.

The corresponding complex spinor is   
\begin{equation}\label{Phisdef}
   \pures = \tfrac{1}{2}\left(\chis + \ii\chihs\right) ,
\end{equation}
which was shown to be a pure spinor in ref.~\cite{GCY}. In what
follows it will be one of the two pure spinors defining the
$\SU(3)\times\SU(3)$ structure.  As a vector field $\chihs\in T\Us$
generates the $U(1)$ action $\pures\mapsto\ee^{\ii\theta}\pures$ where
we view the real and imaginary parts of $\pures$ as coordinates in $U$. 
Furthermore due to the homogeneity of $\hfuncs$ together with the
definition of $\chihs$ given in~\eqref{hatdef} and~\eqref{phicomp} one
infers 
\begin{equation}\label{phiomega}
   \hfuncs(\chis) = \tfrac{1}{2}\sym{\chis}{\chihs}
      = \ii\sym{\pures}{\puresb}\ .
\end{equation}

Now one can show that there is a natural rigid special K\"ahler
structure on $U$. There is already a symplectic structure given by the
spinor inner product. Since the matrix $\omega$ is constant, independent of
$\chis$, we clearly have 
\begin{equation}
\label{kahler}
   \dd\omega = 0 . 
\end{equation}
Actually we have more. A constant $\omega$ gives a flat symplectic
structure and implies that we can introduce Darboux coordinates
$\chis^\alpha=(x^K,y_L)$ on $\Us$ with $K,L=1,\dots,16$ such that  
\begin{equation}\label{Darboux}
   \omega = \dd x^K \wedge \dd y_K\ .
\end{equation}
These will be useful in what follows. 

Next one shows that the complex structure $\mathcal{J}$ on $F\oplus
F^*$ also induces a complex structure $I$ on $T\Us$. Specifically,
viewing $\chihs$ as an element of $T\Us$, one defines $I\in TU\otimes
T^*U$ by     
\begin{equation}
\label{Idef}
   I^\alpha{}_\beta = - \der_\beta\chihs^\alpha 
      = (\omega^{-1})^{\alpha\gamma}\der_\gamma\der_\beta\hfuncs\ ,
\end{equation}
where the second equation follows from~\eqref{phicomp} since
$\omega^{-1}=C$ is constant. In order to see that $I^2=-\id$ we first
note, from the definition~\eqref{hatphi}, that 
$\bar{\hat{\chi}}_\epsilon\Gamma_{\Pi\Sigma}\chihs
=\bar{\chi}_\epsilon\Gamma_{\Pi\Sigma}\chis$ and hence 
$\hfuncs(\chihs)=\hfuncs(\chis)$.  
Together with \eqref{phiomega} this implies 
$\sym{\chis}{\chihs}=\omega(\chihs,\hat{\hat{\chi}}_\epsilon)$
and hence 
%
\begin{equation}
   \hat{\hat{\chi}}_\epsilon = - \chis\ .
\label{hathat}
\end{equation}
Taking  a small variation of \eqref{phicomp} and using \eqref{Idef} we
find $\delta\chihs^\alpha=-I^\alpha{}_\beta\delta\chis^\beta$.
Using this and~\eqref{hathat} we find
$\delta\chis^\alpha=-\delta\hat{\hat\chi}_\epsilon^\alpha 
=I^\alpha{}_\beta\,\delta\hat\chi_\epsilon^\beta
=-I^\alpha{}_\beta\,I^\beta{}_\gamma\,\delta\chis^\gamma$ 
which indeed implies $I^2=-\id$.

To show that $I$ is also integrable one identifies explicit
complex coordinates. 
In the Darboux coordinates~\eqref{Darboux} the complex spinor $\pures$
of~\eqref{Phisdef} can be written as a complex vector on $\Us$ 
\begin{equation}\label{Phicoord}
\begin{split}
   \pures &= \tfrac{1}{2}\left(\chis + \ii\chihs\right) \\
     &= \frac{1}{2}\left(\chis^\alpha 
        - \ii (\omega^{-1})^{\alpha\beta}\der_\beta\hfuncs\right)
         \frac{\der}{\der\chis^\alpha} \\
     &= \frac{1}{2}\left(x^K+\ii\frac{\der\hfuncs}{\der y_K}\right)
        \frac{\der}{\der x^K}
       + \frac{1}{2}\left(y_K-\ii\frac{\der\hfuncs}{\der x^K}\right)
        \frac{\der}{\der y_K} \\
     &\equiv Z^K\, \frac{\der}{\der x^K} - F_K\, \frac{\der}{\der y_K}
\ ,
\end{split}
\end{equation}
where the last line defines the complex functions $Z^K$ and $F_K$ on
$\Us$. 
By definition the components $\pures^\alpha=(Z^K,-F_K)$ satisfy 
$\dd\pures^\alpha=\frac{1}{2}\dd\chis^\alpha-\frac{1}{2}\ii
I^\alpha{}_\beta\dd\chis^\beta$. Hence
$-\ii(I\dd\pures)^\alpha=\dd\pures^\alpha$ and the one-forms 
$\dd\pures^\alpha=(\dd Z^K,-\dd F_K)$ are all of type $(1,0)$ with
respect to $I$. This implies that $Z^K$ and $F_K$ are each separately
complex coordinates on $U$. (They are known as conjugate coordinate
systems.) 

We have shown that on $\Us$ there exists a closed symplectic form
$\omega$ and an integrable complex structure $I$. Furthermore,
from~\eqref{Idef}, one sees that 
$\omega_{\alpha\gamma}I^\gamma{}_\beta 
=\der_\alpha\der_\beta\hfuncs$ is symmetric. This implies $\omega$ and
$I$ are compatible (that is, $\omega$ is a $(1,1)$-form) and hence
together they define a K\"ahler metric. The metric is given by   
\begin{equation}
   G_\text{rigid} = \left(\omega_{\alpha\gamma}I^\gamma{}_\beta\right)
          \dd\chis^\alpha\otimes\dd\chis^\beta
      = \der_\alpha\der_\beta\hfuncs\;
          \dd\chis^\alpha\otimes\dd\chis^\beta\ .
\end{equation}
If we change to complex coordinates $Z^K$, since $G_\text{rigid}$
is by definition Hermitian, one has  
\begin{equation}
   G_\text{rigid} = \der_K\bar{\der}_L\hfuncs\;
      \dd Z^K \otimes \dd\bar{Z}^L \ ,
\end{equation}
where $\der_K=\der/\der Z^K$ and hence one can identify $\hfuncs$ as the
K\"ahler potential. Note that $\dd x^K=\dd Z^K+\dd\bar{Z}^K$ and $\dd
y_K=\dd F_K+\dd\bar{F}_K$. It is then easy to see that the Hermitian
condition (or equivalently the condition that $\omega$ is a
$(1,1)$-form) implies that $\der_{[K}F_{L]}=0$. This implies that
locally we can find a complex function $F$ such that $F_K=\der_K F$.

In summary, one sees, using~\eqref{phiomega} and~\eqref{Phicoord},
that the K\"ahler potential is given by 
\begin{equation}
   K_\text{rigid} \equiv \hfuncs = \ii\sym{\pures}{\puresb} = 
     \ii\left(\bar{Z}^K F_K - Z^K \bar{F}_K \right) ,
\end{equation}
where $F_K=\der_KF$. Comparing with~\eqref{rigidSK} and~\eqref{prepot}
we see that this is the standard form for a rigid special K\"ahler
geometry.\footnote{Anticipating the result we already denoted the
holomorphic section of special K\"ahler geometry in section \ref{SKG}
by $\pure$.} 
One can further show that this is actually a pseudo-K\"ahler geometry: the
signature of the metric $G_\text{rigid}$ is $(30,2)$~\cite{3form}. 

Since we are interested in gravitational theories we really want to
have a local special K\"ahler geometry. Fortunately there is a
straightforward way of obtaining such a structure given the rigid
geometry just described. Recall that $\chihs$ generated a $U(1)$ action on
$\Us$ corresponding to $\pures\to\ee^{\ii\theta}\pures$. The position vector
field $\chis$ generates a scaling $\pures\to\lambda\pures$. Together they
define a $\bbC^*$ action compatible with the complex structure (since
it is generated by the holomorphic vector field $\pures$). Thus one can
define the 30-dimensional quotient moduli space
\begin{equation}
   \mathcal{M}_\epsilon = \Us/\bbC^*\ .
\end{equation}
Under the $\bbC^*$ action the tangent space $T\Us$ descends to a
holomorphic $\Symp(32,\bbR)$ vector bundle $\mathcal{H}$ with
symplectic structure $\omega$. It is a standard result (see for
instance~\cite{freed}) that there is then a local special K\"ahler
structure on $\mathcal{M}$ with K\"ahler potential  
\begin{equation}
   K = - \ln \hfuncs = - \ln\ii\sym{\pures}{\puresb}
      = - \ln\ii\left(\bar{Z}^K F_K - Z^K \bar{F}_K \right) ,
\end{equation}
for some $\pures$ defined up to a K\"ahler transformation
$\pures\to\lambda\pures$ where $\lambda\in\bbC^*$. (In~\cite{freed}
local special K\"ahler manifolds are defined as such quotients.) The
corresponding metric is Euclidean. The $Z^K(z)$ are complex sections
of $\mathcal{H}$ where $z$ are complex coordinates on
$\mathcal{M}_\epsilon$. 
The moduli space corresponds to a space of
$U(3,3)$ structures compatible with the $O(6,6)$ metric, so we can
identify  
\begin{equation}
   \mathcal{M}_\epsilon \simeq O(6,6)/U(3,3) ,
\end{equation}
as the space of complex structures compatible with the
$O(6,6)$-metric. That $\mathcal{M}_\epsilon$ admits a Hermitian metric
is well known (see for instance~\cite{salamonb}). Hitchin's
result~\cite{GCY} is to show that it has a natural special K\"ahler
geometry.

Thus far the discussion has been in terms of the spinor $\chis$. As
such, it appears that the special K\"ahler structures depend on the
choice of volume form $\epsilon$ defining the isomorphism between
forms and spinors. In fact, the final local special K\"ahler structure
is actually independent of the choice of $\epsilon$. All of the
preceding discussion can be repeated in terms of the form $\chi\in\Leo
F^*$. Explicitly, one defines the analog of $\qs(\chis)$, using the
Mukai pairing~\eqref{mukai} rather than the symplectic form $\omega$, 
\begin{equation}
\label{qdef}
   q(\chi) = -\tfrac{1}{4}\mukai{\chi}{\Gamma_{\Pi\Sigma}\chi}
        \mukai{\chi}{\Gamma^{\Pi\Sigma}\chi}\ 
        \in\,  \Lambda^6F^*\otimes\Lambda^6F^* \ ,
\end{equation}
which is now formally the square of a volume form rather than a
scalar. The open set of stable forms, which is isomorphic to $\Us$, is
then given by  
\begin{equation}
\label{Udef}
   U = \left\{ \chi\in \Leo F^* : q(\chi) < 0 \right\} \simeq \Us \ .
\end{equation}
One has the analog of $\hfuncs(\chis)$, 
\begin{equation}
\label{phiforms}
\begin{aligned}
   \hfunc(\chi) 
      &= \sqrt{-\tfrac{1}{3}q(\chi)} \in\,  \Lambda^6F^* \\
      &= \hfuncs(\chis)\,\epsilon 
\end{aligned}
\end{equation}
which is now a volume form. 

The functional $\hfunc(\chi)$ defines a complex structure on $U$ in
complete analogy with the spinor case. One defines the Hamiltonian
vector field $\hat{\chi}$ on $TU\simeq\Leo F^*$ by the action on $TU$
\begin{equation}
   \mukai{\hat{\chi}}{\cdot} = - \dd\hfunc(\cdot)\  ,
\end{equation}
and the corresponding complex vector field
\begin{equation}
\label{phidef}
   \pure = \tfrac{1}{2}\left(\chi + \ii\hat{\chi}\right) .
\end{equation}
The complex structure is then defined as a derivative of $\pure$ as
before. 

The difference arises in the symplectic structure. We can define a map
$\Leo F^*\otimes\Leo F^*\to\bbR$ by $(\chi,\psi)\mapsto\sym{\chis}{\psis}$
but this depends on the choice of $\epsilon$. This means that there is
no canonical rigid special K\"ahler metric on $U$, but only a family
of metrics depending on $\epsilon$ with K\"ahler potentials
$K_\text{rigid}=\hfuncs(\chis)$. However, we can again form the
quotient moduli space 
\begin{equation}
\label{Mdef}
   \mathcal{M} = U/\bbC^* \simeq \mathcal{M}_\epsilon \ .
\end{equation}
The corresponding local special K\"ahler potential given by
$K=-\ln\hfuncs$ is \emph{independent} of the particular choice of
$\epsilon$ in defining the symplectic structure, since rescaling
$\epsilon$ simply shifts $K$ by a constant and corresponds to a
K\"ahler transformation. From this perspective we can introduce the
volume form 
\begin{equation}
\label{kahlerpot}
   \ee^{-K} = \hfunc = \ii\mukai{\pure}{\bar{\pure}}
      = \ii(\bar{Z}^KF_K-Z^K\bar{F}_K) 
      \in \Lambda^6F^* ,
\end{equation}
where we have introduced complex homogeneous coordinates $Z^K(z)$ as
above and $z^a$ are complex coordinates on $\mathcal{M}$ . The
special K\"ahler metric is then given by 
\begin{equation}
   \dd s^2 = \left(
      \frac{\der_a\bar{\der}_b\hfunc}{\hfunc} 
         - \frac{\der_a\hfunc}{H}\frac{\bar{\der}_a\hfunc}{H}\right)
      \delta z^a\delta\bar{z}^b ,
\end{equation}
where $\der_a=\der/\der z^a$ and the powers of $\Lambda^6F^*$ cancel
so that the metric $G$ really is a map $G:\Leo F^*\otimes\Leo F^*\to\bbR$. 

In summary, there is a unique special K\"ahler structure on the
(quotient) space $\mathcal{M}=U/\bbC^*$ of stable forms $\chi\in\Leo
F^*$ where the exponentiated K\"ahler potential $\ee^{-K}$ is
naturally a six-form given by the Hitchin function $H(\chi)$. From now
on we will consider only this structure on the space of forms and not
consider the corresponding spinors.


\subsubsection{Special K\"ahler structure for stable $J$} \label{JSK}

Having discussed the general case, let us now turn to the specific
special K\"ahler structures that arise from $\rho$ and $J$. Let us
start with the symplectic two-form $J$. From Calabi--Yau
compactifications we know that it is naturally paired with the
NS two-form $B$. To match with the discussion of the previous section,
we introduce the $\Spin(6,6)$ spinor 
\begin{equation}\label{eBJdef}
   \chi^+ = 2\re ( c\,\ee^{-B-\ii J} ) \in \Leven F^*
\end{equation}
where  $c\in\bbC-\{0\}$. Note that for general $J$, $B$ and
$c$ the spinor $\chi^+$ is completely generic. 

Using Hitchin's construction we find by inserting \eqref{eBJdef}
into \eqref{phiforms} using \eqref{Gamma-def}
\begin{equation}
   \hfunc(\chi^+) = \tfrac{4}{3}|c|^2 J\wedge J \wedge J\ .
\end{equation}
(One way to see this is to note that the phase of $c$ and the
entire $B$ dependence can be removed by a $O(6,6)$ rotation and hence
both drop out of $\hfunc(\chi)$.) The space of stable $\chi^+$ is then
given by 
\begin{equation}
\label{UJdef}
   U_J = \left\{ \re(c\,\ee^{-B-\ii J}) \in\Leven F^* : 
      J\wedge J \wedge J \neq 0 \right\} \ .
\end{equation}
This is equivalent to the usual condition that $J$ is
non-degenerate. The second spinor $\hat{\chi}$ is found from
\eqref{hatphi} to be  $2\im(c\,\ee^{-B-\ii J})$ and thus the pure complex
spinor reads 
\begin{equation}
   \pure^+ = c\,\ee^{-B-\ii J}\ .
\end{equation}

{}From \eqref{mukai} we see 
that $\Lambda^0F^*\oplus\Lambda^2F^*$ forms a maximal null
subspace of $\Leven F^*$ under the inner product $\mukai{\cdot}{\cdot}$. Thus
we can choose symplectic Darboux coordinates such that the $x^A$ span 
$\Lambda^0F^*\oplus\Lambda^2F^*$. 
In order to distinguish from the special K\"ahler structure for $\rho$
which we will discuss in the next section 
let us denote the complex coordinates on $U_J$ by $X^A$ and $\cF_A$
(instead of $Z^K$ and $F_K$). 
Expanding $\pure^+$ we determine the $X^A$ to be 
\begin{equation}
   X^0 = c , \qquad X^a = - c (B + \ii J)^a 
\end{equation}
where $a=[mn]$ running from $1$ to $15$ denotes the pair of
antisymmetric indices on the two forms. Finally, the $\bbC^*$ action
generated by $\pure^+$ acts by rescaling $c$, that is $c\to\lambda c$
with $\lambda\in\bbC^*$. 

{}From Hitchin's construction we immediately obtain a special K\"ahler
manifold on the quotient space  
\begin{equation}
   \mathcal{M}_J = U_J/\bbC^*\ ,
\end{equation}
with the K\"ahler potential $K_J$ given by 
\begin{equation}
\label{KJ}
   \ee^{-K_J} = \hfunc = \tfrac{4}{3}|c|^2\  J\wedge J \wedge J\ ,
\end{equation}
and we see $\ee^{-K}$ is naturally a six-form as discussed above. On
$\mathcal{M}_J$ one can introduce special coordinates defined in
section~\ref{SKG}  
\begin{equation}
\label{tdef}
   t^a = - X^a/X^0 = (B+\ii J)^a\ .
\end{equation}
(Note we have introduced an extra sign to match the conventional
definition of $t^a$.) In these coordinates the  prepotential $f(t^a)$
introduced in~\eqref{prepotf} then takes the form 
\begin{equation}
  \cF(X^A) = (X^0)^2\, f(t^a) \ , \qquad 
f(t^a)\, \epsilon = t\wedge t \wedge t\ .
\end{equation}
Note that these expressions are exactly analogous to the expressions
for the special K\"ahler geometry on the space of  K\"ahler
deformations on a Calabi--Yau
manifold except that here $J$ and $B$ are functions of all ten
spacetime coordinates and we do not necessarily have a Calabi--Yau manifold.

Finally we must show that the metric defined by $K_J$ corresponds to
the metric for the supergravity kinetic terms
$\dd s^2(J,B)$ given in~\eqref{KEmetric}. Starting from the
K\"ahler potential~\eqref{KJ} we find the K\"ahler
metric
\begin{equation}\label{metriFforms}
   \dd s^2_J = -\frac{3}{2}\left[
      \frac{\delta t \wedge \delta\bar{t} \wedge J}{J^3} 
      - \frac{3}{2}\frac{\delta t\wedge J^2}{J^3}
         \frac{\delta\bar{t}\wedge J^2}{J^3}\right]\ .
\end{equation}
Note that here we see explicitly that both numerator and denominator
are proportional to the volume form which therefore cancels in the ratio. 
Rewriting this expression in terms of contractions with the metric
$g_{mn}$ we have
\begin{equation}
   \dd s^2_J = \tfrac{3}{4}\delta\lambda\delta\lambda
         + \tfrac{1}{2}g^{mn}\delta v_m\delta v_n
         + \tfrac{1}{8}g^{mp}g^{nq}\delta K_{mn}\delta K_{pq}
         + \tfrac{1}{8}g^{mp}g^{nq}\delta B_{mn}\delta B_{pq} , 
\end{equation}
which precisely matches the metric~\eqref{KEmetric} (up to the terms
involving the vector deformation $\delta v$). Again this is similar to the
situation in Calabi--Yau manifolds where the analogous computation can
be found in refs.~\cite{Strominger,CdO}. Recall that the $\delta v$
terms do not correspond to physical deformations but are associated
with different $\SU(3)$ structures which define the same metric. In
the full theory we expect them not to be present. We discuss this in
some more detail in section~\ref{metricSK}. In the limit where we drop
additional spin-$\tfrac32$ multiplets and hence all fields in the
$\rep{3}+\rep{\bar{3}}$ representation of $\SU(3)$, we set $\delta v=0$
and the agreement is exact.


\subsubsection{Special K\"ahler structure for stable $\rho$}
\label{rhoSK}

We now turn to the almost complex structure on $F$ defined by
$\rho\in\Lambda^3F^*$. Unlike the previous case this is not a generic
$\Spin(6,6)$ spinor since it does not contain one-form or five-form
pieces. Nonetheless an analogous construction of the special K\"ahler
moduli space exists. (This was also first given by Hitchin
in~\cite{3form}.) 

One starts by noting that the pairing $\mukai{\cdot}{\cdot}$ defined
in \eqref{mukai} still
defines a symplectic structure on the subspace $\Lambda^3F^*\in\Lodd
F^*$. Then one introduces the (restricted) spinor 
\begin{equation}
   \chi^- = 2\cscale\,\rho \in \Lambda^3F^* \subset \Lodd F^*\ ,
\end{equation}
where the  factor of two is for later convenience and 
$\cscale\in\bbR-\{0\}$ is an arbitrary normalization constant
analogous to the $c$ introduced in~\eqref{eBJdef}. For this restricted
spinor one can still define the forms $q(\chi^-)$ and $\hfunc(\chi^-)$
as in~\eqref{qdef} and~\eqref{phidef}. Explicitly one finds,
using~\eqref{Gamma-def},  
\begin{equation}
\label{q3form}
   q(\chi^-) = 8\, \cscale^4 \,(e^m\wedge i_n\rho\wedge\rho)
      (e^n\wedge i_m\rho\wedge\rho) \
      \in\ \Lambda^6F^*\otimes\Lambda^6F^* \ .
\end{equation}
The space of stable three-forms is 
\begin{equation}
\label{Urdef}
   U_\rho = \left\{ \rho\in\Lambda^3F^* : q(\rho)<0 \right\}\ .
\end{equation}
The key here is that only the $\GL(6,\bbR)\subset\Spin(6,6)$
generators in $\Gamma^{\Pi\Sigma}$ give a non-zero contribution to
$q(\rho)$. From this perspective, rather than defining a
$\SU(3,3)$ substructure of $\Spin(6,6)$, the three-form $\rho$ defines
a $\SL(3,\bbC)$ substructure of $\GL(6,\bbR)$. The spinor stability
then reduces to a notion of three-form stability, that is $U_\rho$
defines an open orbit under $\GL(6,\bbR)$. 
Correspondingly $U_\rho$ is isomorphic to the homogeneous space 
\begin{equation}
   U_\rho \simeq \GL^+(6,\bbR)/\SL(3,\bbC) ,
\end{equation}
where $\GL^+(6,\bbR)$ is the space of real matrices with positive
determinant. Note if we had considered the generic case of
$SU(3) \times SU(3)$ structures, rather than
restricting to $\SU(3)$ structures, $\chi^-$ as
defined~\eqref{su2pure} would be a generic element of $\Lodd F^*$ and
we would be back to the general Hitchin construction. 

Given $\hfunc(\chi^-)$ the construction continues just as above. One
defines $\hat{\chi}^-\in\Lambda^3F^*$, by 
\begin{equation}
\label{rhatdef}
   \mukai{\hat{\chi}^-}{\cdot} = - \dd\hfunc(\cdot)
\end{equation}
so $\hat{\chi}^-=2\cscale\hat{\rho}$ and then the complex spinor
\begin{equation}
   \Omega = \chi^- + \ii\hat{\chi}^- = \cscale\Omegan 
      \in \Lambda^3F^*_\bbC\ ,
\end{equation}
which is the $(3,0)$-form usually used to define a $\SU(3)$
structure. (Recall that $\Omegan$ was defined in \eqref{Jeta} in terms
of normalized spinors $\bar\eta\eta=1$ and 
$\Omega$ differs from  $\Omegan$ by the arbitrary
normalization $\cscale$.)

The $\bbC^*$ action generated by $\Omega$ acts as
$\Omega\to\lambda\Omega$ with $\lambda\in\bbC^*$. One then constructs
the quotient space 
\begin{equation}
   \mathcal{M}_\rho = U_\rho/\bbC^*\ ,
\end{equation}
with the K\"ahler potential $K_\rho$ is defined via the six-form 
\begin{equation}
   \ee^{-K_\rho} = \hfunc = \ii \Omega \wedge \bar{\Omega}\ .
\end{equation}
Note that modding out by the $\bbC^*$ action means that points in
$\mathcal{M}_\rho$ do not distinguish the phase of $\Omega$. Thus
$\mathcal{M}_\rho$ is the moduli space of almost complex structures
rather than $\SL(3,\bbC)$ structures on $F$ and we can identify 
\begin{equation}
   \mathcal{M}_\rho \simeq \GL^+(6,\bbR)/\GL^+(3,\bbC) .
\end{equation}
Choosing a symplectic
basis of $\Lambda^3F^*$ one can introduce complex coordinates
$Z^K$. Crucially, as above, $\Omega$ is a holomorphic section, that is
it depends only on the holomorphic coordinates $z^k$ on
$\mathcal{M}_\rho$. 

As for the case with $J$, one notes that the special K\"ahler geometry
is governed by expressions which are exactly analogous to the
corresponding expressions for the special K\"ahler geometry of complex
structure deformations on a Calabi--Yau manifold~\cite{CdO}. The
difference  here is again that $\Omega$ is a function of all ten
spacetime coordinates and we do not necessarily have a Calabi--Yau
manifold.

Finally we must show that the metric defined by $K_\rho$ indeed
coincides with the 
scalar kinetic term metric $\dd s^2(\rho)$ given
in~\eqref{KEmetric}. To do so it is convenient to note that, using the
Mukai pairing, any complex three-form can be decomposed into a part
along $\Omega$ and an orthogonal piece. In particular, for the
derivative $\der\Omega/\der z^k$ where $z^k$ are holomorphic
coordinates on $\mathcal{M}$, we can always write
\begin{equation}
   \frac{\der\Omega}{\der z^k} = K_k \Omega + \psi_k
\end{equation}
where $\psi_k$ is defined by $\mukai{\psi_k}{\bar\Omega}=0$. The metric on
$\mathcal{M}$ then takes the form 
\begin{equation}
\label{Grho}
   \dd s^2_\rho =
      - \frac{\ii\psi_k\delta z^k\wedge\bar{\psi}_k\delta\bar{z}^k}
        {\ii \Omega\wedge\bar{\Omega}} . 
\end{equation}
Given the expansion of $\rho$ in~\eqref{delJr} we can identify
\begin{equation}
   K_k\delta z^k = \tfrac{3}{2}\delta\lambda - \ii\delta\gamma , 
      \qquad
   \psi_k\delta z^k = 
      \left(\delta M - \delta v\wedge J\right)
         + \ii *\left( \delta M + \delta v\wedge J \right) .
\end{equation}
Substituting into~\eqref{Grho} and rewriting in terms of contractions
with $g_{mn}$ one finds 
\begin{equation}
   \dd s^2_\rho = 
       \tfrac{1}{24}g^{mr}g^{ns}g^{pt}\delta M_{mnp}\delta M_{rst}
            + \tfrac{1}{2}g^{mn}\delta v_m\delta v_n\ ,
\end{equation}
which exactly matches the scalar kinetic term metric
of~\eqref{KEmetric} (up to terms in the vector deformation $\delta v$). 
Again this is analogous to the Calabi--Yau computation which can be
found in ref.~\cite{CdO}. As before the $\delta v$ deformation is
dropped when we remove the extra spin-$\tfrac32$ multiplets by
ignoring all fields in the $\rep{3}+\rep{\bar{3}}$ representation of
$\SU(3)$.


\subsubsection{Special K\"ahler structure for metric
  $\SU(3)\times\SU(3)$-structures}
\label{metricSK}

In the previous two sections we have seen that there are natural
special K\"ahler structures on the space of two forms $B+\ii J$ and
three-forms $\rho$ respectively, each of which agrees with the
corresponding metric in the type II supergravity theory. However
rather than general $J$ and $\rho$ the actual supergravity degrees of
freedom are the metric $g$. In this section, we discuss some of the
issues involved in removing the redundant degrees of freedom. However
a full analysis of the structures which arise will depend on including
the couplings of the additional spin-$\tfrac32$ multiplets and hence
our discussion will not be complete. 

To go from $J$ and $\rho$ to $g$ involves two steps. First one imposes
the conditions 
\begin{equation}
\label{Jrcond}
   J \wedge \rho = 0 , \qquad
   J^3 = \tfrac{3}{2} \rho \wedge \hat{\rho} .
\end{equation}
These imply that together $J$ and $\rho$ define an
$\SU(3)$-structure. Secondly we need to mod out by those degrees of
freedom that are not physical, that is take equivalence classes of
$(J,\rho)$ which describe the same metric $g$ on $F$. In terms of
deformations these are the degrees of freedom parameterized by $v$ and
$\gamma$ in~\eqref{delJr}. Given any stable $\rho$ we can always rescale
$\rho$ so as to satisfy the second condition
in~\eqref{Jrcond}. However modding out by the scale of 
$\rho$ together with the $\gamma$ deformation corresponds precisely to the
$\bbC^*$ action on $U_\rho$ used to define the moduli space
$\mathcal{M}_\rho$. Similarly the $\bbC^*$ action on $U_J$ removes the
non-physical degree of freedom $c$ in~\eqref{eBJdef}. Thus the actual
problem on $\mathcal{M}_J\times\mathcal{M}_\rho$ is to impose the first
condition in~\eqref{Jrcond} and mod out by the vector $v$ degrees of
freedom.

Crucially, one notes that both the constraint $J\wedge\rho=0$ and $v$
transform in the $\rep{3}+\rep{\bar{3}}$ representation under
$\SU(3)$. In the approximation where we are dropping the additional
spin-$\tfrac32$ multiplets we drop all triplet representations. Thus,
in this case, the condition is necessarily satisfied, there are no
$v$ deformations and $(J,\rho)$ define a unique metric. This is
precisely what happens for Calabi--Yau manifolds and also will be our
assumption in section~\ref{d=4N=2} where we will consider truncations
of the general theory. Here we will discuss something of what form the
general theory may take. 

Let us start with the constraint. In fact, it is straightforward to
consider the general case of $\SU(3)\times\SU(3)$ structures. Recall
that a generic stable form $\chi^\pm\in\Leo$ defines an $\SU(3,3)$
structure. Let $U^\pm$ denote the corresponding spaces of stable
$\chi^\pm$, and $\mathcal{M}^\pm=U^\pm/\bbC^*$.  We must identify
those pairs of spinors $(\chi^+,\chi^-)$ which determine an
$\SU(3)\times\SU(3)$ structure, as defined in ref.~\cite{JW}. (In the
case where the structure is integrable, this should presumably be
equivalent to what Gualtieri~\cite{Gualtieri} calls a  ``generalized
Calabi--Yau  metric''.) This condition implies that the spinors define
a metric $g$ and  $B$-field. Taking a generic pair is not sufficient:
we  must impose a condition.  This is the generalization of the
conditions~\eqref{Jrcond} that $J$ and $\rho$ together define a metric
$g$. It defines a subspace 
\begin{equation}
   V \hookrightarrow U^+ \times U^- , 
\end{equation}
which can be identified with the 52-dimensional homogeneous space 
\begin{equation}
   V \simeq O(6,6)\otimes\bbR^+\otimes\bbR^+/(\SU(3)\otimes\SU(3)) .
\end{equation}
The two factors of $\bbR^+$ here correspond to the fact that rescaling
either $\chi^\pm$ by a  non-zero real number defines the
same $\SU(3)\times\SU(3)$ structure. 

The condition can be expressed in a number of ways. Let 
$\mathcal{J}^{\pm\,\Pi}{}_\Sigma$ be the pair of almost complex
structures~\eqref{cJdef} on $F\oplus F^*$ defined by $\chis^\pm$. Two 
equivalent formulations of the condition are
\begin{equation}
\label{mcond}
   [\mathcal{J}^+, \mathcal{J}^-] = 0 \quad \Leftrightarrow \quad
   \mukai{\chi^+}{\Gamma^\Pi\chi^-} = 0 . 
\end{equation}
The first form was first given
in~\cite{Gualtieri}.\footnote{We thank Marco Gualtieri for
discussions confirming the second form of the condition.}  To see
that the second form is a necessary condition one simply notes that
there are no scalars in the decomposition of the vector representation of
$O(6,6)$ under $\SU(3)\times\SU(3)$. From it we learn that $V$ is a
52-dimensional subspace of $U^+\times U^-$. To see explicitly how that
metric and $B$-field arise, one defines a metric $\mathcal{G}$ on
$F\oplus F^*$ as 
\begin{equation}
   \mathcal{G} = \frac{\mathcal{J}^+\mathcal{J}^-}
         {\sqrt{\hfuncs(\chis^+)\hfuncs(\chis^-)}}
      = \frac{\mathcal{J}^-\mathcal{J}^+}
         {\sqrt{\hfuncs(\chis^+)\hfuncs(\chis^-)}} \ .
\end{equation}
The factors of $\hfuncs(\chis^\pm)$ normalize the almost complex
structures, and also ensure that $\mathcal{G}$ is independent of
$\epsilon$. The metric can be written in terms of $g$ and $B$
as~\cite{Gualtieri} 
\begin{equation}
   \mathcal{G}^\Pi{}_\Sigma
      = \begin{pmatrix} -g^{-1}B & g^{-1} \\
         g - Bg^{-1}B & Bg^{-1} \end{pmatrix} ,
\end{equation}
a form familiar from discussions of T-duality \cite{Narain}. If we restrict to the
special case of $\SU(3)$ structures, the condition~\eqref{mcond} is
equivalent to 
\begin{equation}
\label{JBcond}
   J \wedge \rho = B \wedge \rho = 0 
\end{equation}
implying that when restricting to $\SU(3)$ structures it is natural to
restrict $B$ as well. 

The condition~\eqref{mcond} can also be written in terms of the complex
(pure) spinors $\Phi^\pm$ as 
\begin{equation}
\label{33complex}
   \mukai{\Phi^+}{\Gamma^\Pi\Phi^-} = 0 . 
\end{equation}
(For $\SU(3)$ structures the corresponding condition is $(B+\ii
J)\wedge\Omega=0$.) This form is manifestly invariant under the
$\bbC^*$ actions on $U^\pm$ used to define the moduli spaces
$\mathcal{M}^\pm$. Thus we can also view the
condition as defining a subspace of $\mathcal{M}^+\times\mathcal{M}^-$  
\begin{equation}
   \mathcal{N} \hookrightarrow \mathcal{M}^+\times\mathcal{M}^- .
\end{equation}
Equivalently we can think of $\mathcal{N}$ as the quotient of $V$ by
the two $\bbC^*$ actions. Again this gives a homogeneous space, now
48-dimensional,  
\begin{equation}
   \mathcal{N} \simeq O(6,6)/(U(3)\times U(3)) . 
\end{equation}

Since $\Phi^\pm$ are holomorphic functions on $\mathcal{M}^\pm$ the
condition~\eqref{33complex} represents a complex submanifold
$\mathcal{N}$. This means that the special K\"ahler structure on the total
space $\mathcal{M}^+\times\mathcal{M^-}$ induces a special K\"ahler
structure on $\mathcal{N}$. Crucially however $\mathcal{N}$ is no
longer a product manifold where the $\chi^+$ and $\chi^-$ structures
separate into one special K\"ahler manifold for the vector multiplet
scalars and one for (part of) the hypermultiplet scalars. This is
contrary to the usual $d=4, N=2$ form of supergravity. 

Even on the constrained manifold $\mathcal{N}$ we are over-counting
the degrees of freedom since many $\SU(3)\times\SU(3)$ structures
define the same metric and $B$-field. As we already observed in
\eqref{KEmetric} the actual 36-dimensional space
of $g$ and $B$ is isomorphic to the homogeneous Narain moduli space
\cite{Narain} 
\begin{equation}\label{Narainms}
   \mathcal{Q} = O(6,6)/(O(6)\times O(6)) .
\end{equation}
This can be obtained from $\mathcal{N}$ by modding out by those elements
of $O(6)\times O(6)$ not in $\SU(3)\times\SU(3)$. In the
$\SU(3)$ case this corresponds to modding out by the $v$
deformations. This implies that $\mathcal{N}$ is a fibration over
$\mathcal{Q}$
\begin{equation}
\label{fibration}
   \begin{CD}
      \bbC P^3\times\bbC P^3 @>>> \mathcal{N} \\
      && @VVV \\
      && \mathcal{Q}
   \end{CD}
\end{equation}
with fibers 
\begin{equation}
   O(6)/U(3)\times O(6)/U(3) \simeq \bbC P^3\times\bbC P^3 .
\end{equation}
Although the fibers admit a complex structure, it appears that with
respect to the special K\"ahler metric on $\mathcal{N}$ this is not a
complex fibration, and so there is no natural complex structure or
K\"ahler metric on $\mathcal{Q}$. Note that in the restricted case of
$\SU(3)$ structures the corresponding fibration is by a single $\bbC
P^3$ factor. 

We see that in the full theory reducing to the physical degrees of
freedom in $g$ and $B$ takes us away from the usual form of $N=2$
supergravity: first since $\mathcal{N}$ is not a product, and second
in that there is no natural special K\"ahler structure on
$\mathcal{Q}$. Of course this is related to the fact that keeping 
the $SU(3)$ triplets correspond to additional 
spin-$\tfrac32$ gravitinos and additional vectors and scalars which enlarge the
$N=2$ multiplets to multiplets of higher $N$. 
More precisely, if we also include
all scalar fields from the RR-sector the combined field space 
would be the $N=8$ coset $E_{7(7)}/SU(8)$. Here we
concentrated only on the metric and B-field deformations and hence
only discovered the NS-subspace ${\mathcal Q}$ given in \eqref{Narainms}.

In summary, we have seen from Hitchin's results that there are natural
special K\"ahler metrics on the spaces of $\SU(3,3)$ structures given
by spinors $\eta^\pm$. In the special case of $\SU(3)$ structures
$(J,\rho)$ together with the two-form field $B$, we have the K\"ahler
potentials 
\begin{equation}
\label{Kdefs}
\begin{aligned}
   \ee^{-K_J} &= \tfrac{4}{3}|c|^2J\wedge J \wedge J \ , \\
   \ee^{-K_\rho} &= \ii \Omega\wedge\bar{\Omega} \ .
\end{aligned}
\end{equation}
These are the obvious generalizations of the corresponding Calabi--Yau
K\"ahler potentials and correspond to the Hitchin functionals
$H(\chi^\pm)$ where $\chi^+=\re(2c\,\ee^{-B-\ii J})$ and
$\chi^-=\cscale\rho$. The construction generalizes to
$\SU(3)\times\SU(3)$ structures simply by taking $\chi^-$ to be a
generic (stable) odd-form, this is, with one-form and five-form pieces
in addition to $\rho$. If we ignore all triplet representations these
K\"ahler potentials give the metrics calculated directly from
supergravity. In addition all the degrees of freedom in $J$ and $\rho$
are physical. If we drop this condition we must impose the $\SU(3)$
structure constraints~\eqref{JBcond} and mod out by the non-physical
deformation $v$. The resulting structures are not naturally special
K\"ahler manifolds but this is not surprising since in this case we must
include additional multiplets which modify the scalar field sector.


\subsection{Supersymmetry transformation of the gravitinos}
\label{N=2W}


So far we concentrated on the moduli space $\mathcal{M}$ of metric
deformations and showed that it can be universally determined
independently of the specific class of $SU(3)$ structure under
consideration. We have seen that the ``kinetic terms'' in the type II
supergravity are realized in terms of a special K\"ahler sigma model
metric on $\mathcal{M}$ with a K\"ahler potential given in terms of
the corresponding Hitchin functionals. Let us now turn to the
contributions of the ``potential terms'' that is terms without
derivatives in the $T^{1.3}$ bundle.  

In supersymmetric theories the potential is given generically by the
sum of the squares of the scalar part of the 
supersymmetry tranformations of the fermion
fields. Furthermore for the fermions in the vector, tensor and
hypermultiplets this scalar part of the variations is determined by
derivatives of the scalar part of the gravitino variation
\cite{N=2review}. Or in other words the transformation of the
gravitino is the fundamental quantity from which one obtains all terms
in the potential by appropriate derivatives. 
  
In four-dimensional $N=2$ supergravity the transformation of the
gravitinos has the generic form 
\beq 
\label{defS}
   \delta \psi_{A\,\mu} = D_\mu \varepsilon_A
      + i \gamma_\mu S_{AB} \varepsilon^B \ ,
\eeq
where $\varepsilon_A$ with $A=1,2$ are the two supersymmetry parameters
of $N=2$, while $\varepsilon^A$ are the conjugate spinors . In terms
of the ten-dimensional decomposition~\eqref{decompepsilon}, we have
$\varepsilon_A=\varepsilon^A_+$ and
$\varepsilon^A=\varepsilon^A_-$. $S_{AB}$ is an $SU(2)$
matrix\footnote{The four-dimensional $N=2$ theory has a local
  $SU(2)_R$ symmetry which rotates the two (complex)  gravitinos
  $\psi_{A\,\mu}$ into each other. In  ten dimensions it arises from the
  $O(2)$ rotation of the two ten-dimensional Majorana-Weyl fermions into
  each other together with additional generators which, as we will see,
  arise in the decomposition \eqref{decomp}, \eqref{decompIIB}. This
  $SU(2)_R$ is unrelated to the $SU(3)$ or $SU(3)\times SU(3)$ discussed
  so far in this section.}  
which can  equivalently be expressed in terms of three Killing
prepotentials $\cP^x, x=1,2,3$ via
\begin{equation}
\label{4dsusylaws}
   S_{AB} =
      \frac{\ii}2\, \ee^{\tfrac12 K_V}\sigma^x_{AB} \cP^x , 
   \qquad
   \sigma_{AB}^x =
      \begin{pmatrix}
         \delta^{x1} - i \delta^{x2}& -\delta^{x3} \\ 
         -\delta^{x3} & - \delta^{x1} - i \delta^{x2} 
      \end{pmatrix} , 
\end{equation}
where $K_V$ is the K\"ahler potential of the vector multiplets.
The three $\cP^x$ can be viewed as the $N=2$
equivalent of the $N=1$ superpotential and the $N=1$ $D$-term.
Together with its derivatives,
$S$ (or $\cP^x$) determines the scalar potential.

In the spirit of this section we now want to `lift' this discussion to the
full ten-dimensional theory in a background with $SU(3)$ structure.
To do so we simply compute the supersymmetry transformation of the
gravitinos which reside in the gravitational multiplet and write it in a
form analogous to \eqref{defS}. From the result we then
read off the ten-dimensional analogue of $S_{AB}$ or $\cP^x$.

In tables~\ref{FII}, \ref{N=2multipletsA} and \ref{N=2multipletsB}
we determined that the gravitinos in the gravitational multiplet can
be easily identified as the $\SU(3)$-singlet of $\Psi_\mu$. However,
the supersymmetry transformation \eqref{defS} in addition requires 
that its kinetic term is diagonal and no mixed terms with the
spin-$\frac12$ fermions occur. This determines the gravitino as a
particular linear combination among the fermions which now determine. 

We start from the canonical ten-dimensional kinetic term for
${\Psi}_M$. Here it is actually most convenient to start in the
ten-dimensional Einstein frame with metric $g_E$ related to the string
frame metric $g$ by $g_E=\ee^{-\phi/2}g$, since the conventional
gravitino field in this frame has no derivative  coupling to the
dilatino field. The kinetic term then takes the form
\beq
   S_\text{gravitino} = - \frac{1}{\kappa_{10}^2} 
      \int \dd^{10}x\, \sqrt{-g_E} \, 
         \bar{{\Psi}}_M \Gamma^{MNP} D_N {\Psi}_P \ ,
         \qquad M,N,\ldots = 0,\dots,9\ .
\label{10dgravkin}
\eeq
One can immediately see that the simple split
${\Psi}_M=({\Psi}_\mu,{\Psi}_m)$ does not lead to diagonal kinetic
terms. Instead we have to redefine the gravitino according to  
\beq 
\label{red4Dgrav}
   \hat{\Psi}_{\mu} \equiv {\Psi}_{\mu} 
      + \tfrac{1}{2} \Gamma_{\mu}{}^m {\Psi}_m \ ,
\eeq
where $\Gamma_{\mu}{}^m=g_E^{mn} \Gamma_{[\mu m]}$.
When inserted into \eqref{10dgravkin} one easily checks that 
$\hat{\Psi}_{\mu}$ now has a diagonal kinetic term
and therefore should be identified as the field which transforms
as in \eqref{defS}. 

Using \eqref{repdecomp} and the fact that 
we need to focus on the singlet part of $\hat{\Psi}_{\mu}$ we
can express $\hat{\Psi}_{\mu}$ in terms of the $\SU(3)$ 
singlet spinors $\eta_\pm$.
For type IIA we thus have 
\begin{equation}
\label{decomp}
\begin{aligned} 
   \hat \Psi^{\text{IIA}}_{1\,\mu} &=
      \psi_{1\,\mu +} \otimes \eta_+
      + \psi_{1 \, \mu-} \otimes \eta_- + \ldots \ , \\
   \hat \Psi^{\text{IIA}}_{2\, \mu} &= 
      \psi_{2\,\mu -} \otimes \eta_+ 
      + \psi_{2\,\mu+} \otimes \eta_- + \ldots\ ,
\end{aligned}
\end{equation}
where we omitted the triplets.
The indices $1$ and $2$ distinguish the two gravitinos 
which have opposite chirality in IIA.
By slight abuse of notation  we
use plus and minus to indicate both four-dimensonal and
six-dimensional chiralities, respectively. In type IIB both gravitinos
have the same chirality (which we take to be negative) and thus
the appropriate decomposition reads
\beq 
\label{decompIIB}
\hat \Psi^{\text{IIB}}_{A\, \mu} = 
   \psi_{A \, \mu +} \otimes \eta_- 
   + \psi_{A\,\mu-} \otimes \eta_+  +\ldots \ , \qquad A=1,2\ .
\eeq
Recall that there is a similar decomposition of the supersymmetry
parameters given in eqs.~\eqref{decompepsilon}
and~\eqref{decompepIIB}. 
For simplicitly let us consider in the follwing the special case
where we have a $\SU(3)$ structure rather than a $\SU(3)\times\SU(3)$
structure and thus take $\eta_+^1=\eta_+^2=\eta_+$. 
(The computation for a more general $\SU(3)\times\SU(3)$ structure
will be presented elsewhere.)

In type II supergravity the supersymmetry transformation of the 
gravitinos in the Einstein frame is 
\begin{multline}
\label{susygravitinoIIA}
   \delta \Psi_M = D_M \epsilon 
      - \frac{1}{96}\ee^{-\phi/2}\left(
         \Gx_M\,^{PQR} H_{PQR} - 9 \Gx^{PQ} H_{MPQ}\right)P\epsilon  \\
      - \sum_{n} \frac{\ee^{(5-n)\phi/4}}{64\,n!} 
         \left[ (n-1)\Gx_M{}^{N_1\dots N_{n}} 
            - n(9-n) \delta_M{}^{N_1}\Gx^{N_2\dots N_n} \right]
         F_{N_1...N_{n}}\, P_n\, \epsilon  \, ,
\end{multline}
We are using the democratic formulation of~\cite{Bergshoeff}, for
which the sum is over $n=0,2,4,6,8$, $P = \Gamma_{11}$ and $P_n =
-(\Gamma_{11})^{n/2} \sigma^1$ for IIA. For IIB we have instead sum
over $n=1,3,5,7,9$, $P = -\sigma^3$ and $P_n = i \sigma^2$ for
$n=1,5,9$ and $P_n = \sigma^1$ for $n=3,7$. 
Furthermore, 
\begin{equation}\label{Fdef}
F_{n}=dC_{n-1} - H \wedge C_{n-3}
\end{equation}
are the modified RR field
strengths with non standard Bianchi identities. 

Turning on only fluxes that preserve Poincare invariance, and using
the duality relation $F_{n}=(-1)^{\Int[n/2]} *F_{10-n}$, we can write the
supersymmetry transformation in terms of only internal fluxes 
$F_n$, $n=0,\dots,6$. For instance a non-zero $F_4$ with only $\mu$-type
indices is traded for a ``internal'' $F_6$  with $m$-type
indices. In~\eqref{susygravitinoIIA} this gives  twice the contribution
for each flux but now $n$ only takes the values $n=0,\dots,6$. 

To extract the supersymmetry transformation of $\psi_\mu^A$ 
we need to project onto the $\SU(3)$-singlet part of the variation of
the shifted gravitino $\hat{\Psi}^{A}_\mu$. Given the
normalizations of $\eta_\pm$ the relevant projection operators are
\begin{equation}
   \Pi_\pm = \id \otimes 
     2\left(\eta_\pm \otimes\bar{\eta}_\pm \right) \ .
\end{equation}
In order to extract the supersymmetry transformation of $\psi_{\mu +}$,
we should use $\Pi_+$ (or $\Pi_-$) when dealing with a 
10-dimensional positive (or negative) chirality spinor. To match our
definitions in the previous section, we will also rewrite the
variation in terms of the string frame metric $g=\ee^{\phi/2}g_E$ in
the following. This introduces additional factors of the dilaton
into the expressions in~\eqref{susygravitinoIIA}.

Let us first focus on type IIB for which we evaluate, in the string
frame, 
\begin{equation}
\label{susy4DIIB1}
\begin{split}
   \Pi_-\,\delta\hat{\Psi}_{\mu} &= D_{\mu}\varepsilon_+\otimes\eta_- 
      -  \frac{1}{48}\left(
         \bar{\eta}_-\gamma^{mnp}\eta_+ H_{mnp}\right)
         \gamma_{\mu} \sigma^3 \epsilon_-\otimes\eta_- 
         \\ & \qquad\qquad\qquad\qquad\qquad
      - \frac{1}{48}\left(
         \ee^{\phi}\, \bar{\eta}_-\gamma^{mnp}\eta_+ F_{mnp}\right)
         \gamma_{\mu} \sigma^1 \epsilon_-\otimes\eta_- \\     
      &= D_{\mu}\varepsilon_+\otimes\eta_- 
      -  \frac{\ii}{96}\left(
         {\Omegan}^{mnp}H_{mnp}\right)
         \gamma_{\mu} \sigma^3 \epsilon_-\otimes\eta_- 
         \\ & \qquad\qquad\qquad\qquad\qquad
      - \frac{\ii}{96}\left(
         \ee^{\phi} \, {\Omegan}^{mnp}F_{mnp}\right)
         \gamma_{\mu} \sigma^1 \epsilon_-\otimes\eta_- \ ,
\end{split}
\end{equation}
where the second equation used~\eqref{Jeta} and the
decomposition of the ten-dimensional gamma-matrices given in
(\ref{gammadef}). Similarly we compute 
\begin{equation}
\label{susy4DIIB2}
\begin{split}
   \tfrac{1}{2}\Pi_-\,\Gamma_\mu{}^m\delta\hat{\Psi}_{m} &= 
      -\left(\bar{\eta}_-\gamma^mD_m\eta_+\right)
         \gamma_{\mu}\varepsilon_-\otimes\eta_- 
      +  \frac{1}{16}\left(
         \bar{\eta}_-\gamma^{mnp}\eta_+ H_{mnp}\right)
         \gamma_{\mu} \sigma^3 \varepsilon_-\otimes\eta_- 
         \\ & \qquad\qquad\qquad\qquad\qquad
      + \frac{1}{16}\left(
         \ee^{\phi} \, \bar{\eta}_-\gamma^{mnp}\eta_+ F_{mnp}\right)
         \gamma_{\mu} \sigma^1 \varepsilon_-\otimes\eta_- \\     
      &= -\frac{3\ii}{4} {W}_1\gamma_\mu\varepsilon_-\otimes\eta_- 
      +  \frac{\ii}{32}\left(
         {\Omegan}^{mnp}H_{mnp}\right)
         \gamma_{\mu} \sigma^3 \varepsilon_-\otimes\eta_- 
         \\ & \qquad\qquad\qquad\qquad\qquad
      + \frac{\ii}{32}\left(
         \ee^{\phi} \, {\Omegan}^{mnp}F_{mnp}\right)
         \gamma_{\mu} \sigma^1 \varepsilon_-\otimes\eta_- \ .
\end{split}
\end{equation}
In the second equality  we used  $D_m
\eta_+=\frac{i}{4} {W}_1g_{mn}\gamma^n \eta_- +\dots$ (see~\cite{FMT}),
which follows from (\ref{Jeta}) and (\ref{dJdOmega}). We omitted
terms involving the other torsion
classes that vanish when inserted in the bilinear expression
$\bar{\eta}_-\gamma^m D_m\eta_+$. From \eqref{dJdOmega}
we read off $ {W}_1$ to be 
\begin{equation}\label{W1}
   W_1 = - \tfrac{\ii}{36}\, (\dd J)_{mnp} \, {\Omegan}^{mnp} \ .
\end{equation}
Inserting (\ref{susy4DIIB1}) and (\ref{susy4DIIB2})
into~\eqref{red4Dgrav} and comparing to (\ref{defS}) we arrive at 
\begin{equation}
\begin{aligned}
   S_{11}&= \frac{1}{48}  \left(\ii \, \dd J + 
      \,H\right)_{mnp} \, {\Omegan}^{mnp}\ , \\
   S_{22} &=  \frac{1}{48}  \left(\ii \, \dd J - 
      \,H\right)_{mnp} \, {\Omegan}^{mnp}\ , \\
   S_{12} = S^{21} &= \frac{1}{48} e^{\phi}
      \,F_{mnp} \, {\Omegan}^{mnp}\ .
\end{aligned}
\end{equation}

We can write the matrix $S$ in a compact form, which uses the Mukai
pairing defined in~(\ref{mukai}): 
\beq 
\label{SIIBshort}
   S_{AB}(\text{IIB})  = 
      \frac{\ii}{8} \left( \begin{array}{cc} 
            W + P &  Q_\text{B}  \\
            Q_\text{B} &  W -  P
         \end{array} \right)
\eeq
where 
\begin{equation}
\label{defWHF}
\begin{aligned}
     W \epsilon_g & \equiv  \revmukai{\dd e^{\ii J}}{\Omegan} \ , \\
     P \epsilon_g & \equiv  \revmukai{H_3}{\Omegan} \ , \\
     Q_\text{B} \, \epsilon_g & \equiv 
        \ee^{\phi}\revmukai{F_\text{B}}{\Omegan} \ .
\end{aligned}
\end{equation}
Here we have used
\begin{equation}
   \epsilon_g = \tfrac{1}{6}J\wedge J\wedge J 
      = \tfrac{\ii }{8}\,\Omegan\wedge\Omeganb
\end{equation}
denoting the volume form defined by the (string frame) metric
$g_{mn}$, and $F_\text{B}=F_1+F_3+F_5$ is the sum of all IIB RR field
strengths defined in \eqref{Fdef} (out
of which only $F_3$ contributes to the superpotential).

A similar calculation can be done for type IIA, where we need to use both 
$\Pi_+$ and $\Pi_-$ since the two ten-dimensional gravitinos have opposite 
chiralities. For the RR piece, we get the following terms in the
supersymmetry transformation 
\begin{multline}
   \sum_{\text{$n$ even}}\frac{(-1)^{n/2}}{n!}\,\ee^{\phi}
         F_{p_1\dots p_n}\bar\eta_{+}\gamma^{p_1\dots p_{n}}\eta_{+} 
      = \sum_{\text{$n$ even}} \frac{1}{n!}\,\ee^{\phi}
         F_{p_1\dots p_n}\bar\eta_{-}\gamma^{p_1\dots p_{n}}\eta_{-} \\
      = \tfrac{1}{2} e^{\phi} \left( F_0 - \tfrac{\ii}{2} F_{ab}
         J^{ab} - \tfrac{1}{8} F_{abcd}J^{ab}\,J^{cd} 
         + \tfrac{\ii}{48}F_{abcdef}J^{ab}\,J^{cd}\,J^{ef}\right) .
\end{multline}
This term can also be written using the Mukai pairing defined
in~\eqref{mukai} as\footnote{Note that  
  $\revmukai{F_\text{A}}{\ee^{-\ii J}}=[F \wedge\ee^{\ii J}]_6$, where the
  subscript $6$ indicates the top form component, and the change in
  the sign of the exponent accounts for the alternating sign in the
  definition of the pairing, eq.~\eqref{mukai}.}
\beq\label{int}
   \left( F_0 - \tfrac{\ii}{2} F_{ab} J^{ab} 
      - \tfrac{1}{8} F_{abcd} J^{ab} \, J^{cd} 
      + \tfrac{\ii}{48} F_{abcdef} J^{ab} \, J^{cd}\, J^{ef}
      \right) \epsilon_{g} \, 
      = \revmukai{F_\text{A}}{\ee^{-\ii J}} .
\eeq
where $F_\text{A}=F_0+F_2+F_4+F_6$. 

To see explicitly the complexification of $\ee^{-\ii J}$ into
$\ee^{-t}$, where $t=B+\ii J$ as in~\eqref{tdef}, one transforms
the RR terms into another field basis defined by $C=\ee^B A$. This
transforms the field strength defined in~\eqref{Fdef} according to
\begin{equation} 
\label{FA}
\begin{aligned}
   F_{2n}&=\dd C_{2n-1} - H \wedge C_{2n-3} \\
      &= (\ee^B G_\text{A})_{2n} 
      = G_{2n} + B \wedge G_{2n-2} + \dots + B^n \wedge G_0\ ,
\end{aligned}
\end{equation}
where we defined $G_{2n}=\dd A_{2n-1}$ and
$G_\text{A}=G_0+G_2+G_4+G_6$. In this basis the expression~\eqref{int}
is replaced by $\revmukai{F_\text{A}}{\ee^{-\ii J}}=
\revmukai{G_\text{A}}{\ee^{-(B+\ii J)}}$.

Collecting all the pieces together, we get for type IIA the following
matrix $S$:
\begin{equation}
\label{SIIAshort}
   S_{AB}(\text{IIA})=  \frac{\ii}{8} \begin{pmatrix}
      \bar W - \bar P & - \bar Q_\text{A}  \\
      - \bar Q_\text{A} &  W +  P 
      \end{pmatrix} ,
\end{equation}
where $P$ and $W$ are defined as in IIB ($\bar P$, $\bar W$ are their
complex conjugates), and 
\beq \label{defFa}
   \bar Q_\text{A} \, \epsilon_g = 
      \ee^{\phi} \revmukai{G_\text{A}}{\ee^{-(B+\ii J)}} \, .   
\eeq
Using the constraints \eqref{JBcond} $J\wedge\Omegan=B\wedge\Omegan=0$ we can
rewrite the $W$ and $P$ terms as 
\begin{equation}
\begin{aligned}
     W \epsilon_g & \equiv  \revmukai{\dd\Omegan}{\ee^{\ii J}} \ , \\
     P \epsilon_g & \equiv  \revmukai{\dd\Omegan}{\ee^B} \ .
\end{aligned}
\end{equation}

To complete the analysis we would like to write the expressions for
$S_{AB}$ in terms of the K\"ahler potentials $K_J$ and $K_\rho$ and
the unnormalized pure spinor fields $2c\,\ee^{-B-\ii J}$ and
$\Omega$. First we recall that the natural metric~\eqref{4dNS} on
$T^{1,3}$ is $g^{(4)}_{\mu\nu}=\ee^{-2\phi^{(4)}}g_{\mu\nu}$. Since
there is a $\gamma_\mu$ term multiplying $S_{AB}$ in the gravitino
variation~\eqref{defS}, the correctly normalized $S_{AB}$ matrices
should be multiplied by a factor of $\ee^{\phi^{(4)}}$. Given the
definitions~\eqref{Kdefs} of $K_J$ and $K_\rho$ and the
definition~\eqref{4dNS}  of the four-dimensional dilaton $\phi^{(4)}$,
we note that the different six-forms are related by 
\begin{equation}
   \epsilon_g = \frac{1}{8|c|^2}\ee^{-K_J} 
      = \frac{1}{8\cscale^2}\ee^{-K_\rho} 
      = \ee^{-2\phi^{(4)}+2\phi} \ .
\end{equation}
Thus the correctly normalized IIA and IIB matrices $S$ can be written
as 
\begin{align}
\label{SIIA}
   S_{AB}^{(4)}(\text{IIA}) &= 
      \ii\,\ee^{\frac{1}{2}K_J}
         \left(\begin{array}{cc}\dstyle 
            - \ee^{\frac{1}{2}K_\rho+\phi^{(4)}}
               \revmukai{\dd\bar\pure^-}{\pure^+} & \dstyle
            - \tfrac{1}{2\sqrt{2}}\,\ee^{2\phi^{(4)}}
                 \revmukai{G_\text{A}}{\pure^+} 
            \\*[0.5em] \dstyle
            - \tfrac{1}{2\sqrt{2}}\,\ee^{2\phi^{(4)}}
                 \revmukai{G_\text{A}}{\pure^+} & \dstyle
            \ee^{\frac{1}{2}K_\rho+\phi^{(4)}}
               \revmukai{\dd\pure^-}{\pure^+}
         \end{array}\right) , \\*[0.5em]
\label{SIIB}
   S_{AB}^{(4)}(\text{IIB}) &= 
      \ii\,\ee^{\frac{1}{2}K_\rho}
         \left(\begin{array}{cc}\dstyle 
            - \ee^{\frac{1}{2}K_J+\phi^{(4)}}
               \revmukai{\dd\pure^+}{\pure^-} & \dstyle
            \tfrac{1}{2\sqrt{2}}\ee^{2\phi^{(4)}}
               \revmukai{G_\text{B}}{\pure^-} 
            \\*[0.5em] \dstyle
            \tfrac{1}{2\sqrt{2}}\ee^{2\phi^{(4)}}
               \revmukai{G_\text{B}}{\pure^-} & \dstyle
            \ee^{\frac{1}{2}K_J+\phi^{(4)}}
               \revmukai{\dd\bar\pure^+}{\pure^-}
         \end{array}\right) ,
\end{align}
where 
\begin{equation}
   \pure^+ = c\,\ee^{-B-\ii J} , \qquad
   \pure^- = \Omega .
\end{equation}
In deriving the expressions~\eqref{SIIA} and~\eqref{SIIB} we have used
the fact that the constraints $B\wedge\Omega=J\wedge\Omega=0$ imply
that there are no contributions from terms with $\dd c$ or
$\dd\bar{c}$. We have also made a
$U(1)_R\otimes\SU(2)_R$ R-symmetry transformation to remove explicit
dependence on the phase of $c$. These constraints also ensure the
invariance of $S^{(4)}_{AB}$ under rescalings of $c$ and
$\Omega$. Finally, we have also used the fact that $B\wedge\Omega=0$
to replace $F_\text{B}$ with $G_\text{B}$, defined in exact analogy to
$G_\text{A}$ in~\eqref{FA}.\footnote{Later in section~\ref{d=4N=2} 
we will split the field strengths into an exact piece, plus a flux
piece (see, for example,~\eqref{fluxstrength}).} Recalling that
$\ee^{-K_J}$, $\ee^{-K_\rho}$ and $\ee^{-2\phi^{(4)}}$ are all
six-forms, overall, $S_{AB}^{(4)}$ transforms as a section of
$(\Lambda^6F^*)^{-1/2}$. This dependence arises because of the
rescaling by $\ee^{\phi^{(4)}}$. 
 
Comparing~\eqref{SIIA} and~\eqref{SIIB} with~\eqref{4dsusylaws}, and
recalling that for IIA $K_V=K_J$ while for IIB $K_V=K_\rho$, 
we can read off the Killing prepotentials $\cP^x$. 
For type IIA we obtain
\begin{equation}
\label{PIIA}
\begin{aligned}
   \cP^1 &= - 2\ee^{\frac{1}{2}K_\rho+\phi^{(4)}}
               \revmukai{\dd\re\pure^-}{\pure^+} , \\
   \cP^2 &= - 2\ee^{\frac{1}{2}K_\rho+\phi^{(4)}}
               \revmukai{\dd\im\pure^-}{\pure^+} , \\
   \cP^3 &= \frac{1}{\sqrt{2}}\,\ee^{2\phi^{(4)}}
               \revmukai{G_\text{A}}{\pure^+} ,
\end{aligned}
\end{equation}
while for type IIB we find 
\begin{equation}
\label{PIIB}
\begin{aligned}
   \cP^1 &= - 2\,\ee^{\frac{1}{2}K_J+\phi^{(4)}}
               \revmukai{\dd\re\pure^+}{\pure^-} , \\
   \cP^2 &= 2\,\ee^{\frac{1}{2}K_J+\phi^{(4)}}
               \revmukai{\dd\im\pure^+}{\pure^-} , \\
   \cP^3 &= - \frac{1}{\sqrt{2}}\ee^{2\phi^{(4)}}
               \revmukai{G_\text{B}}{\pure^-} .
\end{aligned}
\end{equation}
Note that in both cases that $\cP^x$ transform as scalars since though
the Mukai pairing is an element of $\Lambda^6F^*$, this dependence is
canceled by the K\"ahler potential and $\phi^{(4)}$ factors. 

If we turn off the RR fluxes we see that in both cases only $\cP^1$ and
$\cP^2$ are non-zero. Since the $\cP^x$ are $\SU(3)$ singlets only
singlet terms can contribute. In particular, from~\eqref{dJdOmega}
and~\eqref{constraints} we see they only depend on the singlet torsion
class $W_1\sim \dd J\wedge\Omega$ as we also already observed
in~\eqref{W1}. Furthermore $W_1$ is complexified by $H\wedge\Omega$,
the singlet part of the NS flux. The RR fluxes, which in
both cases enter $\cP^3$, appear only in the singlet representation. 

Eqs.~\eqref{PIIA} and~\eqref{PIIB} are quite generic. The only
restriction was that we assumed there is a single $\SU(3)$ structure
and not the more general situation of an $\SU(3)\times\SU(3)$
structure.
The key difference between these cases is that $\pure^-$ is not generic
but contains only the holomorphic three-form piece $\Omega$. More
generally it would include one- and five-form pieces. 
The precise details of the corresponding derivation of the
prepotentials $\cP^x$ are beyond the scope of this paper and will be
presented elsewhere. Nonetheless, given the simple form the final
expressions in terms of $\pure^\pm$ we conjecture that~\eqref{PIIA}
and~\eqref{PIIB} actually continue to hold also when $\pure^\pm$ are generic
$\SU(3)\times\SU(3)$ pure spinors.

Let us also briefly discuss mirror symmetry which states that 
type IIA and type IIB are equivalent when considered in mirror symmetric
backgrounds. For a pair of mirror Calabi--Yau manifolds even and odd 
cohomologies are interchanged $H^{\rm even} \leftrightarrow H^{\rm odd}$. 
It has been suggested that on manifolds with $SU(3)$ structure this
operation is replaced by an exchange of the two pure spinors $\Omega
\leftrightarrow \ee^{-B-\ii J}$ and the exchange of $\Leven F^*
\leftrightarrow \Lodd F^*$~\cite{FMT}. More generally we expect the
exchange of $\SU(3)\times\SU(3)$ pure spinors and fluxes 
\begin{equation} \label{mirrorspinors}
   \pure^+\leftrightarrow\pure^- , \qquad
   G_\text{A} \leftrightarrow G_\text{B} .
\end{equation}

In the generic $\SU(3)\times\SU(3)$ case, we see that~\eqref{PIIA}
and~\eqref{PIIB} indeed have this symmetry (provided we also exchange
$K_J$ and $K_\rho$).\footnote{For the expressions given we also have to
map $(\cP^2,\cP^3)\to(-\cP^2,-\cP^3)$ in going from IIA to IIB, but
this is just an element of the $\SU(2)_R$ symmetry and is simply an
artifact of our conventions.} However, recall that the expressions
were derived assuming only an $\SU(3)$-structure where $\pure^-$ is
restricted to be the three-form $\Omega$ while $\pure^+=c\,\ee^{-B-\ii
J}$ is generic. 
In this case $\Omega$ and $\ee^{-B-\ii J}$ enter asymmetrically
into~\eqref{PIIA} and~\eqref{PIIB}.
As we already observed in $\cP^1$ and $\cP^2$ only  $\dd (B+\ii
J)\wedge\Omega$ appears. Thus the three-form $\Omega$ fully
contributes while out of $\ee^{-B-\ii J}$ only the two-form  $B+\ii J$
but not its square $(B+\ii J)^2$ appears. This is precisely the issue
of the missing mirror of the magnetic fluxes. In~\cite{Vafa,GLMW} it
was observed that $\dd\Omega$ can be interpreted as a NS four-form
which is the mirror dual of the NS three-form $H$. However in order
to have full mirror symmetry a NS two-form is also necessary. The
four-form can be viewed as giving rise to electric fluxes while the
(missing) two-form would lead to magnetic fluxes. Within the framework 
followed here the NS two-form and thus the magnetic fluxes are still
missing. From the structure of~\eqref{PIIA} and~\eqref{PIIB} it is
clear the NS two-form should be identified with the exterior
derivative of the one-form part of $\pure^-$ which exists in
backgrounds with generic $\SU(3)\times\SU(3)$ structure.\footnote{This
  has also been noticed by T.\ Grimm and we thank him for discussions on this point.} This is also
in accord with similar recent suggestions for example in
ref.~\cite{MR,Hull}. We will return to this issue again in
section~\ref{mirror}.  


\subsection{$N=1$ superpotentials}
\label{N=1W}

In the spirit followed so far in this section we can further constrain
the space-time background to only realize four supercharges
linearly. This results in a further split of the $N=2$ multiplets
which we discussed in section~\ref{fields}. In particular, the $N=2$
gravitational multiplet decomposes into a $N=1$ gravitational
multiplet containing the metric and one gravitino $(g_{\mu\nu},
\psi_\mu)$, and a $N=1$ spin-$\frac{3}{2}$ multiplet containing the
second gravitino and the graviphoton $(\psi_\mu^\prime,
C_\mu)$. Exactly as in $N=2$, the appearance of  a standard $N=1$-type
action requires that we project out the $N=1$ spin-$\frac{3}{2}$
multiplet leaving only the gravitational multiplet together with
vector, tensor and chiral multiplets in the
spectrum.\footnote{Alternatively one can integrate out the
  spin-$\frac{3}{2}$ multiplet and consider an effective action below
  the scale set by the mass of that multiplet~\cite{Louis}.}  
{}From a supergravity point of view such a truncation has been
discussed in refs.~\cite{ADAF} while the truncation occurring in
orientifold compactifications of type II has been studied in
ref.~\cite{BBHL,TGL}. 

We compute the $N=1$ superpotential ${\cal W}$
from the supersymmetry transformation of the linear
combination of the two $N=2$ gravitinos which resides in the $N=1$
gravitational multiplet. 
Exactly which linear combination is kept depends on the specific theory
under consideration.
As a consequence ${\cal W}$ will depend on two
angles which parameterize the choices of embedding an $N=1$ inside $N=2$.

We proceed by choosing the $N=1$
supersymmetry transformation parameter $\varepsilon$ to be any linear
combination of the pair of $N=2$ parameters $\varepsilon^1$ and
$\varepsilon^2$. We parameterize this freedom by writing 
\begin{equation}
   \varepsilon_A = \varepsilon n_A , \qquad
   n_A = \begin{pmatrix}a \\ b \end{pmatrix}\ , \quad
   |a|^2+|b|^2 = 1\ ,
\end{equation}
where $\varepsilon$ is the $N=1$ supersymmetry parameter and $n_A$ is
a vector normalized to one so $\bar{n}^An_A=1$ with $a$
and $b$ complex. Choosing such a linear combination breaks the
$\SU(2)_R$ symmetry of $N=2$ down to a $U(1)_R$ of $N=1$,
corresponding to those rotations preserving $n^A$. The conjugated
spinors $\varepsilon^A$ can be written as $\varepsilon^A=\varepsilon^c
{n^*}^A$ where $\varepsilon^c$ is the conjugate $N=1$ spinor, and
$n^{*\,A} = {\bar a\choose \bar b}$. We can
similarly decompose the gravitinos $\psi_{A\,\mu}$. If $\psi_\mu$ is the
$N=1$ gravitino, the superpotential $\mathcal{W}$ can be extracted
from the supersymmetry variation which has the generic form

\begin{equation}\label{susyN=1}
   \delta\psi_\mu = D_\mu\varepsilon 
       + \ii \ee^{K/2} \mathcal{W}\gamma_\mu \varepsilon^c \ , 
\end{equation}
where $K$ is the total $N=1$ K\"ahler potential. We can use the projector
$\Pi_A{}^B=n_A\bar{n}^B$ to pick out the $N=1$ gravitino and
supersymmetry parameter. Projecting the $N=2$ variation~\eqref{defS} we
find
\begin{equation}\label{susyN=2}
   \delta\psi_\mu = D_\mu \varepsilon 
      + \ii \bar{n}^A S_{AB} n^{*\,B} \gamma_\mu \varepsilon^c\ ,
\end{equation}

Comparing \eqref{susyN=2} with \eqref{susyN=1} using
\eqref{4dsusylaws} 
we arrive at 
\begin{multline}
\label{W}
   \ee^{K/2}\mathcal{W} = \frac{\ii}{2}\ee^{K_V/2}\Big[
      \left(\cos^2\phione\,\ee^{\ii\phitwo}
         - \sin^2\phione\,\ee^{-\ii\phitwo}\right)\cP^1 \\
      -\ii\left(\cos^2\phione\,\ee^{\ii\phitwo}
         + \sin^2\phione\,\ee^{-\ii\phitwo}\right)\cP^2
      - \sin2\phione \cP^3 \Big] ,
\end{multline}
To get this expression, we have used the fact that the superpotential
depends on the sum of the arguments of $a$ and $b$ only by an overall
phase, which can be removed by a $U(1)_R$ transformation. 
We can therefore parameterize $a$ and $b$ using only two angles, 
$\phione, \phitwo$ as
\begin{equation}
a =\cos\alpha\,  \ee^{-\frac{\ii}{2}\beta}\ , \qquad 
b = \sin\alpha\, \ee^{\frac{\ii}{2}\beta}\ .
\end{equation}

{}From \eqref{W} we see that in both cases, type IIA and type IIB, 
the $N=1$ superpotential depends on the fluxes and
the torsion class $W_1$ via $\cP^x$.
In addition it also depends on the two angles $\phione, \phitwo$ which
fix a $U(1)_R$ subgroup inside the $SU(2)_R$ of $N=2$. The
fact that a classification of backgrounds with $\SU(3)$ structure in
type II needs only two angles was indeed anticipated in
\cite{Frey}. 

In order to give the  $N=1$ K\"ahler potential in terms of chiral
multiplets one first needs to
determine the complex structure on the field space which in general is
an involved procedure \cite{HL2,BBHL,TGL}. However, here we do not
need $K$ in any explicit form  but instead it is sufficient 
to give it in terms of $N=2$ quantities. By inspecting the
appropriately normalized (i.e.\ Weyl rescaled)
gravitino mass term one determines 
the generic relation
\begin{equation}\label{Ktot}
K = K_J + K_\rho +2 \phi^{(4)}\ .
\end{equation}
Let us stress once more that \eqref{Ktot} does not express $K$ in 
proper $N=1$ chiral coordinates but in terms of $N=2$ coordinates.
However this is all we need in order to insert
\eqref{Ktot}  into \eqref{W} and using
\eqref{PIIA}, \eqref{PIIB} we arrive at
\begin{multline}
\label{WA}
\ii   \mathcal{W}_{\text{IIA}} = 
     \cos^2\phione\,\ee^{\ii\phitwo}\langle \pure^+, \dd\bar \pure^-\rangle
      -\sin^2\phione\,\ee^{-\ii\phitwo} \langle \pure^+, \dd\pure^-\rangle 
      + {|\cscale|} \sin2\phione\,\ee^\phi \revmukai{G_\text{A}}{\pure^+}  ,
\end{multline}
and
\begin{multline}
\label{WB}
 \ii  \mathcal{W}_{\text{IIB}} = 
    \cos^2\phione\,\ee^{\ii\phitwo}  \revmukai{\dd\pure^+}{\pure^-}
      -\sin^2\phione\,\ee^{-\ii\phitwo}\revmukai{\dd\bar\pure^+}{\pure^-}
      - {|c|} \sin2\phione\,\ee^\phi \revmukai{G_\text{B}}{\pure^-} .
\end{multline}
Note that we have written these expressions in terms of the
ten-dimensional dilaton $\phi$. 

For specific choices of $\phione, \phitwo$ we can reproduce the
$N=1$ superpotentials discussed so far in the literature. 
If we choose $2\phione=-\phitwo=\pi/2$ 
and undo the change of variables
from $G_3$ to $F_3$
in~\eqref{WB} we obtain the Gukov--Taylor--Vafa--Witten
superpotential~\cite{GVW,TV}\footnote{The NS contribution to this
  superpotential was verified in the heterotic theory  using the
  supersymmetry transformation of the gravitino by a calculation
  similar to the one we do to obtain the $N=2$ prepotential
  in~\cite{Beckerconstantin}. This ``GVW choice'' corresponds to the
  relation $a=-ib$, which  gives precisely the $N=1$ conserved spinor
  in compactifications on (conformal) CY with fluxes \cite{GP}.
}
\begin{equation}
\label{WGVW}
   \mathcal{W}_{\text{GTVW}}= 
      \ii \, |c|\, \ee^\phi
\revmukai{\Omega}{ \left( F_3 - \tau H_3\right) } \ ,
\end{equation}
where $\tau = C_0 + \ii\ee^{-\phi}$.

On the type IIA side the mirror symmetric superpotential 
for only  RR-fluxes as suggested in \cite{gukov} 
is obtained from \eqref{WA} for $\alpha = \pi/4$
and $\dd \pure^- =0$
\begin{equation}
\label{WIIAR}    
   \mathcal{W}_{\text{IIA,RR}} = -\ii \, |c|\cscale\, \ee^\phi 
\revmukai{G_\text{A}}{\ee^{-(B+\ii J)}}\  .
\end{equation}

The type IIA mirror superpotential of the NS-fluxes was proposed in~\cite{GLMW}
and it can be recovered from~\eqref{WA} with the choice
$\phione=\pi/2$ and $\phitwo=-\pi/2$
\begin{equation}
\label{Whalf}
\mathcal{W}_{\text{half-flat}} = |c| \revmukai{\dd\Omega}{\ee^{-(B+\ii J)}}\ .
\end{equation}

Finally, the superpotential proposed in ref.\ \cite{Berglund} is expressed in terms
of the periods of a mirror pair of Calabi-Yau threefolds. The resulting
structure is similar to the superpotentials \eqref{WA}, \eqref{WB} and
it would be interesting to establish a precise relationship.

Having obtained the most general $N=1$ superpotential,  
let us go back to $N=2$ and discuss the
truncation to four space-time dimensions.


\section{The $N=2$ effective theory in four dimensions}
\label{d=4N=2}


In the previous sections we studied the ten-dimensional type II
supergravities in spacetimes $M^{1,9}$ where it is possible to single
out eight of the original 32 supercharges. However so far we simply
rewrote the ten-dimensional theory keeping all of the modes in the
theory. In other words we did not yet perform any Kaluza--Klein
reduction. In fact,  we have not even assumed that $M^{1,9}$ is a
product. Generically it only required that $M^{1,9}$ admitted an
$\SU(3)\times\SU(3)$ structure, or, in the special case we considered in
detail, simply an $\SU(3)$-structure. 

In this section we turn to a more restricted situation. First we will
assume that we have topologically a product manifold 
\begin{equation}
\label{productM}
   M^{1,9} = M^{1,3} \times \Y \ .
\end{equation}
Comparing with the generic case~\eqref{Tdecomp}, we identify
\begin{equation}
    T^{1,3} = TM^{1,3}\ , \qquad F = T\Y \ . 
\end{equation}
We then truncate the general ten-dimensional eight-supercharge theory,
keeping only a finite number of ``light'' modes in the spectrum. This
in turn will lead to a four-dimensional effective theory with $N=2$
supersymmetry. For simplicity, we will not discuss the general
situation where there is an $SU(3)\times SU(3)$ structure but instead confine our
attention to backgrounds $\Y$ with only $SU(3)$  structure. In
addition we define the truncation in such a way that apart from the
gravitational multiplet only vector-, tensor- and hypermultiplets are
present in the effective theory. In particular  we project out all
possible spin-$\frac32$ multiplets and in this way end up with a
``standard'' $N=2$ effective action.


\subsection{Defining the truncation}
\label{defreduction}

Generically the distinction between heavy and light modes in a
Kaluza--Klein expansion on the product~\eqref{productM} is not
straightforward. This is in contrast to the situation when $\Y$ is a 
Calabi--Yau manifold. In this case one keeps all the field deformations
which from a four-dimensional point of view are massless
modes. For each supergravity field these correspond to harmonic
forms, and hence the light (massless) modes are finite in number. For
instance, of the metric deformations given in section~\ref{modulispace}, the
massless modes are in one-to-one correspondence with harmonic
deformations of the K\"ahler-form $J$ and the three-form $\rho$. These
give the $h^{1,1}$ K\"ahler moduli and $h^{2,1}$ complex structure
moduli, where $h^{1,1}, h^{2,1}$ are the appropriate Hodge
numbers. Similarly massless deformation of $B_{mn}$ and the
RR-potentials $C_p$ are in one-to-one correspondence with harmonic
two- and $p$-forms respectively. The result is that, for a Calabi--Yau
compactification, rather than considering $J$, $\rho$, $B$ and $C_p$ to
be forms on $\Y$ we truncate to the finite dimensional sub-space of
harmonic forms. 

This suggests that we should take a similar truncation in the generic
case -- not to harmonic forms but some other finite-dimensional
subspace of $\Lambda^*F^*$. Identifying the subspace however is not a
simple task. If we start with a background (like a Calabi--Yau manifold)
which satisfies the equations of motion, we can truncate to the
massless fluctuations. However generically the background is not a
solution. An example of this is a Calabi--Yau compactification with
non-zero $H$ or RR flux. Typically in this case one still keeps the
harmonic deformations even though some of them become massive. This
can be justified in the limit of large manifolds and small fluxes by
the fact that there is a hierarchy between these masses and those of
the Kaluza--Klein modes. Similar arguments can be made for mirror
half-flat manifolds in the large complex structure
limit~\cite{GLMW}. Here, for now, we will simply assume that a
suitable limit can be found where it is consistent to keep only a
finite number of ``light'' modes and not specify how this subset is
defined.

Let us be more specific. We wish to restrict to a set of
finite-dimensional subspaces of $\Lambda^pF^*$. We write these as 
\begin{equation}
   \Lt^p \subset \Lambda^pF^*
   \qquad 
\end{equation}
and assume that all the fields $g$, $B$, $\phi$ and $C_p$ take values
in these subspaces. 
In particular, to describe the metric degrees of freedom, we assume
that we have $O(6,6)$ spinors $\chi^+=2\re(c\,\ee^{-B-\ii J})\in\Lteven$
and $\chi^-=2\cscale\rho\in\Lt^3$. We then define the spaces of stable
forms 
\begin{equation}
\begin{aligned}
   \UtJ &= U_J \cap \Lteven , \\
   \Utr &= U_\rho \cap \Lt^3
\end{aligned}
\end{equation}
where $U_J$ and $U_\rho$ are the spaces of stable forms defined
in~\eqref{UJdef} and~\eqref{Urdef}. We assume that $\UtJ$ and $\Utr$
are open subsets of $\Lteven$ and $\Lt^3$, which is generically the
case. 

Crucially the truncation should not break supersymmetry. This means 
that the special K\"ahler metrics on the spaces $U_J$ and
$U_\rho$ give special K\"ahler metrics on $\UtJ$ and $\Utr$. This is
equivalent to requiring 
\begin{enumerate}
\item the Mukai pairing $\mukai{\cdot}{\cdot}$ is non-degenerate on
      $\Lteo$,  
\item if $\chi^\pm\in\UtJ\text{ or }\Utr$ then
      $\hat{\chi}^\pm\in\UtJ\text{ or }\Utr$, 
\end{enumerate}
where the Mukai pairing is defined in~\eqref{mukai} and
$\hat{\chi}^\pm$ are defined in~\eqref{hatdef}
and~\eqref{rhatdef}. The first condition implies we have symplectic 
structures on $\UtJ$ and $\Utr$, the second that we have complex 
structures. Note that the second condition is equivalently to  
\begin{enumerate}
\item[$2.'$] if $\chi^\pm\in\UtJ\text{ or }\Utr$ then
      $*\chi^\pm\in\UtJ\text{ or }\Utr$, 
      \label{Hodgecond}
\end{enumerate}
where $*$ is the Hodge-star operator defined by the metric $g$ (which
is in turn defined by the particular $\chi^+$ and $\chi^-$). 

As in the Calabi--Yau case we can define $\Lt^p$ in terms of sets of
basis forms. For instance for $\Lt^{2p}\subset\Lambda^{2p}F^*$ we
write 
\begin{equation}
\begin{aligned}
   \Lambda^0F^* \supset \Lt^0 
      &= \big\{\text{constant functions on $\Y$} \big\} \\
   \Lambda^2F^* \supset \Lt^2 
      &= \left\{A^a\omega_a, \ a=1,\dots,\bJ \right\} \\
   \Lambda^4F^* \supset \Lt^4 
      &= \left\{B_a\tilde{\omega}^a, \ a=1,\dots,\bJ \right\} \\
   \Lambda^6F^* \supset \Lt^6 
      &= \left\{C\epsilon \right\}
\end{aligned}
\end{equation}
where $\omega_a$ are a set of basis two-forms,
$\tilde{\omega}^a$ a set of basis four-forms, $\epsilon$ is a volume
form and $A^a$, $B_a$ and $C$ are constant functions on $\Y$. The
condition that the Mukai pairing is non-degenerate on $\Lteven$ is
reflected in the fact that $\Lt^2$ and $\Lt^4$ have the same dimension
$\bJ$. Specifically we choose the basis
$(1,\omega_a,\tilde{\omega}^b,\epsilon)$ such that       
\begin{equation}
\label{dualbasis}
   \mukai{\omega_a}{\tilde{\omega}^b} 
      = - \delta_{a}{}^{b}\epsilon\ , 
      \qquad a, b =1,\dots,\bJ\ ,
\end{equation}
(Recall all other products vanish identically, except for
$\mukai{1}{\epsilon}$ which equals $\epsilon$ by definition.) Note
that this expansion is not quite general in that we have assumed that
$\Lt^0$ is  spanned by constant functions on $\Y$. The reason for this
will be explained below. Finally, in order to impose the second
condition~$2.^\prime$ above, in analogy with the Calabi--Yau case
where $\omega_a$ and $\tilde{\omega}^a$ are harmonic, we allow the
basis vectors in general to depend on the metric $g_{mn}$ so that 
\begin{equation}
   *1\in\Lt^6\ , \qquad
   *\omega_a\in\Lt^4\ , \qquad
   *\tilde{\omega}^a\in\Lt^2\ , 
\end{equation}
for all $a$. 

For the odd forms we choose a more restricted truncation. First for
the three-forms we define, as above,
\begin{equation}
   \Lambda^3F^* \supset \Lt^3 
      = \left\{D^K\alpha_K + E_L\beta^L, \ K, L=0,\dots, \br \right\}
\end{equation}
where $D^K$ and $E_L$ are constant functions on $\Y$ and
$\alpha_K,\beta^L\in\Lambda^3F^*$ are a symplectic set of basis forms
satisfying 
\begin{equation}
\label{sbasis}
   \mukai{\alpha_{K}}{\beta^{L}} =  \delta_K{}^L\epsilon
   \ , \qquad K, L =1,\dots,\br\ ,
\end{equation}
while $\mukai{\alpha_{K}}{\alpha_{L}}=\mukai{\beta^{K}}{\beta^{L}}=0$. 
In addition we require 
\begin{equation}
   *\alpha_K, *\beta^L \in \Lt^3
\end{equation}
for all $K$ and $L$. 

At this point we further simplify the truncation by imposing an
additional condition. For the one- and five-forms we choose to
truncate the spectrum completely, keeping no light modes, so
\begin{equation}
\label{Lt15}
   \Lt^1 = \Lt^5 = 0 . 
\end{equation}
We make this choice because we want to truncate in such a
way that a `standard' $N=2$ gauged supergravity appears which only
contains one gravitational multiplet together with vector, tensor
and  hypermultiplets.  In particular we do not allow the presence of
any spin-$\frac32$ multiplets. In terms of the gravitinos  this
amounts to keeping only the two gravitinos in the gravitational
multiplet but projecting out all other gravitinos which may reside in
spin-$\frac32$ multiplets. 

{}From the decomposition of the ten-dimensional spinors given
in~\eqref{repdecomp} and tables~\ref{NS} and~\ref{FII} we see that the
$\SU(3)$ singlets correspond to the gravitinos in the $N=2$
gravitational multiplet while the $\SU(3)$ triplets lead to gravitinos
which reside in their own spin-$\frac32$ multiplets. Of course the 
triplets are nothing but $(1,0)$-forms on $\Y$ with respect to the
given complex structure. Therefore excluding spin-$\frac32$ multiplets
in the truncation  we are led to project out all modes arising from
one-forms (or triplets) on $\Y$.
In this case one is left with the multiplets given in
tables~\ref{N=2multipletsA} and~\ref{N=2multipletsB}. For consistency
it also implies that we should not be able to construct any
$\rep{3}$-representations from the bases $(\alpha_K,\beta^L)$ and
$(1,\omega_a,\tilde{\omega}^a,\epsilon)$. This means that any
five-form wedge products must vanish, so 
\begin{equation}
\label{no3}
   \omega_a \wedge \alpha_K\ =\ 0\ =\ \omega_a \wedge \beta^K \ , \qquad
   \forall\, a, K\ .
\end{equation}
Thus $J\wedge\rho=0$ holds identically for all $J\in\Lt^2$
and $\rho\in\Lt^3$.\footnote{This also  
removes the subtleties involved in the fact that generically 
multiple $\SU(3)$-structures $(J,\rho)$ on $\Y$ determine the same
metrics $g_{mn}$: here, up to an overall rescaling of
$\Omegan=\rho+\ii\hat{\rho}$ by $\ee^{\ii\alpha}$, each metric
corresponds to a unique pair $(J,\rho)\in\Lt^2\oplus\Lt^3$.}

Our assumption is that, in the truncation, all fields will be expanded
in terms of elements of $\Lt^*$. As we have seen in
section~\ref{N=2W}, the supergravity action also depends on the
intrinsic torsion $\dd\Omega$ and $\dd J$ through the superpotential
terms. Similarly the field strengths $H$ and $F_p$ are written in
terms of exterior derivatives. In order for the truncation to make
sense all such terms also need to be in the truncated set $\Lt^*$. In
other words we require $\Lt^*$ to be closed under $\dd$, that is  
\begin{equation}
\label{dLt}
   \dd : \Lt^p \to \Lt^{p+1} \ . 
\end{equation}
Since $\Lt^1=0$ this means that $\Lt^0$ must contain only constant
functions, as, in fact, we have already assumed.  Since $\Lt^5=0$,
the condition~\eqref{dLt} implies that the rest of the forms in the 
basis of
$\Lteven$  satisfy
\begin{equation}
\label{domega}
\begin{aligned}
   \dd\omega_a &= m_a^K\, \alpha_K + e_{aL}\,\beta^L\ , \\
   \dd\tilde{\omega}^a &= 0 \ ,
\end{aligned}
\end{equation}
where $m_a^K$ and $e_{aL}$ are constant matrices. Since the basis
defined by~\eqref{sbasis} is only specified up to symplectic
transformations the matrices $m_a^K$ and $e_{aL}$ also carry a
representation of the symplectic group 
$\Symp(2\br+2)$  and naturally combine into the symplectic vectors
$V_a:=(e_{aK},m_a^K)$.  

Similarly expanding $\dd\alpha_K$ and $\dd\beta^K$, and
using~\eqref{no3}, have
\begin{equation}
   \mukai{\omega_a}{\dd\alpha_K}
      = - \omega_a\wedge\dd\alpha_K
      = \dd\omega_a\wedge\alpha_K 
      = - \mukai{\alpha_K}{\dd\omega_a}\ . 
\end{equation}
Together with~\eqref{domega} and a similar expression with
$\dd\beta^K$, this implies 
\begin{equation}
\label{dalpha}
\begin{aligned}
   \dd\alpha_K &= e_{aK} \tilde\omega^a\ , \\
   \dd\beta^K &=-m_a^K \tilde\omega^a\ .
\end{aligned}
\end{equation}
Using $\dd^2=0$ yields the consistency condition
\begin{equation}\label{null}
   m_a^K\, e_{bK} - e_{aK}m_b^K= V_a\cdot V_b = 0\  ,
\end{equation}
or, in other words, the symplectic vectors $V_a$ 
have to be null with respect to the symplectic
inner product. The conditions~\eqref{domega},~\eqref{dalpha}
and~\eqref{null} have also been obtained in ref.~\cite{DFTV} using
consistency considerations of $N=2$ gauged supergravity. We are also
going to see their necessity 
from the requirement of four-dimensional gauge
invariance in sections~\ref{IIAreduc} and~\ref{IIBreduc}.

Finally, let us note that when $\Y$ is compact we have a natural map
from $\Lt^6$ to $\bbR$ given simply by integrating the six-form over
$\Y$. In particular, the Mukai pairing leads to a natural
symplectic structure on $\Lteo$ given by  
\begin{equation}
\label{eq:sympt}
    \sym{\psi^\pm}{\chi^\pm} = \int_\Y\mukai{\psi^\pm}{\chi^\pm} .
\end{equation}
If we fix the volume form $\epsilon$ such that
$\int_{\Y}\epsilon=1$ this coincides with the symplectic inner product
$\sym{\psi_\epsilon^\pm}{\chi_\epsilon^\pm}$ on the corresponding
$\Spin(6,6)$ spinors. This is the symplectic structure which naturally
appears in a Calabi--Yau truncation. Note that one then also has a
natural definition of the Hitchin scalar functional
\begin{equation}
\label{HYdef}
   \hfuncY(\chi^\pm) = \int_\Y \hfunc(\chi^\pm) , 
\end{equation}
for forms $\chi^\pm\in\Leo$. 


\subsection{Reducing the Neveu-Schwarz sector}\label{NSred}

Let us first discuss the truncation of the Neveu-Schwarz sector since
it is common to both type II theories. This sector contains the
metric, the two-form $B$ and the dilaton $\phi$. Since we defined the
truncation in such a way that all triplets are projected out we see
from table~\ref{NS} that only $g_{\mu\nu}$, $g_{mn}$, $B_{\mu\nu}$,
$B_{mn}$, and $\phi$ survive in the NS-sector. The $\Spin(1,3)$
singlets $g_{\mu\nu}$, $B_{\mu\nu}$ and $\phi$ trivially descend to
the four-dimensional theory. In complete analogy to the generic
relations~\eqref{4dNS}, we define four-dimensional Einstein frame
metric $g^{(4)}_{\mu\nu}$ and the dilaton~$\phi^{(4)}$ which together
with $B_{\mu\nu}$ becomes a member of a tensor multiplet in both type
II theories as seen from tables~\ref{N=2multipletsA}
and~\ref{N=2multipletsB}. The difference now is that these fields
depend only on the four space-time coordinates of $M^{1,3}$.

We already argued that instead of $g_{mn}$ we can discuss the theory
more conveniently in terms of $J$ and $\rho$. Our definition of the
truncation assumed that $J$ (and $B_{mn}$) and $\rho$ have an
expansion in terms of a basis of $\Lt^2$ and $\Lt^3$ respectively. 
The conditions on the spaces $\Lt^p$ arose because we wanted the
special K\"ahler metrics on $U_J$ and $U_\rho$ to descend to $\UtJ$
and $\Utr$. As a consequence we can take the results of section~\ref{SKM}
to give expressions for the corresponding K\"ahler potentials. 

Consider first the metric on $\UtJ$. In the truncated space the
complex $O(6,6)$ spinor $\pure^+$ has the expansion
\begin{equation}\label{Phiexp}
   \pure^+ = c\,\ee^{-B-\ii J} 
      = X^0 + X^a\omega_a - \mathcal{F}_a\tilde{\omega}^a
         - \mathcal{F}_0\epsilon
\end{equation}
where if we expand 
\begin{equation}
\label{Jexp}
   B+\ii J = t^a \omega_a
\end{equation}
then we have complex coordinates on $\UtJ$ 
\begin{equation}
   X^A = (X^0, X^a) = (c,-ct^a) .
\end{equation}
The Hitchin functional $\hfuncY(\pure^+)$ is given by 
\begin{equation} \label{HitchinJ}
\begin{aligned}
   \hfuncY 
      &= \ii\left(\bar{X}^A\mathcal{F}_A 
          - X^A\bar{\mathcal{F}}_A\right) \\ 
      &= \tfrac{4}{3}|c|^2\int_\Y J\wedge J\wedge J
      = \tfrac{1}{6}\ii|X^0|^2\kappa_{abc}
         (t-\bar t)^a (t-\bar t)^b (t-\bar t)^c\ ,
\end{aligned}
\end{equation}
where
$\kappa_{abc}\equiv\int_\Y\omega_a\wedge\omega_b\wedge\omega_c$ and
${\mathcal{F}}_A=({\mathcal{F}}_0,{\mathcal{F}}_a)$. 

For $\Utr$, the complex three-form $\pure^-$ has the expansion
\begin{equation}
\label{Omegaexp}
   \pure^- = \Omega = Z^K\, \alpha_K - F_L\,\beta^L\ ,
\end{equation}
now defining complex coordinates $Z^K$ on $\Utr$. The corresponding
Hitchin functional is 
\begin{equation}
   \begin{aligned}
   \hfuncY\   
      =\ \ii\left(\bar{Z}^K F_K - Z^K \bar{F}_K\right) \
      =\ - \ii \int_\Y \Omega\wedge\bar{\Omega}\  .
\end{aligned}
\end{equation}

Since we have truncated in such a way to remove the triplet degrees of
freedom, we necessarily satisfy the $\SU(3)$ conditions
$J\wedge\rho=B\wedge\rho=0$, as a result of
eqn.~\eqref{no3}. Furthermore, the extra deformations which modify the
$\SU(3)$ structure but leave $g$ invariant are also projected
out. Thus none of the subtleties discussed in
section~\ref{metricSK} is of any concern here. 
Instead the moduli space of metric
deformations is simply 
\begin{equation}
   \Mt = \MtJ \times \Mtr
\end{equation}
where $\MtJ$ and $\Mtr$ are the spaces $\UtJ$ and $\Utr$ modulo
rescalings of $c$ in $\pure^+$ and the magnitude and phase of $\Omega$
in $\pure^-$, that is 
\begin{equation}
   \MtJ = \UtJ/\bbC^*\ , \qquad
   \Mtr = \Utr/\bbC^*\ .
\end{equation}
These spaces can be parametrized by the local ``special'' coordinates
$t^a=X^a/X^0$ and $z^k=Z^K/Z^0$. In the latter case we isolate one
(labeled $\alpha_0$) of the $\alpha_K$, and assume that we can
consistently scale its coefficient to unity when expanding
$\Omega$. The corresponding K\"ahler potentials are
\begin{equation}
\begin{aligned}
   K_J &= - \ln\, \hfuncY 
      = - \ln \int_\Y \tfrac{4}{3}|c|^2J\wedge J\wedge J
      = - \ln \ii\left(\bar{X}^A\mathcal{F}_A 
          - X^A\bar{\mathcal{F}}_A\right)\ , \\ 
   K_\rho &= - \ln\, \hfuncY 
      = - \ln \ii \int_\Y \Omega\wedge\bar{\Omega} 
      = - \ln \ii\left(\bar{Z}^K F_K - Z^K \bar{F}_K\right)\ . 
\end{aligned}
\end{equation}
Note the natural scalar Hitchin functionals~\eqref{HYdef}, allow us to
define scalar K\"ahler potentials, as should appear in a
four-dimensional theory, rather than the six-forms $\ee^{-K_J}$ and
$\ee^{-K_\rho}$ that appear in the general ten-dimensional theory of
sec.~\ref{SKM}.

These expressions coincide with the K\"ahler potentials for the moduli
space of Calabi--Yau manifolds. This is a consequence of the fact that
the presence of torsion does not affect the kinetic terms for the
fields but only the superpotential as we already argued in
section~\ref{SKM}.  

Using~\eqref{Jexp},~\eqref{Omegaexp},~\eqref{domega}
and~\eqref{dalpha} we see that both $J$ and $\Omega$ are not closed
but obey  
\begin{equation}
\label{dJdOsum}
\begin{aligned}
   \dd\Omega &= 
      \left(Z^K e_{aK} + F_K m^K_a\right)\tilde\omega^a\ , \\
   \dd J &= 
      \IM t^a \left(m_a^K \alpha_K + e_{aL} \beta^L\right)\ , \\
   \dd (J\wedge J) &= 0\ ,   
\end{aligned}
\end{equation}
where the last equation is a direct consequence of the fact that there
are no one- or five-forms in the truncated
subspace. Comparing~\eqref{dJdOsum} with~\eqref{dJdOmega} we infer
that the torsion classes $W_1$, $W_2$ and $W_3$ can be non-trivial
while $W_4=W_5=0$. Of course, this is precisely due to the fact that
we are dropping all $\rep{3}$ and $\rep{\bar{3}}$ representations and
both $W_4$ and $W_5$ are triplets. 

{}From \eqref{dJdOsum} we see that the non-zero torsion
is parameterized by the (constant) matrices $e_{aK}$ and $m_a^K$. 
They can be chosen arbitrarily and only have to satisfy~\eqref{null}.
Ref.~\cite{GLMW} considered the special case $m_a^K =0 =e_{ak}$ or in
other words kept only $e_{a0}\neq0$. From \eqref{dJdOsum} we learn
that this implies $d\Omega=Z^0e_{a0}\tilde\omega^a$.
Put differently, for $m_a^K =0=e_{ak}$, $\Omega$ satisfies
additionally $\dd\,\IM\Omega =0$. This in turn implies that the torsion
class $W_1 \oplus W_2$ is real and such $SU(3)$ manifolds are called
half-flat~\cite{CS}.

Let us now turn to the Ramond--Ramond sector and discuss the truncation
of the ten-dimensional fields. As we will see local gauge invariance
gives a separate argument for the relations~\eqref{domega}
and~\eqref{dalpha}. Since the Ramond sector differs for type IIA and
type IIB we discuss both case in turn. Let us start with type IIA.

\subsection{The reduction of the type IIA RR-sector}
\label{IIAreduc}

The RR-sector of the ten-dimensional type IIA supergravity
contains a one-form $A_1$ and a three-form $A_3$.\footnote{As we
  already discussed in section \ref{N=2W}
there are commonly two different field basis for the $p$-form gauge
potentials used. They are related by $C_p=e^B A_p$ and as a consequence
the definition of their field strength differs. In type IIA it is
more convenient to use the $A$-basis.}  
Since we are projecting out the triplets we see from table~\ref{RRIIA} that 
$A_1$ only contains a singlet which again trivially decends to the
four-dimensional theory. This four-dimensional vector field 
is commonly denotes by $A_1^0$ since it is related to the graviphoton
in the gravitational multiplet.\footnote{Once additional vector
multiplets are present the graviphoton is only defined up to
symplectic rotations and thus, as we will see, 
$A_1^0$ is a component in a symplectic vector.}

The three-form gauge potential $A_3$ is expanded into the 
basis $\omega_a$ and $(\alpha_{K},\beta^{L})$ 
introduced in \eqref{sbasis} and \eqref{dualbasis} as
\begin{equation}
\label{Aexpansion}
\begin{aligned}
   A_3 &= A_1^a\wedge\omega_a + \xi^K\, \alpha_{K} 
      +\tilde\xi_L\, \beta^{L}\ . 
\end{aligned}
\end{equation}
As before the coefficents in this expansion correspond to dynamical fields
in the four-dimensional effective action. The $A_1^a$ denote
$b_J$ four-dimensional  vectors (or one-forms) while
$\xi^K, \tilde\xi_L$ are $2b_\rho+2$ scalars.
Together with the fields from the NS-sector discussed in the previous
section they assemble
into $N=2$ multiplets as shown in  table~\ref{spectrum}.
This table is the four-dimensional `effective' version of
table~\ref{N=2multipletsA}.
\begin{table}[h]
\begin{center}
\begin{tabular}{| c | c |c|} \hline
   \rule[-0.3cm]{0cm}{0.8cm} 
multiplet & multiplicity & bosonic field content\\ \hline  
\rule[-0.3cm]{0cm}{0.8cm} 
gravity multiplet  &1& {\small $(g_{\mu \nu},A_1^0)$}
   \\ \hline
   \rule[-0.3cm]{0cm}{0.8cm} vector multiplets & $b_J$ &
  {\small $(A_1^a, t^a)$}\\ \hline
   \rule[-0.3cm]{0cm}{0.8cm} hypermultiplets  & $b_\rho$&
   {\small $(z^k,\xi^k,\tilde\xi_k)$
}\\ \hline
\rule[-0.3cm]{0cm}{0.8cm} tensor multiplet &1&
   {\small $(B_2^{(4)}, \phi, \xi^0,\tilde\xi_0)$
}\\ \hline

\end{tabular}
\caption{\small 
\textit{ $N=2$ multiplets for Type IIA supergravity compactified on
  $\Y$.}}\label{spectrum}
\end{center}
\end{table}

The spectrum looks the same as the $N=2$ spectrum obtained
in Calabi--Yau compactifications \cite{BCF,LM}. The difference here is that
the expansion \eqref{Aexpansion}
is no longer in terms of harmonic forms on $\Y$ but instead in terms
of the forms which obey \eqref{domega}, \eqref{dalpha}. As a consequence 
the fields are no longer massless or in other words the
forms $\omega^a, \alpha_K, \beta^K, \tilde\omega^a$ are no longer
zero modes of the Laplace operator. Instead they are eigenvectors
of $\Delta$ with eigenvalues given by their masses. 

The field strength of the ten-dimensional 
gauge potentials $A_1, B, A_3$ are defined as
\beq\label{fstrength}
G_2 = \dd A_1\ , \qquad H = \dd  B\ , \qquad G_4 = \dd A_3 \ ,\qquad
F_4 = G_4 +  B\wedge G_2\ .
\eeq
Since we want to include background fluxes we split the field strengths
into an exact piece plus a flux term. Explicitly we have
\begin{equation}\label{fluxstrength}
\begin{aligned}
G_2 &= G_2^\flux + \dd  A_1\ , \qquad H = H^\flux 
+ \dd  B\ , \\
F_4 &=  G_4^\flux - H^\flux  \wedge   A_1 +   B \wedge G_2^\flux
 + \dd  A_3 +   B\wedge \dd  A_1\ .
\end{aligned}
\end{equation}

The background fluxes can also be expanded in the truncatd basis as
\beq\label{IIAfluxes}
G_2^\flux = m_{\rm RR}^a\, \omega_a \ ,\qquad
G_4^\flux = e_{{\rm RR}\,a}\, \tilde\omega^a \ ,\qquad
H^\flux = m_{0}^K \,\alpha_K + e_{{0}\,K}\,\beta^K \ ,
\eeq
which defines the RR-flux parameter $e_{{\rm RR}\,a}, m_{\rm RR}^a$
and the NS-flux parameter $e_{0\,K}, m_{0}^K $.
We have choosen this notation for the NS-fluxes since, as we will see, 
they naturally combine
with the torsion parameters $e_{a\,K}, m_{a}^K $ to form the matrices
\beq\label{defme}
e_{AK} = (e_{0\, K}, e_{aK})\ ,\qquad 
m_{A}^K = (m_0^K, m_{a}^{K})\ , \qquad A=0,\ldots, b_J\ .
\eeq
In analogy with \eqref{null} the fluxes \eqref{IIAfluxes} also have to satisfy 
a consistency condition. They should be choosen such that 
$\dd F^\flux = \dd^\dagger F^\flux =0$ for all fluxes.
This results in the conditions
\begin{equation}
\begin{aligned}
m_{\rm RR}^a\,  m_{a}^{K} = 0 = m_{\rm RR}^a\, e_{aK} \ , \qquad
e_{\rm RR}^a\, m_{a}^{K} = 0 = e_{\rm RR}^a\, e_{aK}\ ,\qquad
V_0 \cdot V_a = 0 \ ,
\end{aligned}
\end{equation}
where the symplectic vectors $V$ and their symplectic inner product are
defined in \eqref{null}.

Let us come back to the field strength \eqref{fluxstrength} and
discuss their four-dimensional gauge invariance. 
The four-dimensional theory has a standard gauge invariance associated
with the $(b_J + 1)$ Abelian gauge bosons $A^A_1 = (A_1^0,A_1^a)$ 
and a two-form gauge invariance associated with the four-dimensional
NS two-form $B_2^{(4)}$
\beq\label{4dgt}
A_1^A \to A_1^A+\dd\La^A\ ,\qquad B_2^{(4)} \to B_2^{(4)}+\dd\La_1 \ ,
\eeq
where $\La^A$ are scalar gauge parameters while $\La_1$ is an
independent
one-form gauge parameter.
{}From \eqref{fluxstrength} we see that both $G_2$ and $H$
are gauge invariant but,
at first sight, $F_4$ is not. Furthermore, also 
$\dd A_3$ is  naively not invariant under \eqref{4dgt}.
This can be seen by 
inserting the expansion \eqref{Aexpansion} into $\dd A_3$ which yields
\begin{equation}\label{FAexpansion}
\begin{aligned}
\dd A_3 &=  \dd A_1^a\wedge\omega_a + A_1^a\wedge\dd \omega_a
+ \dd\xi^K\wedge \alpha_{K} +\xi^K \dd\alpha_{K}
+ \dd\tilde\xi_L\wedge\beta^{L} +  \tilde\xi_L  \dd\beta^{L} \ ,\\
&= \dd A_1^a\wedge\omega_a 
+ (\dd\xi^K +  A_1^a\, m_a^K)\wedge\alpha_{K} 
+ \xi^K\,  \dd\alpha_{K} \\ & \hspace*{6cm}
+ (\dd\tilde\xi_L +  A_1^a  e_{aL})\wedge\beta^{L} 
+  \tilde\xi_L\,   \dd\beta^{L}\ ,
\end{aligned}
\end{equation}
where in the second equation we inserted \eqref{domega}.
The terms including $A_1^a$ violate the gauge invariance \eqref{4dgt}.

In order to recover gauge invariance we have to modify the
transformation laws. 
The two-form gauge invariance can be maintained by assigning
the transformations
\beq\label{Agauge}
A_1^a \to A_1^a - m_{\rm RR}^a \La_1
\eeq
to the vectors. This transformation implies  that one linear
combination  of vectors
is pure gauge or in other words this linear combination
can be `eaten' by the two-form 
$B_2^{(4)}$. As a consequence $B_2^{(4)}$ becomes massive by a
Stueckelberg-type mechanism as already observed in \cite{LM}.

The local one-form  gauge invariance of  $A_1^A$ can be recovered
by assigning  Peccei-Quinn type transformations
to the RR-scalars $\xi^K, \tilde\xi_L$
\beq\label{xigauge}
\xi^K \to \xi^K - \La^A m_A^K \ , \qquad 
\tilde\xi_L\to \tilde\xi_L - \La^A e_{AL}\ .
\eeq
In this case an appropriate fraction of the scalars  $\xi^K,\tilde\xi_L$
can be eaten by the gauge bosons or in other words $\xi^K$ and $\tilde\xi_L$
are the appropriate Goldstone bosons which render some of the 
$A_1^A$ massive.

Note that gauge invariance can only be maintained  if we impose
\eqref{domega} with 
$m_A^K$ and $e_{AL}$ being constant matrices.
We could repeat the same argument in the dual formulation 
of type IIA \cite{Bergshoeff} where instead of $A_3$ the dual gauge potential
$A_5$ appears. In this case gauge invariance leads to the constraints
\eqref{dalpha} together with $\dd \tilde\omega^a  = 0$ of \eqref{domega}.
Thus also from this point of view one can motivate the differential
relations \eqref{domega}, \eqref{dalpha} which is essentially the
argument given in \cite{DFTV}.

{}From \eqref{xigauge} we also see that the torsion
and the NS-fluxes precisely play the role of Killing vectors which gauge 
translational isometries of the RR scalars $\xi^K$ and $\tilde\xi_L$
in the hypermultiplet sector. Via the standard relations of gauged
supergravity \cite{N=2review} this induces a scalar potential as was worked out
explicitly, for example, in refs.\ \cite{LM,GLMW}. 
The magnetic RR-fluxes instead render the four-dimensional 
antisymmetric tensor $B_2^{(4)}$ massive and also induce a scalar 
potential \cite{LM}.\footnote{Although not visible form the
discussion, here let us add that the electric RR-fluxes play the role
of Green-Schwarz-type couplings~\cite{LM}.}

Before turning to the reduction of type IIB let us 
summarize the situation so far.
We truncated the ten-dimensional spectrum
by insisting that there are only two gravitinos in the gravitational
multiplet and a finite number of modes in the effective action. 
This corresponds to expanding the ten-dimensional fields in 
terms of basis defined in \eqref{sbasis} and \eqref{dualbasis}.
In addition we required local gauge invariance in the effective 
action which independently 
led to the conditions \eqref{domega} and \eqref{dalpha}.

The same analysis can be repeated for type IIB supergravity to which we
now turn. 

\subsection{The reduction of the type IIB RR-sector} \label{IIBreduc}

In IIB supergravity the RR sector contains a scalar $l\ (= C_0)$, a
two form  $C_2$ and a four-form $C_4$. 
Exactly as in the previous section these fields 
are expanded in terms of the finite basis defined in 
\eqref{sbasis} and \eqref{dualbasis} as follows
\begin{equation}\label{Bexpansion}
\begin{aligned}
C_2 &= C_2^{(4)} + c^a \,\omega_a\ ,  \\
C_4 &= D_2^a \wedge \omega_a + V^{K} \wedge
               \alpha_{K} - U_{K} \wedge \beta^{K} +
               \rho_a\, \tilde \omega^a\ .
\end{aligned}\end{equation}
{}From a four-dimensional point of view  $C_2^{(4)}, D_2^a$ are two-forms,
$ V^{K},  U_{K}$ are one-forms and 
$\rho_a$ are  scalars.
The field strength $F_5$ of $C_4$ is self-dual which 
eliminates half of the degrees of freedom
in the expansion of $C_4$. Conventionally one chooses to eleminate
$D_2^a$ and $U_K$ in favour of  $\rho_a$ and 
$V^K$.
Together with the fields from the NS sector all fields
assemble into the $N=2$ multiplets given in table~\ref{tableIIB}
which is the four-dimensional `effective' version of
\ref{N=2multipletsB}.\footnote{If we choose to eliminate the scalars  $\rho_a$
in favor of the four-dimensional tensors $D_2^a$ the $b_J$
hypermultiplets are replaced by $b_J$ tensor multiplets $(t^a,c^a,D_2^a)$.} 
\begin{table}[h]
\begin{center}
\begin{tabular}{| c | c |c|} \hline
   \rule[-0.3cm]{0cm}{0.8cm} 
multiplet & multiplicity & bosonic field content\\ \hline  
 \rule[-0.3cm]{0cm}{0.8cm}
gravity multiplet  & 1&{\small $(g_{\mu \nu},V^0)$} 
   \\ \hline
   \rule[-0.3cm]{0cm}{0.8cm} vector multiplets & $b_\rho$&
  {\small $(V^k, z^k)$}\\ \hline
   \rule[-0.3cm]{0cm}{0.8cm} hypermultiplets  & $b_J$ &
   {\small $(t^a,c^a,\rho_a)$ 
}\\ \hline
\rule[-0.3cm]{0cm}{0.8cm} double-tensor multiplet &1&
   {\small $(B_2^{(4)}, C_2^{(4)}, \phi, l)$
}\\ \hline
\end{tabular}
\caption{\small 
\textit{ $N=2$ multiplets for Type IIB supergravity compactified on
  $\Y$.}}
\label{tableIIB}
\end{center}
\end{table}

The field strengths of the ten-dimensional gauge potentials $C_2, B$ and
$C_4$ are defined as 
\begin{equation}\begin{aligned}\label{fieldstr}
  H &= \dd  B\ , \qquad 
 F_3\ =\ \dd  C_2 -  l \dd B\ ,\qquad
   F_5 = \dd  C_4 - H \wedge C_2 \ . 
\end{aligned}
\end{equation}
Separating again the fluxes from the exact piece
we arrive at
\begin{equation}\begin{aligned}\label{fieldflux}
  H &= H^\flux + \dd  B\ , \qquad 
 F_3\ =\ G_3^\flux -  l H^\flux + \dd C_2 -  l \dd B\ ,\\
   F_5 &= B \wedge  G_3^\flux - H^\flux \wedge C_2 
+\dd  C_4 - H \wedge C_2 \ . 
\end{aligned}
\end{equation}
As before the fluxes are expanded in terms of the truncated basis
as
\beq\label{IIBfluxes}
G_3^\flux = \mf_{\rm RR}^K \,\alpha_K + \ef_{{\rm RR}\,K}\,\beta^K \ ,\qquad
H^\flux = \mf_{0}^K \,\alpha_K + \ef_{{0}K}\,\beta^K \ ,
\eeq
where we use $\ef, \mf$ to denote the fluxes in type IIB.
In this case consistency requires $\dd F_3^\flux = \dd H^\flux = 0$
which translates into
\beq
V_0 \cdot V_a = 0 \ , \qquad V_{\rm RR}\cdot V_a = 0 \ ,
\eeq
where we defined the symplectic vector 
$V_{\rm RR} = (\ef_{{\rm RR}\, K}, \mf_{\rm RR}^K)$.

In type IIB the four-dimensional theory has a 
standard one-form gauge invariance associated
with the $(b_\rho + 1)$ Abelian gauge bosons $V^K$ and their
magnetic duals $U_K$. In addition there are  $(b_J + 2)$
two-form gauge transformations.\footnote{The issue of gauge invariance
  is best discused in terms of the tensor multiplets $(t^a,c^a,D_2^a)$
where the $\rho_a$
  are eliminated in favor of the four-dimensional tensors $D_2^a$ and
  thus the $b_J$ hypermultiplets are traded for $b_J$ tensor
  multiplets.} 
Together they read 
\beq\begin{aligned}\label{4dgtB}
V^K &\to V^K + \dd\La^K_V \ , \qquad U_K\to U_K + \dd\La_K^U\ ,\\
B_2^{(4)} &\to B_2^{(4)} + d\La^{(B)}_1\ , \qquad 
C_2^{(4)} \to C_2^{(4)} + d\La^0_1\ , \qquad 
D_2^a\to D_2^a + d\La^a_1 \ .
\end{aligned}
\eeq
{} From \eqref{fieldflux} we see that $H$ and $F_3$ are invariant
but $F_5$ is not. Exactly as in type IIA also $\dd C_4$ is not gauge invariant.
Inserting \eqref{Bexpansion} into $\dd C_4$  we arrive at
\begin{equation}
\label{FBexpansion}
\begin{aligned}
   \dd C_4 &= \dd D_2^a \wedge\omega_a + D_2^a\wedge\dd\omega_a 
      + \dd V^{K}\wedge\alpha_{K} - V^K\wedge \dd\alpha_K 
         \\ & \hspace*{2cm}
      - \dd U_K \wedge\beta^K + U_K \wedge\dd\beta^K
      + \dd\rho_a \wedge\tilde \omega^a\ + \rho_a \dd\tilde \omega^a \\
   &= \dd D_2^a \wedge\omega_a 
      + (D_2^a \mf_a^K +\dd V^{K})\wedge\alpha_{K}
      + (D_2^a \ef_{aK} +\dd U_K)\wedge\beta^K 
         \\ & \hspace*{2cm}
      + (\dd\rho_a - \ef_{aK} V^K - \mf_a^K U_K)\wedge\tilde \omega^a\ ,
\end{aligned}
\end{equation}
where in the second equation we inserted \eqref{domega}, \eqref{dalpha}.
The terms in the last line involving $V^K, U_K$ explicitly violate the
gauge invarianve \eqref{4dgtB}.

As in IIA, local gauge invariance necessitates \eqref{domega}  and
(\ref{dalpha}) in which case the transformation laws \eqref{4dgtB} 
can be modified according to
\begin{equation}\begin{aligned}
V^K & \to V^K + \mf_A^K\, \La^A_1\ , \qquad
U_K\to U_K + \ef_{AK}\La^A_1\ ,\\
V^K & \to V^K + \mf_{\rm RR}^K \La^{(B)}_1\ , \qquad
U_K\to U_K + \ef_{{\rm RR}\, K}\La^{(B)}_1\ ,\\
\rho_a &\to \rho_a + \ef_{aK}\La^K_V  + \mf_{a}^L\La_L^U \ .
\end{aligned}\end{equation}
With these transformations gauge invariance is recovered.
We see that for a non-vanishing magnetic matrix $\mf_A^K $ or a
non-vanishing magnetic RR-flux $\mf_{\rm RR}^K$
some of the electric gauge bosons $V^K$ become pure gauge
degrees of freedom which can be absorbed (eaten) into 
$C_2^{(4)}, B_2^{(4)}$ or $D_2^a$.
Put differently, these antisymmetric tensors
gain longitudinal degrees of freedom
by a Stueckelberg mechanism or in other words they become massive.
Note that on the type IIA side this only happens for magnetic RR-fluxes but not
for NS-fluxes or torsion. Instead in type IIB it occurs for all
magnetic fluxes and torsion. This situation is summarized in
table~\ref{results}. 

For electric fluxes some of the scalars $\rho_a$ become Goldstone
bosons and are eaten by the corresponding vectors while for  electric
RR-fluxes a Green-Schwarz type coupling is induced \cite{LM}. 
We will also see these facts in the supersymmetric
transformation law of the gravitinos to which we now turn.
\begin{table}[h]
\begin{center}
\begin{tabular}{| c | c |c|} \hline
   \rule[-0.3cm]{0cm}{0.8cm} 
& IIA& IIB \\ \hline  
 \rule[-0.3cm]{0cm}{0.8cm}
electric RR-flux $e_{\textrm{RR}}$ & Green--Schwarz coupling &Green-Schwarz coupling
   \\ \hline
   \rule[-0.3cm]{0cm}{0.8cm} 
magnetic  RR-flux $m_{\textrm{RR}}$ & massive tensor $B_2^{(4)}$ & massive tensor $B_2^{(4)}$\\ \hline
   \rule[-0.3cm]{0cm}{0.8cm} 
electric NS-flux $e_0$ & massive $A_1^0$ & massive $A_1^0$ \\ \hline
\rule[-0.3cm]{0cm}{0.8cm} 
magnetic NS-flux  $m_0$ &  massive $A_1^0$ &massive tensor $C_2^{(4)}$ \\
\hline  \rule[-0.3cm]{0cm}{0.8cm} 
electric torsion $e_{aK}$ & massive $A_1^a$ & massive $A_1^K$ \\ \hline
\rule[-0.3cm]{0cm}{0.8cm} 
magnetic torsion $m_a^K$ & massive $A_1^a$ & massive tensors $D_2^a$ \\ \hline
\end{tabular}
\caption{\small 
\textit{Effect of fluxes and torsion.}}
\label{results}
\end{center}
\end{table}


\subsection{Relation to gauged supergravity}\label{sugra}

In section~\ref{N=2W} we computed the transformation of the
gravitinos which reside in the gravitational multiplet.
{}From this we read off the matrix $S_{AB}$ and via \eqref{4dsusylaws}
the three Killing prepotentials ${\cal P}^x$ which
can be viewed as the $N=2$ version of the superpotential and the $D$-term.
The purpose of this section is to truncate these results to the finite
basis \eqref{sbasis} and \eqref{dualbasis} and then
compare  with the
formulas of gauged $N=2$ supergravity.

In section~\ref{N=2W} we gave $S^{(4)}_{AB}$ in two different forms.
In \eqref{SIIBshort} and \eqref{SIIAshort} we used a ten-dimensional
field basis while in  \eqref{SIIA} and \eqref{SIIB} we essentially
already used a four-dimensional
field basis. To be more precise we performed a Weyl rescaling in the
ten-dimensional action and introduced the four-dimensional dilaton
$\phi^{(4)}$ but kept the dependence on all ten space-time coordinates.
Now we merely need to truncate  \eqref{SIIA} and \eqref{SIIB}
(or equivalently \eqref{PIIA} and \eqref{PIIB}) to the finite-dimensional
subspace of light modes and integrate over the compact manifold $\Y$.

Using the truncation defined in \ref{defreduction}--\ref{IIBreduc}
we infer from \eqref{PIIA} for type IIA
\beq\label{PAint}
\begin{aligned}
(\cP^1 +i \cP^2) \
& = \ -2\, \ee^{\tfrac12 K_\rho+ \phi^{(4)}} \int_\Y
  \revmukai{\dd\pure^-}{\pure^+}\ ,\\
 \cP^3\ & =\ \frac{1}{\sqrt 2}\, \ee^{2\phi^{(4)}}\int_\Y
 \revmukai{G_\text{A}}{\pure^+} \ ,
\end{aligned}\eeq
where $\pure^+ = c\,\ee^{-t},\ \pure^- = \Omega$. 
Inserting the separation \eqref{fluxstrength} 
of $G_A$ into flux terms and field strength we arrive after partial
integration at 
\beq\label{PAfinal}
\begin{aligned}
(\cP^1 +i \cP^2) \
& =\ -2c\, \ee^{\tfrac12 K_\rho+ \phi^{(4)}} \int_\Y
\big(\dd t + H^\flux\big)\wedge\Omega\ ,\\
\cP^3\  & = \ \frac{c}{\sqrt 2}\, \ee^{2\phi^{(4)}}\int_\Y e^{-t} \wedge (G_{2n}^\flux + \dd A_{2n-1} - H^\flux\wedge A_{2n-3})   \ ,
\end{aligned}\eeq
where $t= B+\ii J$ and 
$e^{t} \wedge G_{2n}^\flux = \tfrac16 G_0^\flux t^3 + \tfrac12 G_2^\flux t^2
+ G_4^\flux t + G_6^\flux$. We can go one step further by
inserting the expansions \eqref{dJdOsum} and \eqref{IIAfluxes} 
into \eqref{PAfinal} and performing the integrals. This yields 
\beq\begin{aligned}\label{Pexp}
(\cP^1 + \ii \cP^2)\ & = \  -2\, \ee^{\tfrac12 K_\rho+ \phi^{(4)}} \,
X^A(e_{AK}Z^K + m_A^K F_K)\ ,\\
 \cP^3\ & =\ -  \frac{c}{\sqrt 2}\, \ee^{2\phi^{(4)}}
\big[X^A  (e_{AK}\xi^K  + m_A^K \tilde\xi_K)
+  (X^A e_{{\rm RR}\, A } + \cF_A m_{\rm RR}^A)\big]\ ,
\end{aligned}\eeq
where we used \eqref{defme} and $X^A=(c,-ct^a)$.
\eqref{Pexp} can be compared with the generic structure of $\cP^x$ as
dictated by $N=2$ supergravity. In this case one has the generic expression 
\cite{N=2review,DSV}
\begin{equation}\label{Px}
\cP^x = X^A P^x_A +\omega_\alpha^x (e^\alpha_A X^A - m^{\alpha A}
\cF_A)\ ,
\end{equation} 
where $P^x_A$ are the Killing prepotentials which only depend on
scalars in $N=2$
hypermultiplets. The second term in \eqref{Px} arises when tensor multiplets
are present and $e^\alpha_A$ denotes possible Green-Schwarz-type
couplings while $m^{\alpha A}$ are related to mass terms of
antisymmetric two-tensors. ($\omega_\alpha^x$ is an appropriate
$SU(2)$ connection on the hypermultiplet moduli space, see \cite{DSV}
for further details.)
Comparing \eqref{Pexp} and \eqref{Px} we see that they are  consist
with each other and furthermore 
for $m_{\rm RR}^A\neq 0$ we necessarily have massive
tensor fields as was already observed in section~\ref{IIAreduc}
and refs.~\cite{LM,DSV}.

We can repeat the same analysis for type IIB supergravity
starting from \eqref{PIIB} and obtain
\begin{equation}
\label{PfinalB}
\begin{aligned}
   (\cP^1 - \ii \cP^2) & = -2\, \ee^{\tfrac12 K_J+ \phi^{(4)}} \int_\Y
       \revmukai{\dd\pure^+}{\pure^-}
     = -2c\, \ee^{\tfrac12 K_J+ \phi^{(4)}} \int_\Y
        \left(\dd t + H^\flux\right)\wedge\Omega\ ,\\
     &= -2\, \ee^{\tfrac12 K_J+ \phi^{(4)}}
        \left(Z^K \ef_{AK} X^A + F_K \mf^K_A X^A\right)\ , \\
    \cP^3 & = \frac{1}{\sqrt 2}\, \ee^{2\phi^{(4)}}\int_\Y
       \revmukai{G_\text{B}}{\pure^-} 
     = \frac{1}{\sqrt 2}\, \ee^{2\phi^{(4)}}\int_\Y   
        \left(G_3^\flux - l H^\flux +\dd C_2 -l \dd B\right) \wedge \Omega \\
     &= \frac{1}{\sqrt 2}\, e^{2\phi^{(4)}} 
        \left(Z^K (\ef_{{\rm RR}\,K}- \ef_{AK}\xi^A) 
           + F_K (\mf_{\rm RR}^K  - \mf_A^K \xi^A)\right) \ ,
\end{aligned}
\end{equation}
where we defined
\begin{equation}\label{xidef}
\xi^A = (\xi^0, \xi^a)=(l,c^a - l b^a)\ ,\qquad  \ef_{AK} = (\ef_{0K}, \ef_{aK})\ ,\quad 
\mf_{A}^K = (\mf_0^K, \mf_{a}^{K})\ .
\end{equation}

The comparison with $N=2$ supergravity essentially uses again
\eqref{Px} 
but since
in type IIB the scalars $t^a$ and $z^k$ in vector and hypermultiplets
are interchanged compared to type IIA \eqref{Px} has be replaced by
\begin{equation}\label{PxIIB}
\cP^x = Z^K P^x_K +\omega_\alpha^x (e^\alpha_K Z^K - m^{\alpha K}F_K)\ .
\end{equation}

Comparing \eqref{PfinalB} with \eqref{PxIIB}
we see that for electric fluxes  $\ef_{{\rm RR}\,K}, \ef_{AK}$ we have a standard
gauged supergravity while for magnetic fluxes $\mf_{\rm RR}^K, \mf_A^K$
we have in addition massive antisymmetric tensors \cite{LM,DSV,DFTV}.
This was already observed at the end of section~\ref{IIBreduc} and is 
summarized in table~\ref{results}.


\subsection{Mirror symmetry}
\label{mirror}

Now that we have discussed the supersymmetry transformation
in terms of fluxes and torsion let us return to the issue 
of mirror symmetry. 
For Calabi--Yau manifolds the mirror conjectures states
that for a given Calabi--Yau $Y$ there exists a mirror manifold $\tilde
Y$, with the property that the Hodge numbers and K\"ahler and complex
structure deformations are exchanged in passing from $Y$ to
$\tilde{Y}$. In particular the moduli spaces $\cM_J$ for one manifold
and $\cM_\rho$ for the other (as well as their respective
prepotentials) are identified. In string theory mirror symmetry
manifests itself in the equivalence of type IIA compactified on $Y$
and type IIB compactified on the mirror manifold $\tilde Y$. In
particular this states that the respective low energy effective
Lagrangians are identical for compactifications on mirror manifolds.
In the notation used in the previous section this amounts to the
identification \cite{BGHL}
\beq\label{mirrorex}
X^A \leftrightarrow Z^K\ , \qquad
\cF^A \leftrightarrow F^K\ ,\qquad \xi^A \leftrightarrow \xi^K\ ,
\eeq
on mirror pairs.

Of course it is an interesting question to see what happens to this
symmetry for manifolds with $SU(3)$ structure and in particular 
to what extent the exchange \eqref{mirrorex} also holds 
on a generalized pair of mirror manifolds. 
It was proposed in \cite{FMT} that mirror symmetry for manifolds of
$SU(3)$ structure amounts to the exchange of the two pure spinors 
together with an exchange of even and odd RR-forms
\beq \label{genmirror}
e^{-B-\ii J}\ \leftrightarrow\ \Omega \ ,\qquad G_{\rm even}\ \leftrightarrow\
G_{\rm odd}\ .
\eeq
As we already discussed at the end of section~\ref{N=2W},
\eqref{genmirror} is a special case of the more general map 
(\ref{mirrorspinors})
which we expect to hold if instead of 
a $SU(3)$ structure one repeats the computation 
for a $\SU(3)\times\SU(3)$-structure.
Using the expansions \eqref{Phiexp}, \eqref{Omegaexp},
\eqref{Aexpansion}, \eqref{Bexpansion} and \eqref{xidef} one readily
verifies that  the map \eqref{genmirror} implies \eqref{mirrorex} if
no fluxes are turned on.

In Calabi--Yau
compactifications with only electric and magnetic RR-fluxes 
arising in \eqref{IIAfluxes} from the IIA RR-field
strength $G_{2n}^\flux$ and in \eqref{IIBfluxes} from IIB $G_3^\flux$,
it was indeed observed in refs.\ \cite{LM} that  at the level of the
effective action 
mirror symmetry is straightforwardly realized as the exchange
(\ref{mirrorex}) together with an exchange of the flux parameters 
\begin{equation}\label{RRmirror}
e_{RR} \leftrightarrow \ef_{RR}\ ,\qquad  m_{RR} \leftrightarrow \mf_{RR} \ .
\end{equation}
Furthermore for the case of RR fluxes on manifolds with $SU(3)$
structure, it was shown in \cite{GMPT} that the
supersymmetry equations are symmetric under the exchange
(\ref{genmirror}).  
This is also reflected in the Killing prepotentials derived in this
paper. From  Eqs.\ (\ref{PIIA}) and
(\ref{PIIB}) we see that the two (ten-dimensional) ${\cal P}^3$ 
are symmetric under (\ref{genmirror}). Similarly, Eqs.\
\eqref{PAfinal} and \eqref{PfinalB} show that in the truncated theory
the two $\cP^3$s are exchanged under the map (\ref{mirrorex}) together
with \eqref{RRmirror}, i.e. exactly in the same way as in Calabi--Yau
compactifications with RR fluxes. 

The situation for NS-fluxes is more involved. 
On the one hand, the expressions for the Killing prepotentials 
$\cP^1$ and $\cP^2$ in Eqs (\ref{PIIA}) and
(\ref{PIIB}), look perfectly mirror symmetric, i.e. respect the
exchange (\ref{mirrorspinors}). Note however that this exchange does
not map $H$-flux (which appears in $\dd \pure^+$) to itself, but
rather to torsion components in $\dd \pure^-$. Nevertheless, as we
mentioned at the end of section \ref{N=2W}, these expressions were
obtained for the particular case of a single $SU(3)$ structure, where
$\pure^-$ contains only a 3-form and not all odd forms. In this case,
from $\dd \pure^+$, the 3-form $\dd (B+iJ)$ contributes to the 
superpotential, while the five-form $\dd (B+iJ)^2$ does not,
since there is no 1-form in $\pure^-$.   

After performing the truncation, we 
can see explicitly by comparing \eqref{Pexp} and \eqref{PfinalB}
that $H^\flux$ is not mapped to itself. 
Indeed,
the type IIA NS-fluxes $e_{0K}, m^0_K$ arising from $H^\flux$ in
\eqref{IIAfluxes} and the type IIB NS-fluxes
$\ef_{0K}, \mf^0_K$ arising from $H^\flux$ in \eqref{IIBfluxes}
are not interchanged under 
mirror symmetry. Instead for 
electric fluxes they are mapped
to the torsion coefficients $e_{A0}, \ef_{A0}$ of half-flat manifolds
\cite{GLMW,GM,Tomasiello}. 
In fact this immediately generalizes 
for the entire electric matrices $e_{AK}, \ef_{KA}$ defined in
\eqref{defme} and \eqref{xidef}. We find perfect
agreement under mirror symmetry if we perform the map
\eqref{mirrorex} and simultaneously exchange the electric matrices 
\begin{equation}
e_{AK} \leftrightarrow \ef_{KA}\ .
\end{equation}

For the magnetic matrices $m^K_A, \mf^K_A$ no mirror symmetry is observed
in eqs.\ \eqref{Pexp} and \eqref{PfinalB}. 
In this case antisymmetric tensor fields become massive
on the IIB side while on the IIA side this is not the case.
As we already stated before we expect that such a contribution naturally arises when the
backgrounds with $SU(3)$ structure are replaced by the more general
backgrounds with $SU(3)\times SU(3)$ structure.
Work along these lines is in progress.


\section{Conclusions}\label{conclusions}


In this paper we  have shown that type II supergravities in
space-time backgrounds which admit an $SU(3)\times SU(3)$ structure
share many features with four-dimensional $N=2$ gauged supergravities.
The reason for this is that in such backgrounds the ten-dimensional
Lorentz symmetry $SO(1,9)$  can be replaced by a symmetry
$SO(1,3)\times SU(3)\times SU(3)$ and one can consistently write the
theory in terms of only eight out of the 32 original supercharges. 
Following the approach pioneered in~\cite{deWN}, this in turn allowed
us to rewrite the ten-dimensional theory in a form which strongly
resembles the four-dimensional $N=2$ gauged supergravities but without
the need of any Kaluza--Klein truncation. For simplicity, in many
parts of the paper we concentrated on the special situation where
there is a single $\SU(3)$ structure (or in other words where the two
$\SU(3)$ structures coincide). However, given the covariant form of
the final expressions we expect that they hold for general
$\SU(3)\times\SU(3)$ structures. 

In particular we showed, using results in~\cite{GCY}, that the
metric deformations together with $B$ can be viewed as coordinates of
a product of special K\"ahler manifolds. The corresponding K\"ahler
potentials can be expressed in terms of a pure spinor $\pure^\pm$ of
$\Spin(6,6)$ and its complex conjugate 
and they coincide with the corresponding Hitchin functionals. In the
$\SU(3)$ case the two pure spinors are $\pure^+=e^{-B-\ii J}$ and
$\pure^-=\Omega$ and are constructed as bispinors of the
$SU(3)$-singlet spinors $\eta_{\pm}$.

We also computed the supersymmetry transformation of the gravitinos
and determined the three Killing prepotential $\cP^x$ in the case of
type IIA and IIB. They are expressed (see eqns.~\eqref{PIIA}
and~\eqref{PIIB}) in terms of $\Spin(6,6)$ invariant inner products of
the pure spinors $\pure^\pm$ with the RR fluxes and with $\dd\pure^\mp$
which encode the intrinsic torsion and $H$-flux.

By further breaking the $SU(2)_R$ symmetry of the $N=2$ down to a
$U(1)_R$ , we were able to find the D-term and superpotential of an
$N=1$ theory. They depend on two angles that parameterize the $N=2\to
N=1$ reduction. In this way we obtained the most general $N=1$
superpotential for manifolds admitting an $SU(3)$ structure which
contains in particular all the cases that have been studied so far. 
The case of the heterotic string compactified on manifolds
with $\SU(3)$ structure will be presented in~\cite{Iman}.

An important caveat to the reformulation was that we explicitly
dropped the multiplets containing additional spin-$\frac32$
fields, present because that underlying theory really has $N=8$
supersymmetry. In particular, including these multiplets should modify
the scalar field moduli space. This is related to the fact that there
is actually no natural special K\"ahler geometry on the physical scalar
degrees of freedom, but only on the larger space of $\SU(3)\times\SU(3)$
structures. In the full formulation, it must be possible to gauge away
some of the degrees of freedom in $\pure^\pm$. Equally, there should
actually be an underlying $N=8$ type theory, where all the
supersymmetries are kept and the scalar fields parameterize a
$E_{7(7)}/\SU(8)$ coset. 

In the second part of the paper we demanded that the ten-dimensional
manifold had a product structure and performed a truncation of the
spectrum. This reduced theory was then shown to be consistent with a
four-dimensional $N=2$  gauged supergravity. In particular the gauged
isometries are translational isometries of the hypermultiplet and
tensor multiplet moduli space with gauge charges or mass parameters
given by the fluxes and the torsion. Electric fluxes (RR, NS or
torsion) give masses to the vectors coming from the three- or
four-form RR potential for IIA or IIB, resulting in a standard $N=2$
gauged supergravity, while magnetic fluxes give mass to the antisymmetric
tensors by a Stueckelberg-type mechanism.

The truncation of the spectrum done in section~\ref{d=4N=2} also
excluded spin-$\frac32$ multiplets which amounted to projecting out
$\rep{3}+ \rep{\bar 3}$ representations. This sets the torsion classes
${W}_4$ and ${W}_5$ to zero. Additionally, no warp-factor was
allowed. It turns out that allowing for these torsion classes or a  
warp factor is not straightforward. A first step in that direction was
taken  in refs.~\cite{GDW, deAlwis}, which considered
KK-reductions in warped products with a conformal Calabi--Yau factor.
In particular, it was claimed in \cite{GDW} that the warp factor
affects the $N=1$ K\"ahler potential, but not the superpotential. It
would be interesting to generalize our results allowing for such
warped compactifications.  

As we mentioned, for simplicity we mostly considered the case of a
single $\SU(3)$ structure instead of the more general situation of a
$\SU(3) \times \SU(3)$ structures.  
It should not be too difficult to generalize our results to the
generic situation. The main difference is that in this case there are
globally defined vectors (given by bilinears involving the two spinors
with one gamma matrix) which are nowhere vanishing if there is a
common $\SU(2)$ substructure.
We expect that the K\"ahler potential~\eqref{kahlerpot} and the
Killing prepotentials~\eqref{PIIA} and~\eqref{PIIB} take the same form
in this generalized set-up but are now evaluated with $\pure^\pm$
defining $\SU(3)\times\SU(3)$ structures. These formulas further
suggest that in the framework of $\SU(3)\times\SU(3)$ structures
mirror symmetry with fluxes is restored and the missing magnetic
fluxes can be located. We will return to these questions
elsewhere. 

One important question we have not addressed directly is the
connection between the formulation we have discussed here and
topological string theories. The target space theories of the
topological A and B models are theories of deformations of
complexified K\"ahler structure $B+\ii J$ or complex structure
$\rho$ called K\"ahler and Kodaira--Spencer gravity respectively. 
It has recently been argued~\cite{Dijkgraaf,GS,Nekrasov} that these
are equivalent to theories based on the Hitchin functionals
$H(\chi^+)$ and $H(\chi^-)$ where $\chi^+=\re(c\,\ee^{-B-\ii J})$ and
$\chi^-=\cscale\rho$. In ref.~\cite{PW} the equivalence was shown to
hold at one-loop for the B model, provided one considered the full 
functional~\eqref{phidef} for odd forms rather than that based
on~\eqref{q3form} restricted to three-forms. In these cases the
Hitchin functional is taken as the action of the theory. Crucially one
must also assume that $\rho$ and $B+\ii J$ are closed and one only
takes variations by exact forms. Hitchin's result~\cite{3form} is that
the equations of motion then imply that the relevant structure is
integrable. Here, considering the effective theory of the physical
string, we have seen that the Hitchin functionals naturally appear as
K\"ahler potentials without the restrictions that the forms are
closed. $\cN=2$ supersymmetric vacua require integrability of the 
two structures \cite{JW}, while $\cN=1$ vacua impose integrability
of only one of the two \cite{GMPT2}. This led \cite{GMPT2} to conjecture
that there is a topological model associated to any $\cN=1$ vacuum. 
It would be interesting to understand the connection between
the topological and the physical theories in more detail. 

Let us end by noting that the approach presented in this paper of
rewriting the ten-dimensional supergravities in a $\SU(3)\times\SU(3)$
background can equally well be applied to other theories and structure
groups. 
Of particular interest is eleven-dimensional supergravity
rewritten in a $\SO(1,3)\times G_2$ background. In this case the
ten-dimensional theory should resemble a four-dimensional $N=1$ theory
with a K\"ahler potential, a superpotential and D-terms. There is also 
a Hitchin functional for the three-form describing the $G_2$ and it is
natural to conjecture that this will give the exponential of the
K\"ahler potential. 

\vskip 1cm


\subsection*{Acknowledgments}

This work is supported by DFG -- The German Science Foundation, the
European RTN Programs HPRN-CT-2000-00148, HPRN-CT-2000-00122,
HPRN-CT-2000-00131, MRTN-CT-2004-005104, MRTN-CT-2004-503369 and the
DAAD -- the German Academic Exchange Service. The work of M.G.\ is
supported by European Commission Marie Curie Postdoctoral
Fellowship under contract number MEIF-CT-2003-501485. D.W.\ is supported
by a Royal Society University Research Fellowship. 

We want to thank especially Ruben Minasian and Alessandro Tomasiello
for collaboration in early stages of this work. We have greatly
benefited from conversations and correspondence with Thomas Grimm, 
Marco Gualtieri, Jan
Gutowski,  Chris Hull, Hans Jockers, Peter Mayr, Andrei Morianu, George
Papadopoulos, Michela Petrini, Sakura Sch\"afer-Nameki, Uwe
Semmelmann, Pierre Vanhove and Silvia Vaula. 

J.L.\ thanks E.\ Cremmer and the LPTENS\ in Paris for hospitality
and financial support during the inital stages of this work.


\end{document}